\documentclass{aa}  

\usepackage{graphicx}
\usepackage[colorlinks,urlcolor=blue,citecolor=blue,linkcolor=blue]{hyperref}
\usepackage{adjustbox}
\usepackage{subfig}
\usepackage{natbib}
\usepackage{hyperref}

\usepackage{siunitx}

\usepackage{txfonts}
\usepackage{pifont}
\usepackage{placeins}

\def\R23{\mbox{$\rm R_{23}$}}

\def\arcsec{\hbox{$^{\prime\prime}$}}

\newcommand{\HI}{\ion{H}{I}}
\DeclareSIUnit{\mas}{mas}
\DeclareSIUnit{\dex}{dex}

\usepackage{ulem}

\begin{document}

\title{MUSE-DARK}
\subtitle{~IV.  Comparing Dark Matter Densities to Star Formation Histories in 375 intermediate-$z$ galaxies}
\titlerunning{Comparing DM inner density slopes to SFHs }

\author{B. I. Ciocan\inst{1}
        \and N. F. Bouché\inst{1}
        \and J. Fensch\inst{1}
        \and D. Krajnović\inst{2}
        \and   J. Freundlich \inst{3}
        \and H. Desmond \inst{4}
         \and A. Jeanneau \inst{1}
        \and B. Famaey \inst{3}}

\institute{Université Lyon 1, ENS de Lyon, CNRS, CRAL, UMR 5574, Saint-Genis-Laval, France\\
\email{bianca-iulia.ciocan@univ-lyon1.fr}
\and
Leibniz-Institut für Astrophysik Potsdam (AIP), An der Sternwarte 16, 14482 Potsdam, Germany
\and
 Univ. de Strasbourg, CNRS, Observatoire astronomique de Strasbourg, 11 rue de l’Université, 67000, Strasbourg, France
   \and
    Institute of Cosmology \& Gravitation, University of Portsmouth, Dennis Sciama Building, Portsmouth, PO1 3FX, UK}

\date{}

\abstract{
\textit{Context:} 
In the $\Lambda$CDM paradigm,  the inner shape of baryonless dark matter (DM) haloes is expected to be steep (cuspy), whereas flat profiles (cores) are often observed, and baryonic feedback is expected to reconcile observations with this expectation by dynamically heating DM profiles into cored density distributions.
However,  the link between feedback and DM profiles remains limited. \\
\textit{Aims:} We aim to investigate the DM halo properties of a sample of intermediate-$z$ star-forming galaxies in order to investigate and quantify the connection between halo inner slopes
and baryonic assembly histories.\\
\textit{Methods:}  We analyse a large sample of 375 galaxies at $0.1 < z < 1.43$ spanning  $7 < \log(M_{\star}/M_{\odot}) < 11.2$ using 3D forward modelling of MUSE kinematics and disk–halo decompositions with mass-dependent $\alpha\beta\gamma$ (DC14-based) and generalised  Navarro-Frenk-White (gNFW)  halo profiles.  Stellar population parameters are inferred from multi-band Spectral Energy Distribution fitting of UV-to-MIR photometry using a non-parametric star formation history (SFH)  model. \\
\textit{Results:} Bayesian model comparison shows that the mass-dependent $\alpha\beta\gamma$ parametrisation (DC14-based) of halo density profiles is statistically preferred over generalised  Navarro-Frenk-White profiles (gNFW) for a substantial fraction of the sample. The inferred inner slopes from gNFW fits nevertheless follow the DC14 expectation and reproduce the expected dependence on the stellar-to-halo mass ratio, despite this not being explicitly imposed. Partial correlation coefficient and random forest regression analyses consistently identify stellar mass (or, equivalently, the stellar-to-halo mass ratio) as the primary driver of the inner density slope, 
a result that holds robustly even when the analysis is restricted to the gNFW fits. In contrast, SFH-related quantities (e.g. formation time and 'burstiness' indicators) exhibit no independent predictive power on the DM halo inner slope once the dependence on stellar mass is accounted for. Cored systems preferentially occupy lower-mass, more bursty bins with later assembly times, whereas cuspy systems are associated with higher stellar masses, lower star-formation activity, and earlier assembly epochs, in agreement with theoretical expectations of stellar-feedback-induced cores. \\
\textit{Conclusions:} The inner slope of DM haloes in the MUSE-DARK sample is primarily set by the baryon-to-halo mass scale, with no detectable independent dependence on SFH diagnostics, once this scaling is accounted for.  }

\maketitle



\setcounter{section}{0}
\section{Introduction}
Dark matter (DM) is a fundamental component of the current cosmological paradigm and dominates the matter content of the Universe. One of the earliest and most direct pieces of evidence for DM came from galaxy rotation curves (RCs), which remain approximately flat at large galactocentric radii (\citealt{vdh}, \citealt{Rubin}, \citealt{1974ApJ...193..309R}, \citealt{1975ApJ...201..327R}, \citealt{1978PhDT.......195B}), in contrast to the Keplerian decline expected from the observed baryonic matter alone.

Within the $\Lambda$ cold dark matter ($\Lambda$CDM) framework, DM halos provide the scaffolding for structure formation, with this model successfully reproducing the observed large-scale distribution of matter and its evolution with cosmic time. However, tensions between theoretical predictions and observations persist on sub-Mpc scales (\citealt{bull17}).

In particular, DM-only simulations predict halos with steep central density cusps, typically described by $\rho(r)\propto r^{-\gamma}$ with $\gamma \sim 0.8$--$1.4$, and well characterised by a Navarro–Frenk–White (NFW) profile (\citealt{nfw}, \citealt{Klypin}). In contrast, it has long been known that observations of many low surface brightness galaxies indicate significantly shallower inner density profiles, consistent with constant-density cores ($\gamma \sim 0$--$0.5$; e.g. \citealt{fp}, \citealt{sal}, \citealt{Weldrake}, \citealt{sal2}, \citealt{oh}). This long-standing discrepancy is known as the core--cusp problem \citep{debok10}.

A range of baryonic mechanisms have been proposed to reconcile observations with expectations within $\Lambda$CDM, most notably, rapid gas removal due to feedback processes. Numerical and theoretical work has shown that repeated, bursty episodes and stellar feedback can drive large-scale outflows that lead to fluctuations in the gravitational potential. These fluctuations dynamically heat up the DM halo and lead to the formation of central density cores (\citealt{nav}, \citealt{rg}, \citealt{gov}, \citealt{pon}, \citealt{tey}, \citealt{br}, \citealt{2016MNRAS.461.2658D}, \citealt{2016MNRAS.461.1745E}, \citealt{free}, \citealt{lifre}, \citealt{Jackson}, \citealt{aza}).

In this framework, the efficiency of core formation is expected to depend strongly on the stellar-to-halo mass ratio, $M_{\star}/M_{\rm halo}$. Hydrodynamical simulations suggest that core formation is maximised at $M_{\star}/M_{\rm halo} \sim 3$--$5 \times 10^{-3}$, corresponding to $10^9 \lesssim M_{\star}/M_{\odot} \lesssim 10^{10}$, while both lower- and higher-mass systems tend to retain cuspy profiles due to insufficient feedback energy or deeper potential wells, respectively (\citealt{dc14}; \citealt{read}, \citealt{Lazar}, \citealt{aza}). This relation appears to hold in the NIHAO simulations, at least approximately, up to $z \sim 2$ \citep{toll}, although it is not universally reproduced across many other hydrodynamical simulations (e.g. \citealt{Bose}). 

More recently,   \cite{Sarrato} investigated the connection between star formation histories (SFHs) and the inner DM density profiles of simulated galaxies from the NIHAO and FIRE-2, finding that the burstiness and duration of the SFH can explain the scatter in the relation between the inner DM slopes and stellar-to-halo mass ratio. Within this framework, the authors find that cored galaxies in the bright dwarf regime tend to be more bursty and have a longer SFH duration than their more cuspy counterparts. Similarly, \cite{nu} studied the DM density profiles of galaxies from the Mochima suite of cosmological zoom-in simulations, finding that galaxies that form most of their stars early tend to retain steep cusps, whereas those with more extended or recent star formation (SF) develop oscillating inner slopes driven by bursty feedback and tidal perturbations.

Indeed, both observations and simulations suggest that lower-mass galaxies tend to exhibit more stochastic and bursty SFHs than their massive counterparts. Studies based on nebular-to-UV SF indicators and spectral energy distribution (SED) fitting have shown that low-mass galaxies typically experience more time-variable SF and assemble their stellar mass over more extended timescales \citep[e.g.][]{2008ApJ...675..234P, Weisz, Guo, 2019ApJ...881...71E, Mehta, ml}. Cosmological simulations and empirical models similarly predict increasingly bursty SF toward lower stellar masses as a consequence of strong stellar feedback and rapid gas cycling \citep[e.g.][]{Neistein, Sparre, Faucher, Iyer, 2020MNRAS.497..698T}. 

RCs remain one of the most direct probes of the mass distribution in star-forming galaxies (SFGs), providing key constraints on the inner structure of DM halos \citep{Rubin2}. Observational evidence for cored DM density profiles has now been reported both in the local Universe \citep[e.g.][]{katz, alle, li20}, and out to $z \sim 2$ \citep[e.g.][]{genz, Sharma2, shach, nicolas, ciocan}. However, establishing a direct connection between core formation and the underlying baryonic processes remains challenging. In particular, only a limited number of studies have attempted to link DM cores to bursty SFHs (\citealt{read3}) or to elevated star formation rates (SFRs, \citealt{nicolas}), and current constraints remain hampered by small sample sizes.

For example, \cite{read3} combine stellar kinematics and H\,\textsc{i} RCs to infer the inner DM density slopes of a sample of 16 local dwarf spheroidal and dwarf irregular galaxies spanning a wide range of SFHs. They find that systems which ceased SF more than $\sim$6 Gyr ago are consistent with cuspy DM profiles. In contrast, galaxies with more extended and sustained SFHs preferentially exhibit constant-density cores. 


Overall, while some studies point toward a potential connection between SF activity and DM halo structure, a robust observational link between inner density slopes and detailed SFHs remains elusive. With this fourth project from the MUSE-DARK series (see also \citealt{ciocan, ciocan2,Jeanneau}), we aim to address this knowledge gap by investigating the connection between the DM density profiles and SFHs in a large statistical sample of 375 SFGs at $0.1 < z < 1.43$, spanning a wide stellar mass range ($7< \log(M_{\star}/M_{\odot}) < 11.2$).  We apply the disk--halo decomposition methodology introduced in \citet{nicolas} and validated in \cite{ciocan} to a substantially larger sample, enabling a statistically robust exploration of (1) DM scaling relations and (2) any correlation between DM inner slopes, $\gamma$, and SFHs. Previous observational studies on DM halo properties at intermediate-to-high $z$ have been limited to samples typically comprising $\leq100-150$ galaxies (e.g. \citealt{shach, ciocan, Danhaive}), as opposed to the large samples that will become available at $z=0$ (e.g. 4000 galaxies in BIG-SPARC, \citealt{2026IAUS..392..119H}). The sample assembled here therefore represents one of the largest samples to date with disk--halo decompositions at $z\sim1$, providing an unprecedented opportunity to statistically constrain the interplay between galaxy evolution and DM halo structure.

This paper is organised as follows. In Sect.~\ref{data}, we present the spectroscopic and photometric data used in this work, as well as the sample selection. In Sect.~\ref{methods}, we describe the methodology adopted for modelling the kinematics and performing the 3D disk--halo decomposition, while Sect.~\ref{SED} presents the SED-fitting procedure. We present our main results in Sect.~\ref{results} and discuss the implications of our results, as well as the limitations of our modelling in Sect.~\ref{discussion}. Finally, in Sect.~\ref{concl}, we summarise our main findings and present our conclusions.

Throughout this paper, we adopt a \cite{Planck} cosmology with $\Omega_M = 0.307$, $\Omega_\Lambda = 0.693$, and $H_0 = 67.7\,\mathrm{km\,s^{-1}\,Mpc^{-1}}$, for which 1\arcsec\ corresponds to $\sim$8.23 kpc at $z=1$. All logarithms are base 10. Stellar masses and SFRs are derived assuming a \cite{chabrier03} initial mass function (IMF).


\section{Data}
\label{data}

\subsection{MUSE spectroscopic observations}
This work combines three complementary VLT/MUSE \citep{bacon10} datasets: (i) the MUSE Hubble Ultra Deep Field survey (hereafter MHUDF; \citealt{udf1,udf2}); (ii) the MUSE Cosmic Assembly Survey Targeting Extragalactic Legacy fields (hereafter MUSCATEL; PIs: L. Wisotzki, R. Bacon, T. Contini); and (iii) the MUSE Atlas of Lensing Cluster survey (hereafter Lensing Clusters; \citealt{Richard, Claeyssens}), which are briefly described below:

(i)  MHUDF comprises three datasets: (1) a $3\times 3\: \mathrm{arcmin}^2$ mosaic of nine MUSE fields with a typical depth of 10 hours (MOSAIC); (2) a single $1\times 1\:\mathrm{arcmin}^2$ pointing reaching 31 hours of exposure (UDF10); and (3) the adaptive optics-assisted \citep{AO} MUSE eXtremely Deep Field (MXDF), with a total integration time of 140 hours. This compilation includes the full sample analysed in the first and third projects of the MUSE-DARK series (\citealt{ciocan,ciocan2}); we refer the reader to those works for further details.

(ii) MUSCATEL consists of parallel-field observations of four Hubble Frontier Fields clusters (HFF; \citealt{Lotz}), accessible from the VLT: Abell 2744, MACS J0416.1-2403, Abell S1063, and Abell 370. The survey follows a tiered-depth strategy. The wide and shallow tier covers a $3'\times 3'$ mosaic with 100-minute exposures, while the intermediate (“medium”) tier spans a $2'\times 2'$ mosaic with 5-hour integrations. The deepest tier consists of a single $1'\times 1'$ MUSE pointing centred on the medium field, reaching a total exposure time of 25 hours. All observations were obtained with the VLT adaptive optics facility, except for one shallow-field pointing in Abell S1063. A full description of the data reduction, redshift measurements, and catalogue construction will be presented in a dedicated MUSCATEL survey paper (Urrutia et al., in prep.), see also \cite{Pessa} for more details.

(iii) Lensing Clusters is a legacy programme targeting the central regions of 19 massive galaxy clusters with MUSE. In this work, we use the subset of observations targeting the four HFF clusters (Abell 2744, MACS J0416.1-2403, Abell S1063, and Abell 370), focusing on the cluster foreground galaxies.  The final data cubes combine exposures obtained in both standard and adaptive optics modes, yielding effective integration times ranging from 2 to 15 hours per pointing.  We note that \cite{Jeanneau} (in MUSE-DARK-II) performed a morpho-kinematic analysis using the same data set; however, they focused on the background lensed galaxies, whereas here we analyse only galaxies with $z<z_{\rm cluster}$. 

The point spread functions (PSFs) of all three data sets are modelled as circular Moffat profiles, with a median full width at half maximum (FWHM) of $\sim 0.65''$ at $\lambda \sim 7000$--7050\,\AA. More details are given in \cite{udf2, Richard} and Urrutia et al., (in prep.).

We started from science-ready MUSE data cubes and processed all derived products in a homogeneous manner. For each galaxy, we extracted sub-cubes centred on the emission line of interest using the spectroscopic redshift, adopting a typical spectral window of $\sim 30$\,\AA, and subtracted the underlying continuum. In crowded regions, we applied spatial masking to minimise contamination from neighbouring sources. We then measured emission-line properties in each spaxel using the 2D line-fitting code \texttt{CAMEL} \citep{camel}, which yielded resolved flux, signal-to-noise ratio, velocity, and velocity dispersion maps. Finally, we forward-modelled the kinematics with \textsc{GalPaK$^{\rm 3D}$} \citep{galpak}, see Sect. \ref{methods}.

\subsection{Ancillary photometric data}
\label{phot}
In addition to the MUSE spectroscopic observations, this work also used the wealth of photometric observations targeting the Hubble Ultra Deep Field (UDF) and the FF clusters. 

\subsubsection*{Photometric catalogues}
For the MHUDF dataset, we exploited the extensive ancillary data available in the UDF, in particular the JADES photometric catalogues \citep{Eisenstein, Robertson, Johnson}. These provide homogenised multi-wavelength photometry in 23 broad- and medium-band filters spanning the ultraviolet to mid-infrared (UV--MIR), combining deep HST and JWST imaging. The dataset includes JWST/NIRCam observations (F090W, F115W, F150W, F182M, F200W, F210M, F277W, F335M, F356W, F410M, F430M, F444W, F460M, F480M), complemented by HST/ACS optical bands (F435W, F606W, F775W, F814W, F850LP) and HST/WFC3 near-infrared bands (F105W, F125W, F140W, F160W).

For the MUSCATEL and lensing-cluster fields, we relied on the HFF-DeepSpace catalogues \citep{ship}, which combine HST/ACS (F435W, F606W, F814W) and HST/WFC3 (F105W, F125W, F140W, F160W) imaging with ground-based $K_s$-band data from the KIFF survey \citep{Brammer}, as well as Spitzer/IRAC observations at 3.6 and 4.5~$\mu$m (and at 5.8 and 8.0~$\mu$m where available). These datasets provide multi-wavelength coverage from the optical to the MIR in up to $\sim$17 filters. We use these catalogues to infer stellar population parameters and SFHs via SED fitting, see Sect.~\ref{SED}.

\subsubsection*{Morphology}
For the MHUDF data set, we used the F160W morphological analysis of \cite{morph}, performed with \texttt{GALFIT} \citep{galfit}, which provides measurements of the half-light radius ($R_{\rm e}$), Sérsic index ($n$), axis ratio ($b/a$), and position angle (PA). We adopt the F160W-based structural parameters, as this band corresponds to the reddest available HST filter and therefore best traces the underlying stellar mass distribution.

For the MUSCATEL and lensing-cluster fields, we used the morphological measurements of \cite{Nedkova}, derived using the multi-wavelength fitting tools \texttt{GalfitM} and \texttt{Galapagos-2}, developed within the MegaMorph project \citep{Hausler}. As for the MHUDF sample, we adopt the structural parameters obtained in the F160W band to ensure consistency across the data sets, and use the inclinations from these catalogues as priors in our dynamical modelling (Sect. \ref{methods}). 


\subsection{Sample of intermediate-$z$ SFGs}
\label{sample}

We select a sample of intermediate-redshift SFGs following the criteria described in \citet{ciocan}, and provide details on the selection criteria and quality cuts in Appendix~\ref{app:snr}. Briefly, we select SFGs with spatially resolved emission-line kinematics suitable for disk--halo decomposition, with high signal-to-noise optical emission-line detections, sufficient spatial resolution (see Fig.~\ref{snr}), inclinations $i>30^{\circ}$,  and rotation-dominated velocity fields ($v_{\rm max}/\sigma>1$).  

The final  sample consists of 375 galaxies spanning $0.1<z<1.43$ and $7<\log(M_{\star}/M_{\odot})<11.2$ (Fig.~\ref{mstar_z}). The sample is considered to be 'mass-complete', i.e., having the same minimum $M_\star$ above $\log(M_{\star}/M_{\odot})=8.8$ over the redshift range probed. 

\section{Dynamical modelling}
\label{methods}

We analysed the galaxy kinematics using the 3D forward-modelling code \textsc{GalPaK$^{\rm 3D}$} \citep{galpak,nicolas1}, following the same methodology introduced in \citet{nicolas} and extensively validated in \citet{ciocan}. 

First, we performed morpho-kinematic modelling (Sect.~\ref{gpk1}) to select rotation-dominated SFGs with $v_{\rm max}/\sigma > 1$, well suited for  RC decomposition (Sect. \ref{sample}). Second, we carried out a disk--halo decomposition using the approach described in \cite{nicolas} and \cite{ciocan} (Sect.~\ref{gpk2}). Within our forward modelling framework, we account for observational effects by convolving the 3D disk models with a circular Moffat PSF and a Gaussian line spread function (LSF) model from \citet{udf1}.  

Throughout this work, we followed the modelling assumptions, priors, and fitting strategy introduced in \citet{ciocan}, unless otherwise specified. A detailed description of the methodology, together with the associated consistency checks and validation tests, is provided in that work. We briefly summarise the main steps of the analysis below.

\subsection{3D morpho-kinematic modelling}
\label{gpk1}

The ionised gas emission was modelled as a 3D rotating disk with a Sérsic surface brightness profile, $\Sigma(r)$, characterised by a Sérsic index $n_{\rm gas}$ and effective radius $R_{\rm e}$. The disk is allowed to vary in inclination, $i$, and position angle, PA, while its vertical structure is described by a Gaussian profile with scale height $h_{\rm z} = 0.15\,R_{\rm e}$ \citep{Genzel06,cresci09}.

The velocity profile, $v_{\rm c}(r)$,  was parametrised by the universal rotation curve (URC; Eq. 8 from \citealt{persic}),  
while the dispersion profile, $\sigma(r)$, consists of the combination of a thick disk $\sigma_{\rm{thick}}$, defined as $\sigma_{\rm{thick}}(r)/v_c(r)=h_{\rm{z}}/r$, and a dispersion floor, $\rm{\sigma_{0}}$, added in quadrature (for more details see \citealt{galpak}).  

As demonstrated in \cite{galpak,nicolas1,ciocan}, this 3D forward-modelling approach enables robust recovery of rotation curves out to $\sim2$--$3$ effective radii by fully exploiting the information contained in the datacube, including low-S/N spaxels. The resulting modelled velocity fields are in  good agreement with the observed kinematics.

\subsection{3D disk--halo decomposition}
\label{gpk2}

We decomposed the circular velocity as:
\begin{equation}
 v_{\rm{c}}^2(r) = v_{\rm{DM}}^2(r) + v_{\rm{disk}}^2(r) + v_{\rm{\HI}}(r) |v_{\rm{\HI}}(r)|,  \label{eq:vdecomp}
  \end{equation} and corrected for pressure support via $v_{\rm c}^2=v_{\perp}^2+v_{\rm AD}^2$, with $v_{\rm AD}$ following the turbulent disk approximation of \citet{Dalcanton}. We adopted the same parametric models for the disk and HI component as in \citet{ciocan} and defer the reader to that work for more details.

 For the DM component, $v_{\rm DM}(r)$, we used three different halo density profiles, which are detailed below:
 
 (1) DC14: As in \cite{ciocan}, we adopted the mass-dependent generalised $\alpha$--$\beta$--$\gamma$ profile \citep{Hernquist,zhao}, calibrated on hydrodynamical MaGICC simulations spanning dwarf- to Milky-Way-mass galaxies by \cite{dc14}:
\begin{equation}
\rho(r)=\frac{\rho_{\rm s}}{(r/r_{\rm s})^{\gamma}\left[1+(r/r_{\rm s})^{\alpha}\right]^{(\beta-\gamma)/\alpha}}.
\end{equation}
Here, $r_{\rm s}$ and $\rho_{\rm s}$ are the scale radius and density, while $\alpha$, $\beta$, and $\gamma$ control the transition, outer slope, and inner slope, respectively. The DC14 model links these shape parameters to $X=\log(M_{\star}/M_{\rm halo})$, through the simulation-calibrated relations:
\begin{eqnarray} \alpha &=& 2.94 - \log[(10^{X+2.33})^{-1.08} + (10^{X+2.33})^{2.29}] , \label{alpha}\\ \beta &=& 4.23 + 1.34X + 0.26X^{2}, \label{beta}\\ \gamma &=& -0.06 + \log[(10^{X+2.56})^{-0.68} + 10^{X+2.56}]. \label{gammaeq} \end{eqnarray}
We adopted these relations in our modelling and restricted $X\in[-3.0,-1.2]$, to remain within the model's well-defined core--cusp regime.

(2) DC14-scatter:  To account for potential intrinsic deviations from the mean DC14 mapping, we introduced an extended model in which stochastic scatter is allowed in the inner density slope. While in the fiducial DC14 prescription described above, the parameters $\alpha, \beta, \gamma$ are deterministic functions of  $X$, here we allow the inner slope $\gamma$ to vary at fixed $X$. This is motivated by hydrodynamical simulations, which show that feedback-driven core formation exhibits significant halo-to-halo variation at fixed $X$, particularly in the inner slope $\gamma$ \citep[e.g.][]{toll,Sarrato}. We therefore defined a modified model in which  the inner slope, defined in Eq. \ref{gammaeq} is allowed to fluctuate: $
\gamma(X) \rightarrow \gamma(X) + \delta_s,$  where $\delta_s$ represents an additional intrinsic scatter parameter we fitted for. Our model assumes $\delta_s \in [-0.3, 0.3]$, following \cite{Sarrato}, who found a $1\sigma$ scatter of $\sim 0.25-0.27$ dex in the $\gamma(X)$ relation for NIHAO and FIRE-2 simulated galaxies, respectively. All other parameters retain the fiducial DC14 parametrisation, so that $\delta_s$ is the only additional free parameter introduced in the fit for each galaxy.

(3) gNFW:  Following \cite{genz}, \cite{genz2}, and \cite{shach}, we also tested the generalised Navarro–Frenk–White (gNFW) halo profile in our modelling. The classical \cite{nfw} profile is defined as a special case of the general $\alpha$–$\beta$–$\gamma$ model   (e.g. \citealt{Hernquist}, \citealt{zhao}) with $(\alpha, \beta, \gamma) = (1, 3, 1)$. In contrast, the gNFW profile fixes $(\alpha, \beta) = (1, 3)$ while allowing the inner slope $\gamma$ to vary freely, without imposing any mass dependence, as opposed to  DC14, which links $\alpha$, $\beta$ and $\gamma$ to  $M_\star/M_{\rm halo}$. The resulting density profile is given by:

\begin{equation}
\rho(r)=\frac{\rho_{\rm s}}{(r/r_{\rm s})^{\gamma}\left[1+(r/r_{\rm s})^{3-\gamma}\right]}.
\end{equation}

We note that for the 126 galaxies previously analysed with the URC and DC14 models in \citet{ciocan}, we adopted their published results. For the remaining galaxies, we performed the URC and DC14 fits following the same modelling procedure. The DC14-scatter and gNFW models are newly fitted in this work, and the fits were applied consistently to the full sample (described in Sect.~\ref{sample}).

\subsection{Model parameters and priors}
\label{gpkparamssect}

Following \citet{ciocan}, we adopted broad, uniform priors for most model parameters, except for the inclination, stellar mass, and virial velocity. For the inclinations, we assigned flat priors centred on the values derived from the HST Sérsic fits (Sect.~\ref{phot}), adopting $i = i_{\rm HST} \pm 5^\circ$ for all models. For the gNFW model, we constrained the stellar masses using flat priors based on the SED-fitting results (Sect.~\ref{SED}), $\log(M_\star/M_\odot)=\log(M_\star/M_\odot)_{\rm SED}\pm0.15$. In contrast, we imposed no prior on $M_\star$ for the DC14 and DC14-scatter models, since the baryon--halo coupling is already encoded through the parameter $X$, which helps alleviate the disk--halo degeneracy. 
\footnote{As shown by \citet{ciocan} and confirmed in this work, the stellar masses inferred dynamically with the DC14 model are in good agreement with the SED-based estimates. We note that within the DC14 framework, the disk-halo degeneracy is mitigated by fitting for $\log(M_\star/M_{\rm halo}$) together with the virial velocity, which allows the stellar mass to be constrained directly by the kinematics.}
For the gNFW fits, we additionally imposed the stellar-to-halo mass relation (SHMR) of \citet{cosmos2020} as a prior. The mean relation is given by their Eq.~9:
$\log(M_{\rm halo})=
\log(M_1)
+
\beta \log\left(\frac{M_\star}{M_{\star,0}}\right)
+
\frac{\left(M_\star/M_{\star,0}\right)^{\delta}}
{1+\left(M_\star/M_{\star,0}\right)^{-\gamma}}
-\frac{1}{2},$ and we adopt the best-fitting parameters from their Table~F1 for the appropriate redshift bins ($0.2<z<1.5$). We account for the intrinsic scatter of the SHMR, $\sigma_{\log M_{\rm halo}}$, which ranges from $\sim0.16$ to $0.29$ dex across the redshift bins, by assuming a Gaussian distribution in $\log M_{\rm halo}$ around the mean relation. The prior is evaluated at the redshift and stellar mass of each galaxy, with the uncertainty in $M_\star$ propagated through the relation. Thus, the resulting prior on $M_{\rm halo}$ accounts for both the measurement uncertainty in $M_\star$ and the intrinsic scatter of the SHMR. Since $M_{\rm vir}$ is parameterised through $v_{\rm vir}$ in our gNFW models, this translates into an effective prior on the virial velocity.
\footnote{When the fits are performed without SMHM-priors, both the DC14 and DC14-scatter models recover stellar and halo masses consistent with the expected SMHM relations \citep{Behroozi, cosmos2020}. In contrast, owing to its greater flexibility, the gNFW parametrisation can admit solutions that deviate substantially from this relation (in some cases $M_{\rm vir}<M_\star$), motivating the inclusion of $\Lambda$CDM-based priors, as is commonly done in low-redshift disk--halo decompositions (e.g. \citealt{alle,katz}). We note that the inferred DM inner density slopes remain consistent between the gNFW fits with and without SMHM priors for $\sim80\%$ of the sample.} 
The full set of parameters and priors are summarised in Appendix \ref{par}, Table~\ref{gpkparams}.

We optimised the best model parameters using the multimodal nested sampling algorithm \texttt{pyMultiNest} \citep{feroz,Buchner}. For the fits, we adopted 400 live points, a sampling efficiency of 0.8, and an evidence tolerance of 0.5, ensuring robust and efficient posterior sampling.

Parameter estimates were then taken as the medians of the marginalised posterior distributions, while the quoted 95\% confidence intervals correspond to the 2.5th and 97.5th percentiles. As demonstrated in \citet{ciocan}, parameter covariances are generally weak, a result that we confirm for the present sample through inspection of the posterior distributions.
 
\section{Characterising the star formation history}
\label{SED}

\subsection{Spectral Energy Distribution fitting}

Stellar population properties were derived from SED fitting using the ancillary photometric data described in Sect.~\ref{phot}. Prior to the modelling, we removed invalid flux measurements and imposed a minimum relative uncertainty of 5\%, following \cite{Carnall18,Carnall19}. This error floor mitigates the underestimation of uncertainties in the high signal-to-noise regime, where systematic calibration errors dominate over statistical photometric uncertainties.

The SED fitting was performed with the Bayesian code \texttt{BAGPIPES} \citep{Carnall18}. For the fits, we fixed the redshift to the spectroscopic value. We adopted the \cite{BC03} stellar population synthesis models, assuming a \cite{kroupa} IMF, and the \cite{CF00} dust attenuation law, parameterised following the two-parameter formulation with $A_V \in [0,3]$ and slope $n \in [0.3,2.5]$, where a Gaussian prior is imposed on $n$  ($\mu=0.7$ and $\sigma=0.3$). We included nebular emission assuming a constant ionisation parameter in the range $\log \mathcal{U} \in [-4,-2]$, and a fixed birth cloud dispersal timescale of $t_{\rm BC}=10$ Myr.  We adopted a log-uniform prior for the stellar metallicity, with $Z \in [0.01,5]Z_\odot$.

The  SFH was modelled using a non-parametric continuity model following \cite{Leja19} and \cite{Leja19b}. The SFH was discretised into seven lookback-time bins, within which the SFR is assumed to be constant and is coupled via a continuity prior that disfavours abrupt variations between adjacent bins. Following \cite{Leja19} (their equation 2), the bins were defined at 0--30, 30--100, 100--330, and 330--1100 Myr, as well as 0.85\,$t_{\rm univ}$--$t_{\rm univ}$, where $t_{\rm univ}$ is the age of the Universe at the observed redshift. The last two bins were equally spaced logarithmically between 1100 Myr and 0.85\,$t_{\rm univ}$. Variations in the SFH are controlled through differential SFR parameters in each bin, each following a Student-t prior with scale 0.3 and degrees of freedom $\nu=2$, promoting smooth but flexible SFHs. 

We show the full set of model parameters and priors in Appendix \ref{appendix:sed}, Table \ref{SEDpara}. Furthermore, in  Fig. \ref{SED_fit}, we show examples of the SED fits for two galaxies from our sample (top), together with their reconstructed  SFHs (bottom).

We optimised the model parameters using the \texttt{pyMultiNest} \citep{feroz,Buchner} sampler and adopted the posterior medians as their best-fit values. We note that the stellar masses and SFRs were converted to a \citet{chabrier03} IMF by applying corrections of $-0.034$ dex and $-0.027$ dex, respectively \citep{2014ARA&A..52..415M}.

\subsection{SFH indicators at the observed epoch}
\label{SFHdiagn}
In addition to the SED-inferred $M_\star$ and SFR, we used complementary diagnostics to characterise the SFHs of the sample, which are detailed below.

\begin{figure}
   \centering
    \includegraphics[width=0.4\textwidth,angle=0,clip=true]{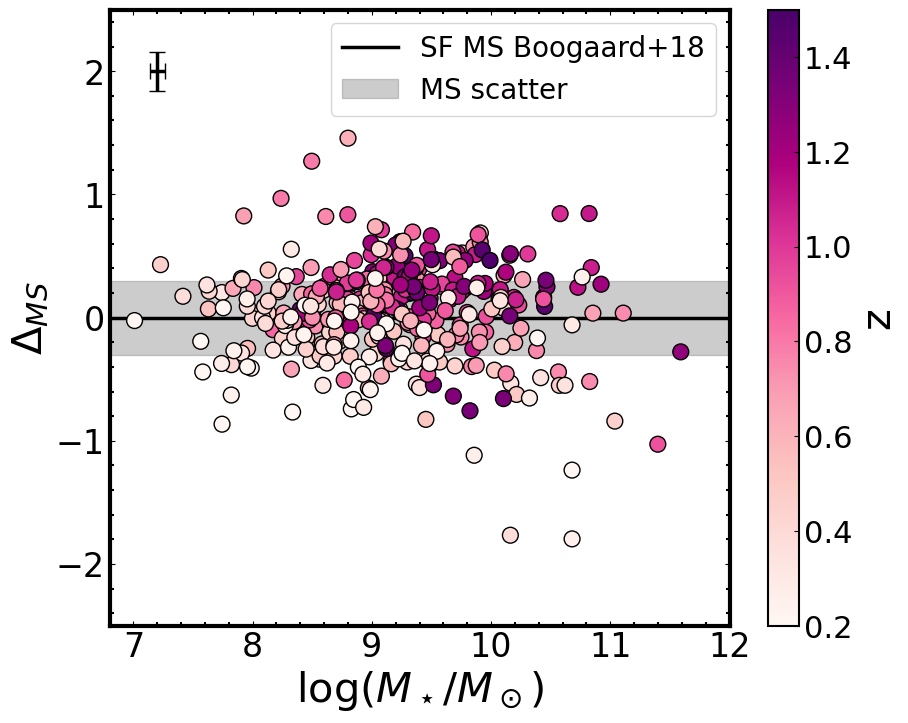}
   \centering   
\caption{Offset from the star-forming main sequence of \citet{bog},  evaluated at the stellar mass and redshift of each galaxy from the sample. The stellar masses and SFRs (averaged over the past 100 Myr) are inferred from SED fitting (Sect.~\ref{SED}).  Data points are colour-coded by redshift. The black solid line shows the main sequence from \citet{bog}, while the grey shaded region indicates the 0.3 dex  scatter around the relation. The black cross in the upper right corner represents the median $1\sigma$ uncertainties in $M_{\star}$ and SFR  from \texttt{BAGPIPES}.}
\label{sfr}
\end{figure}

First, we defined the offset from the star-forming main sequence of \cite{bog} as:
\begin{equation}
\Delta_{\rm MS} = \log \mathrm{SFR}_{100\,\mathrm{Myr}} - \log \mathrm{SFR}_{\rm MS}(M_\star, z), \label{ms}
\end{equation}
which quantifies the deviation of each galaxy from the population-averaged SF relation at fixed $M_\star$ and $z$. Note that the SFRs used in Eq. \ref{ms} are averaged over the past 100 Myr. We show in Fig.~\ref{sfr} $\Delta_{\rm MS}$ as a function of $M_\star$ for our sample. We find that the majority of galaxies lie within the intrinsic $\pm0.3$ dex scatter of the main sequence, confirming that the sample predominantly consists of typical SFGs, with a smaller fraction showing enhanced or suppressed SF activity, indicative of starbursts and quenching, respectively.

Second, we used $n_{\rm SFR}$, which  quantifies the ratio between the instantaneous SFR (averaged over the past 100 Myr) inferred from the model SFH and the time-averaged SFR over the  lifetime of the galaxy:
\begin{equation}
n_{\rm SFR} =\frac{ \rm{SFR}_{\rm 100\:Myr}}{\langle \rm{SFR} \rangle}= \frac{\rm{SFR_{\rm 100\:Myr}}\cdot t_0}{M_{\rm formed}}.
\end{equation}
Here, $t_0$ represents the cosmic time at the galaxy's observed redshift, while  $M_{\rm formed}$ is the total stellar mass formed up to $t_0$ (see  Eq. 12 in \citealt{Carnall18}). Thus, $n_{\rm SFR} $  measures the deviation of the current SF activity from the long-term average growth of the galaxy. We caution that because $n_{\rm SFR}$ compares the SFR averaged over the past 100 Myr to the lifetime-averaged SFR, elevated values do not uniquely trace bursty behaviour and may also arise naturally from secular evolution along rising SFHs, making this parameter a rather ambiguous "burstiness" indicator.

Third, we defined a variability (or burstiness) indicator following, e.g., \cite{Sparre, Cole}, which compares SFRs sensitive to different timescales:
\begin{equation}
n_{\rm burst} = \frac{\mathrm{SFR}_{10\,\mathrm{Myr}}}{\mathrm{SFR}_{200\,\mathrm{Myr}}}.
\end{equation}
Throughout this work, both ${\rm SFR}_{10\,\mathrm{Myr}}$ (i.e. SFR averaged over the past 10 Myr) and ${\rm SFR}_{200\,\mathrm{Myr}}$ (i.e. SFR averaged over the past 200 Myr)  were derived from the posterior SFHs inferred with \texttt{BAGPIPES}. The 10 Myr estimate is expected to be more sensitive to modelling assumptions, but the broad wavelength coverage and high-S/N photometry provide significant leverage on recent SF activity.

As an independent validation of the recent ${\rm SFRs}_{10\,\mathrm{Myr}}$ inferred from the SED fitting, we also derived ${\rm SFR}_{10\,\mathrm{Myr}}$ from optical emission-line tracers measured from the integrated MUSE spectra. In all cases, SFRs were converted to a common \citet{chabrier03} IMF by applying the appropriate systematic offsets to the adopted calibrations. For the  [OII] emitters, we used the \cite{Gilbank} calibration, which empirically accounts for the average dependence of [OII] luminosity on metallicity and dust attenuation. We caution that although [OII] traces SF on short ($\sim10$ Myr) timescales comparable to H$\alpha$, it is a less direct SFR indicator because of its residual sensitivity to ionisation conditions. 
For the remaining galaxies, we estimated the ${\rm SFR}_{10\,\mathrm{Myr}}$ from the extinction-corrected H$\alpha$ luminosity using the calibration of \cite{Kennicutt}. The H$\alpha$ luminosities were corrected for dust attenuation using the Balmer decrement following \cite{Brocklehurst}. 
We find that the emission-line and SED-derived estimates of ${\rm SFR}_{10\,\mathrm{Myr}}$ are consistent within their uncertainties, providing confidence that the SED fitting reliably recovers the recent SF activity. We adopted the SED-derived SFRs throughout this analysis. 

Under these assumptions, 
values of $n_{\rm burst}>1$ correspond to systems currently experiencing elevated recent SF relative to their 200 Myr average, while $n_{\rm burst}<1$ indicates systems whose SF has declined over $\sim$200 Myr timescales. 
We note that this parameter does not trace the variability of SF throughout the full SFH, and refer to \cite{Rezaee} for a discussion of the limitations of such diagnostics in constraining bursty SFHs.


\subsection{SFH indicators throughout the lifetime of the galaxy}
\label{SFHdiagn2}

To more accurately isolate the stochastic component of the inferred SFHs, we removed the average evolution expected for galaxies evolving along the star-forming main sequence and measured the residual fluctuations around this relation as a function of cosmic time. For each posterior SFH realisation, we first reconstructed the stellar mass assembly history and evaluated the corresponding main-sequence SFR at every epoch. The stellar mass assembly history was obtained by integrating the SFH from the onset of SF, $t_{\rm start}$, to a given cosmic time $t$, and normalising it to the stellar mass at the observation epoch, $M_{\star,0}$:
\begin{equation}
M_\star(t) =
M_{\star,0}
\frac{
\int_{t_{\rm start}}^{t} {\rm SFR}(t')dt'
}{
\int_{t_{\rm start}}^{t_0} {\rm SFR}(t')dt'
}.
\end{equation}
Here, $t_{\rm start}$ denotes the beginning of the time interval covered by the reconstructed SFH, while $t_0$ is the cosmic time corresponding to the observed redshift. This procedure preserves the relative mass assembly encoded by the inferred SFH while ensuring consistency with the inferred final stellar mass. The expected main-sequence SFR was then computed using the redshift-dependent relation of \citet{bog},
$ {\rm SFR}_{\rm MS}(M_\star(t),z(t))$, where the stellar mass entering the relation is the reconstructed stellar mass assembly history. To avoid extrapolating the main-sequence calibration beyond its validity, we restricted the analysis to epochs satisfying $M_\star(t) > 0.1\,M_{\star,0}
\quad {\rm and} \quad
{\rm SFR}(t) > 10^{-5}\,{\rm M_\odot\,yr^{-1}}$. The instantaneous offset from the evolving main sequence is defined as:
\begin{equation}
\Delta_{\rm MS}(t) =
\log({\rm SFR}(t))- \log({\rm SFR}_{\rm MS}(M_\star(t), z(t))).
\end{equation}

We characterised these fluctuations using two complementary diagnostics. First, we computed the mass-weighted root-mean-square deviation,
\begin{equation}
\sigma_{\rm burst} =
\left[
\sum_i w_i
\Delta_{\rm MS}(t_i)^2
\right]^{1/2},
\end{equation}
 where $t_i$ denotes the cosmic time associated with timestep $i$ and
\begin{equation}
w_i =
\frac{{\rm SFR}(t_i)\Delta t_i}
{\sum_j {\rm SFR}(t_j)\Delta t_j},
\end{equation}
with $\Delta t_i$ denoting the time interval associated with timestep $i$.
The weights correspond to the fraction of stellar mass formed during each timestep, such that $\sigma_{\rm burst}$ preferentially emphasises fluctuations occurring during the phases responsible for the bulk of stellar mass assembly.  

We additionally quantified the contribution of strongly enhanced SF episodes to the total stellar mass formed. Burst phases were defined as epochs satisfying $\Delta_{\rm MS} > 0.3~{\rm dex}$, 
and the burst mass fraction is given by:
\begin{equation}
f_{\rm burst} =
\frac{
\sum_{\Delta_{\rm MS}>0.3}
{\rm SFR}(t_i)\Delta t_i
}{
\sum_i
{\rm SFR}(t_i)\Delta t_i
}.
\end{equation}
This quantity measures the fraction of stellar mass formed during elevated SF episodes and provides a complementary indicator of burst-dominated stellar mass assembly.

Finally, we considered three complementary indicators of stellar mass assembly: the mass-weighted formation time, $t_{\rm form}$, defined as the average cosmic time at which the stellar mass is formed,
\begin{equation}
t_{\rm form} =
\frac{
\displaystyle\int_{t_{\rm start}}^{t_0}
t\cdot{\rm SFR}(t)dt
}{
\displaystyle\int_{t_{\rm start}}^{t_0}
{\rm SFR}(t)dt
}=t(z_{\rm form}),
\end{equation}
where $t$ is the cosmic time measured from the Big Bang, $t_{\rm start}$ denotes the onset of the reconstructed SFH, and $t_0$ is the cosmic time at the observation epoch. Additionally, we also considered the formation times $t_{50}$ and $t_{90}$, which are defined as the cosmic times at which 50\% and 90\% of the total formed stellar mass have been assembled, respectively. All three quantities are expressed as cosmic times measured from the Big Bang, with larger values corresponding to later stellar mass assembly.

\section{Results}
\label{results}

\subsection{Disk-halo decomposition}
 \subsubsection{Model comparison}



Following \cite{ciocan,nicolas}, we compared the DC14, DC14-scatter, and gNFW disk--halo decompositions using the Bayesian evidence, $\log(\mathcal{Z})$, computed with \texttt{pyMultiNest} \citep{Buchner}. We note that we rescale the evidence by -2 such that it is on the same scale as the usual information criterion  (following \citealt{kr}). We quantified the preference between two models, $M_1$ and $M_2$, using : $
\Delta \log(\mathcal{Z})=\log(\mathcal{Z}_{M_1})-\log(\mathcal{Z}_{M_2})$, and interpreted $\Delta \log(\mathcal{Z})$  following \citet{kr}. Specifically, $\Delta \log(\mathcal{Z}) >2$  (<-2) indicates positive evidence for $M_2$ ($M_1$). Values in the range $-2 < \Delta \log(\mathcal{Z}) < 2$ indicate that both models are comparable. Values of  $\Delta \log(\mathcal{Z}) > 6 (20) $ indicate strong (very strong) evidence in favour of $M_2$, while $\Delta \log(\mathcal{Z}) < -6 (-20) $ indicate strong (very strong) evidence in favour of $M_1$.

\begin{figure}
   \centering
    \includegraphics[width=0.5\textwidth,angle=0,clip=true]{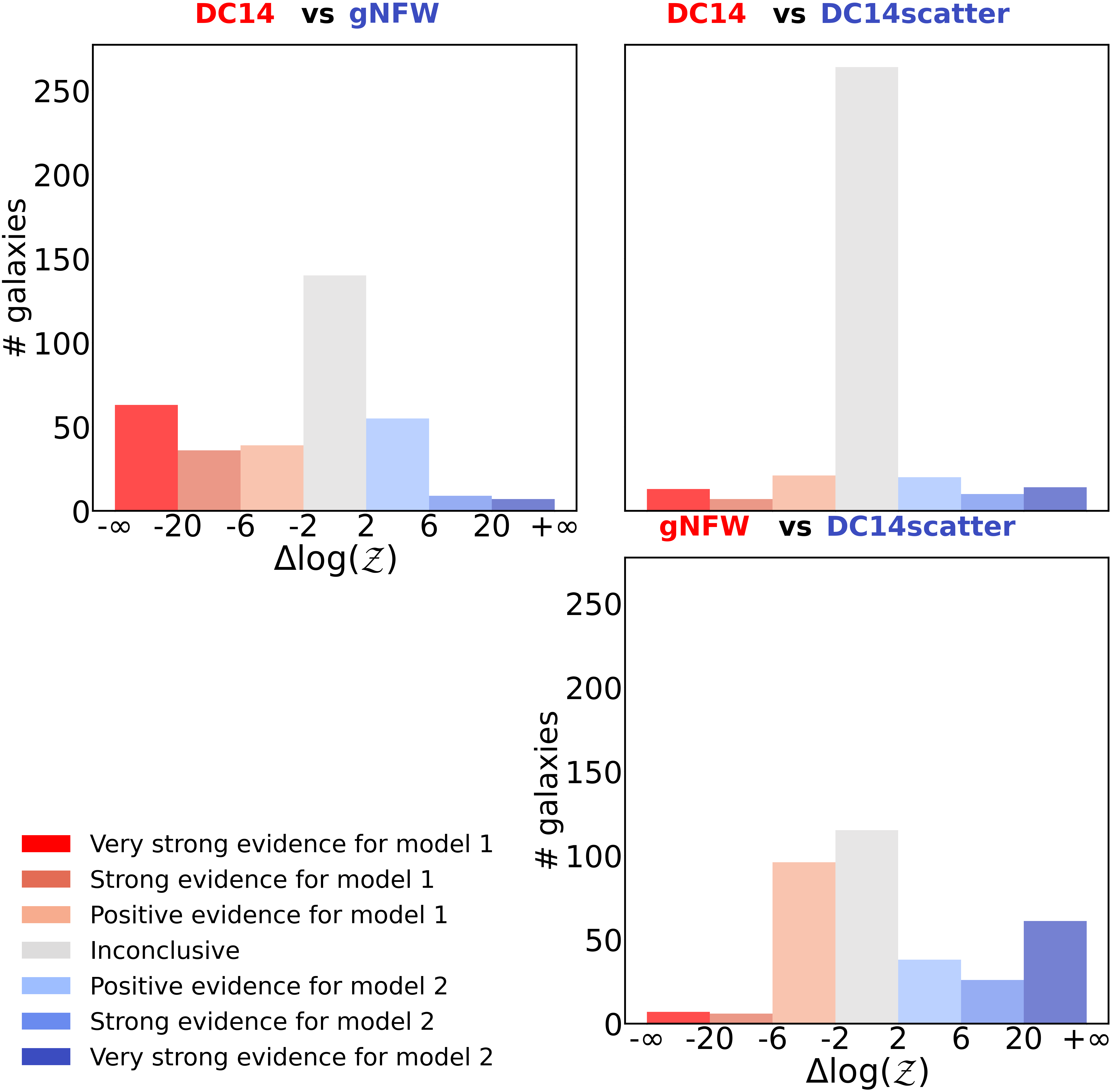}
   \centering   
\caption{Histograms showing the Bayes factor for all pairs of halo models used in this study. Red bins indicate galaxies with strong to very strong evidence in favour of the first model, while blue bins correspond to galaxies showing strong to very strong evidence in favour of the second model. Intermediate red and blue bins show the galaxies with moderate evidence for either model. Grey bins show systems where both models perform comparably. In each panel, model 1 and model 2 refer to the first and second models listed in the title, respectively.}
\label{bayes}
\end{figure}

Figure~\ref{bayes} presents the distribution of Bayes factors for the three halo models considered in the disk--halo decomposition: DC14, DC14-scatter, and gNFW.
The first panel shows that, for 147 galaxies (39\% of the sample), the most likely model given the data is DC14 when compared to gNFW, while for 73 galaxies (20\%), gNFW is the most likely model given the data. For the remaining 41\% of galaxies, the two models provide comparable fits.
For the DC14 versus DC14-scatter comparison, the second panel indicates that the two models are largely indistinguishable.  For 41 (44) galaxies ($\sim11-12$\%), DC14 (DC14-scatter) is the most likely model given the data, while for the remaining $\sim80\%$ of the sample, both models provide statistically comparable fits.
Finally, the bottom panel shows that for 126 galaxies (34\% of the sample), DC14-scatter is the most likely model given the data when compared to gNFW, while for 105 galaxies (28\%), gNFW is the most likely model given the data when compared to DC14-scatter; for the remaining 38\% of galaxies, both models provide comparable fits.


In Appendix \ref{appendix:decomp}, we show the same Bayesian model comparison in Fig.~\ref{bayes2}, but when restricting our sample to the highes-S/N galaxies ($\rm{S/N_{eff}}>50$). We find that, in this case, DC14(-scatter) is more likely than gNFW, given the data, with a considerably higher number of galaxies showing positive evidence for the former. 

Overall, these results indicate that, in terms of posterior probability, mass-dependent halo profiles based on the DC14 framework are more strongly supported by the data than the gNFW parametrisation. Also, introducing an additional scatter parameter in the DC14 framework does not lead to a statistically significant improvement for most galaxies compared to the fiducial DC14 fits. Similar results were obtained in \cite{ciocan}, where six different DM halo profiles were compared, and the DC14 model was found to have the highest posterior probability among the tested models.

In Appendix \ref{appendix:decomp},  Fig.~\ref{decomp}, we also show examples of the RC decomposition and resulting density profiles obtained with the fiducial DC14 model for two galaxies from our sample. We release all catalogues and data products from the disk-halo decomposition for the full sample on the DARK website \footnote{\url{https://dark.univ-lyon1.fr/data-releases/}}.

\subsubsection{Halo Scaling Relations}

In this section, we examine three fundamental halo scaling relations using the best-fitting parameters from the fiducial DC14 disk--halo decomposition, and investigate whether the trends reported by \cite{ciocan} are recovered in the significantly larger  sample.

\begin{figure*}
   \centering
    \includegraphics[width=0.35\textwidth,angle=0,clip=true]{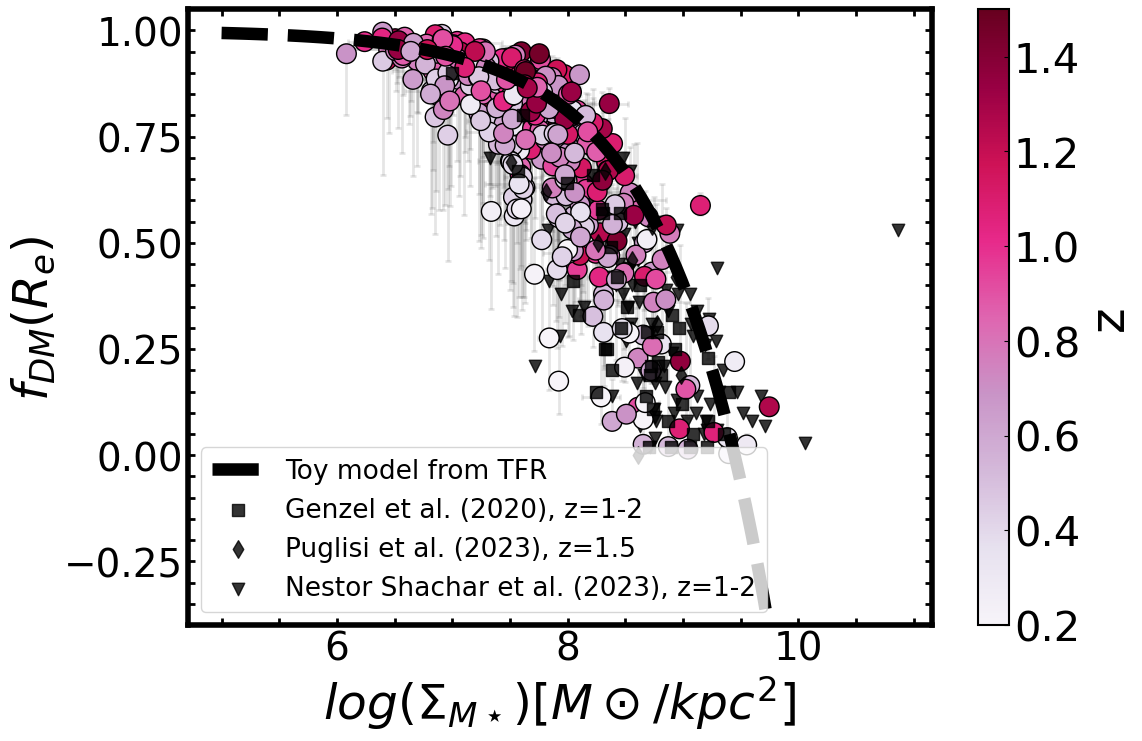}
   \centering  
      \centering
    \includegraphics[width=0.335\textwidth,angle=0,clip=true]{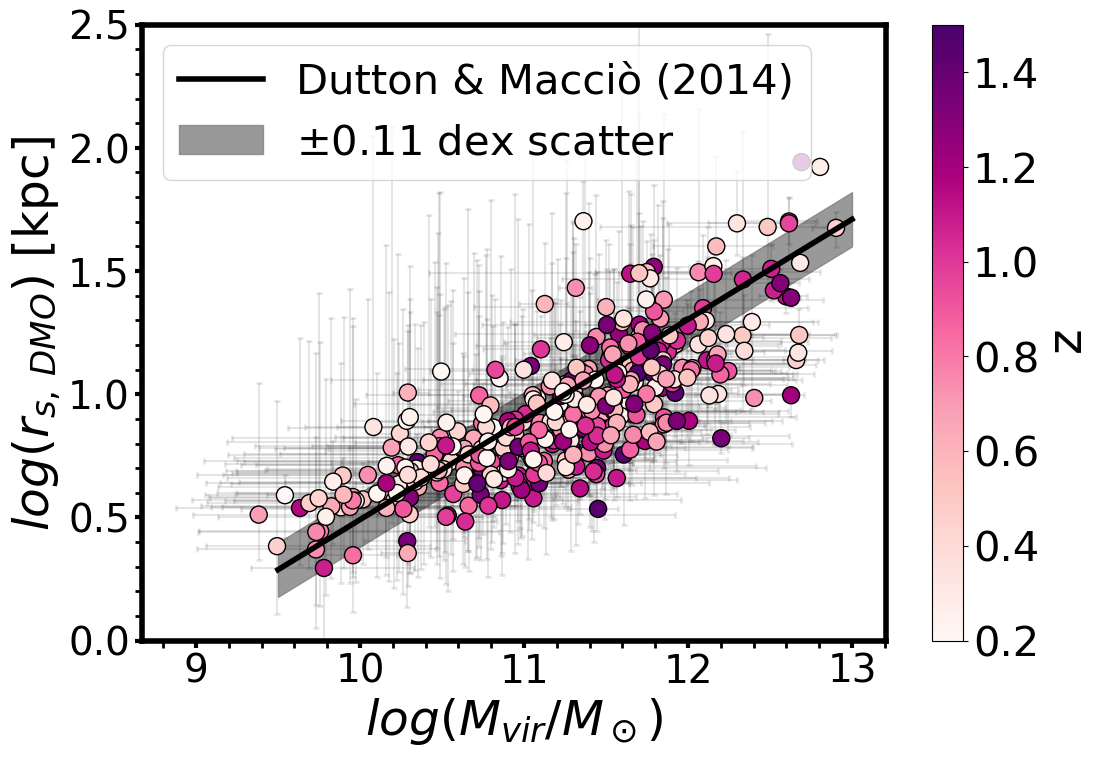}
   \centering   
         \centering
    \includegraphics[width=0.3\textwidth,angle=0,clip=true]{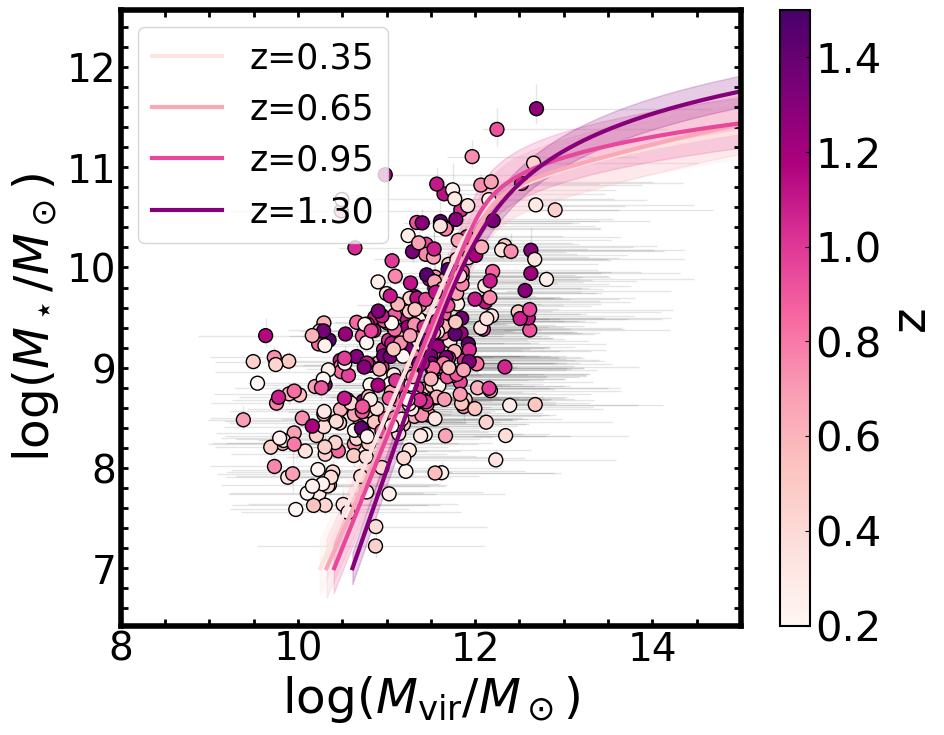}
   \centering   
\caption{DM halo scaling relations derived from the DC14 fits. 
Left: DM fraction within the effective radius, $f_{\rm DM}(<R_{\rm e})$, as a function of the stellar mass surface density. The black diamonds show the DM fractions inferred by \cite{pug} for $z\sim1.5$ SFGs, the black squares represent the $z=1-2$ sample from \cite{genz2}, and the black triangles correspond to the RC100 sample from \cite{shach}. The black dotted line indicates the toy model derived from the \cite{1977A&A....54..661T} relation (see \citealt{nicolas}). 
Middle: DM-only scale radius, $r_{\rm{s,DMO}}=r_{\rm vir}/c_{\rm vir}$ [kpc] as a function of virial mass. The solid black line shows the predictions from \cite{dutton}, while the gray shaded region shows the expected scatter of 0.11 dex around the relation. 
Right: Stellar mass--halo mass relation. The coloured lines and shaded regions show the relations and scatter from \cite{cosmos2020} at different redshifts,  respectively, as indicated by the legend. 
In all panels, the sample is colour-coded according to galaxy redshift. Error bars represent the 95\% confidence intervals.}
\label{halos}
\end{figure*}

Figure~\ref{halos} (left) shows the DM fraction within the effective radius, $f_{\rm DM}(<R_{\rm e})$, as a function of stellar mass surface density, $\Sigma_{M\star}$, for the  sample. Consistent with the results of \cite{ciocan}, we find that $\sim77\%$ of galaxies are DM dominated within $R_{\rm e}$. We recover a tight anti-correlation between $f_{\rm DM}(<R_{\rm e})$ and $\Sigma_{M\star}$, in agreement with previous observational studies \citep{genz2, pug, shach, 2026arXiv260112315S}. As discussed by \cite{ciocan}, this relation can naturally arise from the coupling between galaxy rotation velocities and stellar mass through the   \cite{1977A&A....54..661T} relation, as illustrated by the simple toy model at $z\sim1$ (black dotted line in Fig.~\ref{halos}, right).

Figure~\ref{halos} (middle) presents the relation between the virial mass, $M_{\rm vir}$, and the halo scale radius, $r_{\rm{s,DMO}}=r_{\rm vir}/c_{\rm vir}$, where $r_{\rm vir}$ is the virial radius and $c_{\rm vir}$ the concentration. The galaxies broadly follow the halo structural relations predicted by DM-only (DMO) simulations \citep{dutton}. In particular, we find no significant evolution of the scale radius with redshift, consistent with theoretical predictions which show that $r_{\rm s}$  is $z$-invariant (e.g. \citealt{bull, dutton}). 

Finally, Fig.~\ref{halos} (right) shows the stellar-to-halo mass relation for our sample, comparing the SED-based stellar masses with the virial masses from our 3D disk--halo decomposition. Our results are qualitatively consistent with the \cite{cosmos2020} SMHM relation \footnote{The \cite{cosmos2020} stellar-mass-halo-mass relation is obtained from the COSMOS2020 catalogue using a joint modelling of the observed stellar mass functions and galaxy clustering with a phenomenological halo-occupation model.}, albeit with substantially larger scatter ($\sim0.6$ dex compared with $\sim0.15$--$0.3$ dex from \citealt{cosmos2020}). This difference is expected because the scatter in \cite{cosmos2020} describes the intrinsic dispersion in stellar mass at fixed halo mass, whereas our scatter also includes measurement uncertainties in the individually inferred stellar and halo masses. To conclude, the DC14 disk--halo decomposition recovers the main halo scaling relations expected from $\Lambda$CDM-based models, in agreement with \cite{ciocan}, and with observational studies focusing on local galaxy samples (e.g \citealt{alle}, \citealt{katz}).   

 \subsubsection{ DM inner slopes}
\label{gamma}
\begin{figure*}
   \centering
    \includegraphics[width=0.47\textwidth,angle=0,clip=true]{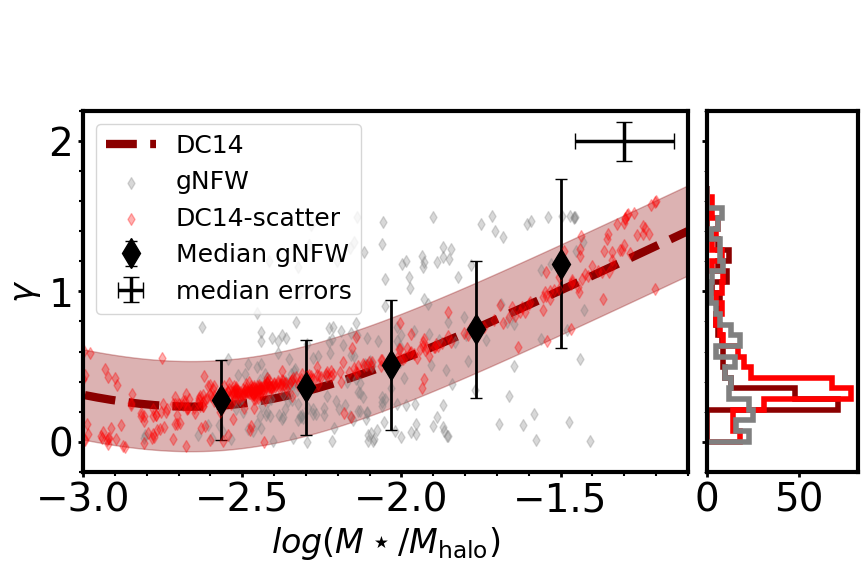}
   \centering  
      \centering
    \includegraphics[width=0.32\textwidth,angle=0,clip=true]{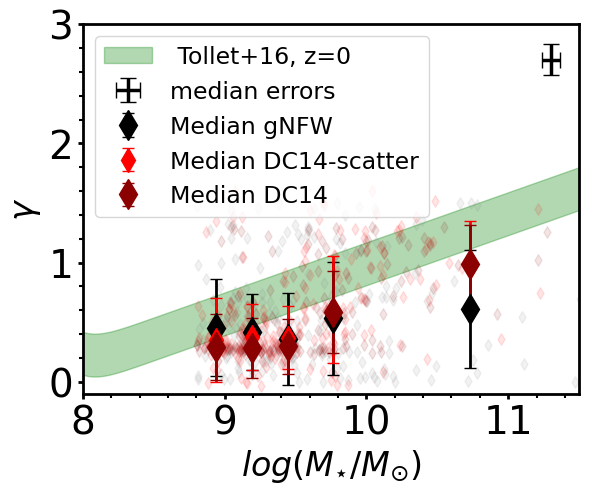}
   \centering   
\caption{Left: DM inner slope as a function of the stellar-to-halo mass ratio for the entire sample. The dark red curve shows the DC14 fits, while the dark red shaded area shows the 0.3 dex scatter around the relation. The red diamonds show the DC14-scatter fits. The grey diamonds show the gNFW fits. The black diamonds show the median values from the gNFW fits in five bins, while the error bars represent the standard deviation within each bin.  The histogram on the right shows the distribution of DM inner slope values from our DC14, DC14-scatter and gNFW fits, in dark red, red and grey, respectively. Right: DM inner slope as a function of the stellar mass for the complete sample with $\log(M_\star/M_{\odot})>8.8$. The dark red (red; black) diamonds show the median values for the DC14 (DC14-scatter; gNFW) fits in five equal-occupancy mass bins, and the error bars depict the standard deviation within each bin. The green curve shows the predictions from \cite{toll} at $z=0$. In both panels, the crosses from the upper right corners depict the  68\% confidence intervals ($\sim1\sigma$) measurement uncertainties.  }
\label{gammaf}
\end{figure*}

Figure \ref{gammaf} (left) shows the inferred DM inner slope as a function of the stellar-to-halo mass ratio for the DC14 fits (dark red), DC14-scatter fits (red), and gNFW fits (grey). 
The DC14-scatter results closely follow the DC14 relation, with only a subsample of galaxies ($\sim30\%$) deviating by more than 0.1 dex from the $\gamma$–$\log(X)$ relation (dark-red curve).  The black diamonds show the median trend from the gNFW fits. Although the gNFW profile imposes no intrinsic dependence of $\gamma$ on the stellar-to-halo mass ratio (unlike the DC14-based models),  a correlation with $\log(M_{\star}/M_{\rm halo})$ emerges, with the gNFW medians broadly consistent with the DC14 expectations. Importantly, this correlation is not driven by the adopted stellar-to-halo mass relation prior: repeating the gNFW analysis without the \cite{cosmos2020} $M_{\star}$--$M_{\rm halo}$ prior yields the same $\gamma$--$\log(M_{\star}/M_{\rm halo})$ trend. This indicates that the data support the existence of a genuine connection between the DM inner slope and the stellar-to-halo mass ratio, in line with predictions from hydrodynamical simulations (e.g.   \citealt{toll}, \citealt{fre}, \citealt{Lazar}, \citealt{aza}).  
 
In terms of the inner DM density slopes, we find that $\sim$70\% of the sample exhibits cored profiles ($\gamma < 0.5$) when adopting the DC14 model, as shown by the red histogram in Fig.~\ref{gammaf}. The DC14-scatter and gNFW fits yield slightly lower fractions of $\sim$60\% cored systems, shown by the dark-red and grey histograms, respectively. Despite these differences, the inferred inner slopes are broadly consistent across the three parametrisations, with no significant systematic differences, particularly for the highest-S/N galaxies and for cases where the Bayesian comparison does not favour one model over another.
 
 Figure \ref{gammaf} (right) shows $\gamma$  from the DC14 fits (dark red), DC14-scatter fits (red), and gNFW fits (grey) as a function of the stellar masses for the sample. The green band shows the $z=0$ prediction from \citet{toll} using the NIHAO simulations. While the simulations predict a dependence of $\gamma$ on stellar mass, our data exhibit a much larger spread and only a weak correlation in which $\gamma$ increases with stellar mass, indicating progressively steeper inner density profiles toward more massive galaxies. However, the relation shows large scatter, with substantial intrinsic dispersion within each mass bin, making the trend significantly noisier than the corresponding $\gamma$-$M_\star/M_{\rm halo}$ relation.  This behaviour is, nonetheless,  consistent with the results of \citet{fre}, who showed, using the same simulation suite, that the inner halo slope is more strongly coupled to the stellar-to-halo mass ratio than to stellar mass alone, reflecting the key role of the SF efficiency in regulating halo structure. 

\subsection{Dependence of SFH Diagnostics on Galaxy Properties}
 \label{s}
To assess how the inferred SFHs vary across our galaxy sample, we examined the dependence of the various SFH diagnostics used throughout this work on stellar mass and redshift. We defer the reader to Appendix~\ref{sfh} for more details. Briefly, we find that $n_{\rm SFR}$, $\sigma_{\rm burst}$, and $f_{\rm burst}$ increase towards higher redshift and lower stellar masses, while $n_{\rm burst}$ primarily evolves with redshift and shows no significant dependence on $M_\star$ in our sample. We also find that more massive galaxies formed the bulk of their stellar populations at earlier cosmic times (i.e. $t_{\rm form}$ decreases with $M_\star$), consistent with the galaxy downsizing scenario. These trends are consistent with previous observational studies and predictions from hydrodynamical simulations \citep{2019ApJ...881...71E, Mehta, Faucher, Iyer, 2020MNRAS.497..698T}.

\subsection{Linking SFHs to DM inner slopes}
\label{link}
\subsubsection{What drives the inner DM slope?}

The inner DM slope, $\gamma$, exhibits a weak systematic dependence on $M_\star$ (Section~\ref{gamma}), with more massive galaxies being typically more cuspy. Similarly, the SFH diagnostics show clear trends with $M_\star$ (Section~\ref{s}; see also Appendix~\ref{sfh}). In particular, $n_{\rm SFR}$, $\sigma_{\rm burst}$, $f_{\rm burst}$, and $t_{\rm form}$ reflect the well-established dependence of galaxy SFHs on stellar mass: lower-$M_\star$ galaxies tend to assemble their stellar populations at later cosmic times and exhibit more bursty SFHs than their higher-$M_\star$ counterparts. In addition, several of the SFH diagnostics show significant evolution with redshift, reflecting the changing SF  properties of galaxies across cosmic time.

Because both $\gamma$ and several SFH diagnostics are correlated with $M_\star$  (and the SFH diagnostics also evolve with $z$), any apparent relation between them may arise from their shared dependence on these quantities. To isolate the relation between halo structure and galaxy SFHs, a partial correlation coefficient (PCC) analysis was performed in Appendix~\ref{ppcana}. PCC quantifies the correlation between two variables after statistically removing the influence of one or more additional correlated quantities. We applied this analysis to $\gamma$ and all SFH diagnostics considered in this work ($n_{\rm SFR}$, $f_{\rm burst}$, $\Delta_{\rm MS}$, $t_{\rm form}$, $t_{50}$, $t_{90}$, $\sigma_{\rm burst}$, and $n_{\rm burst}$), while treating both stellar mass and redshift as control variables.

For the DC14 and DC14-scatter fits, we find that stellar mass shows the strongest correlation with $\gamma$, whereas all SFH diagnostics exhibit only weak residual correlations with the inner DM slope once the dependence on both $M_\star$ and redshift is removed. For the gNFW fits, the PCC results are less conclusive, with stellar mass and the different SFH diagnostics showing similarly weak partial correlations with $\gamma$. We refer the reader to Appendix~\ref{ppcana} for the complete PCC results.

Our PCC analysis specifically probes linear associations between pairs of variables after controlling for $M_\star$ and redshift. Because the partial correlations are obtained by removing linear dependencies on the conditioning variables and computing Pearson correlations between the resulting residuals, this approach does not capture more complex nonlinear or non-monotonic dependencies. Moreover, this analysis cannot simultaneously assess the relative predictive contribution of multiple, potentially correlated SFH diagnostics. To address these limitations, we use Random Forest regression in the next Section, which can capture nonlinear relationships and assess, simultaneously, the relative importance of $M_\star$ and the full set of SFH diagnostics in predicting $\gamma$.

\subsubsection{Random Forest Regression}
\label{RF}

To more robustly assess the relative importance of the different physical parameters in predicting $\gamma$, we performed a Random Forest (RF) regression analysis \citep{Breiman}. RF is a non-parametric ensemble learning method widely used to infer non-linear relationships between galaxy properties (e.g. \citealt{Bluck}, \citealt{2024A&A...681A.121S}, \citealt{maria}). The method constructs an ensemble of decision trees trained on bootstrap resamples of the data and random subsets of features, thereby reducing variance and mitigating overfitting. This makes it particularly well-suited to our problem, where the input variables are mutually correlated and may contribute non-linearly to the target variable.

For this, we used the \texttt{RandomForestRegressor} implemented in \texttt{scikit-learn} \citep{Pedregosa}. The feature set includes $M_\star$ (and $M_\star/M_{\rm vir}$ for the gNFW fits), as well as the SFH diagnostics $n_{\rm SFR}$, $n_{\rm burst}$, $\Delta_{\rm MS}$, $t_{\rm form}$, $t_{50}$, $t_{90}$, $\sigma_{\rm burst}$, and $f_{\rm burst}$. The data were divided into training and test subsets using a fixed $70\%/30\%$ split, with the latter kept as an independent held-out sample. The RF hyperparameters were optimised using 5-fold cross-validation on the training set, exploring the number of trees, minimum samples per leaf and split, maximum tree depth, and the number of features considered at each split. The optimisation favours a configuration with $n_{\rm estimators}=1000$, $\texttt{min\_samples\_leaf}=10$, $\texttt{min\_samples\_split}=20$, and \texttt{max\_features}=\texttt{None}, the latter allowing all features to be considered at
each split. Similar hyperparameters were adopted in \cite{Bluck}.  As a check on the choice of regression algorithm, we also performed the same cross-validation procedure using an Extra Trees regressor. The two methods yielded very similar cross-validated performance, and we therefore adopted the RF as our fiducial model.

To propagate observational uncertainties in the input features, we performed a Monte Carlo (MC) procedure in which the training data were repeatedly perturbed according to Gaussian noise defined by the measured uncertainties. For each of $N_{\rm MC}=100$ realisations, an independent RF model was trained using the optimised hyperparameters, yielding an ensemble that captures the effects of measurement uncertainties and model stochasticity. We note that this approach does not constitute a full Bayesian treatment of the posterior uncertainty in $\gamma$, but provides an empirical estimate of predictive variability.

Model performance was evaluated using the coefficient of determination ($R^2$), computed from the ensemble-averaged predictions. Across the different halo models, the RF achieves relatively low predictive power, with $R^2$ values of order $\sim0.3$. The limited predictive performance may arise from intrinsic scatter in the relation and/or physical dependencies not captured by the input features. Thus, while the RF can identify the relative predictive power of the input features, the results should not be interpreted as indicating that these parameters fully explain the observed scatter in $\gamma$.  We quantified feature relevance using (i) the mean decrease in impurity (MDI) from each forest and (ii) permutation importance evaluated on the held-out test set \citep{Breiman}. Both metrics were computed for each MC realisation, and we report their mean and standard deviation across the ensemble.

\begin{figure*}
   \centering
   \includegraphics[width=0.45\textwidth,angle=0,clip=true]{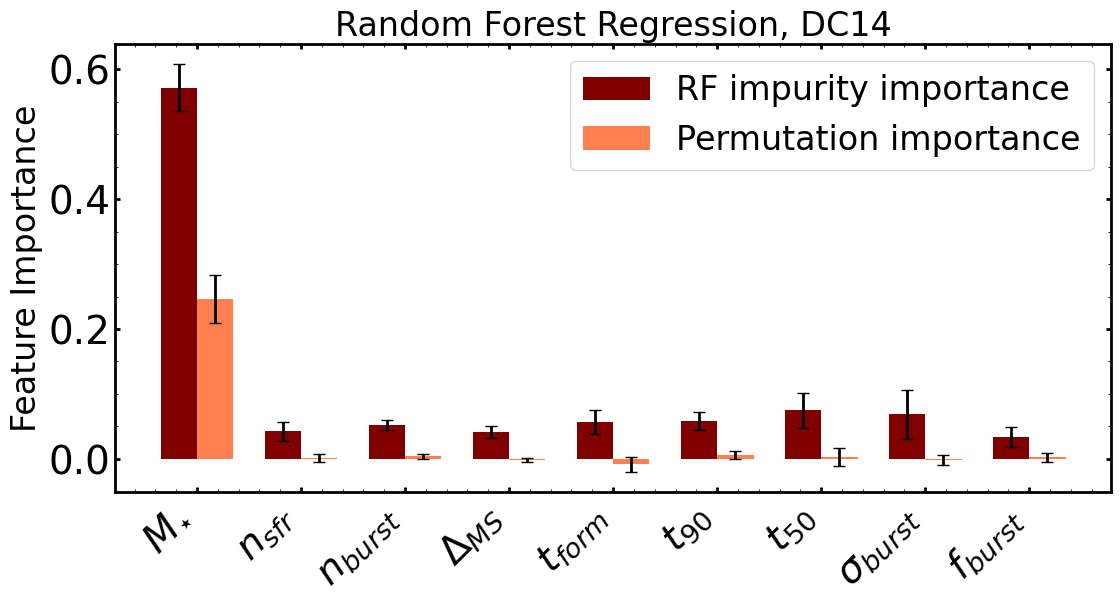}
   \centering   
       \includegraphics[width=0.45\textwidth,angle=0,clip=true]{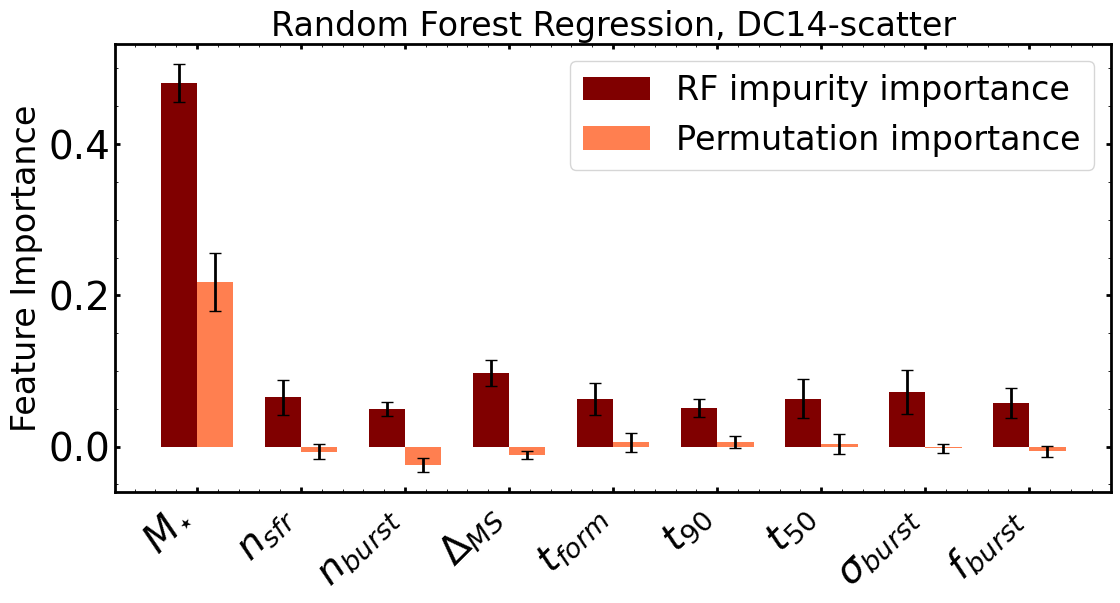}
   \centering   

      \includegraphics[width=0.45\textwidth,angle=0,clip=true]{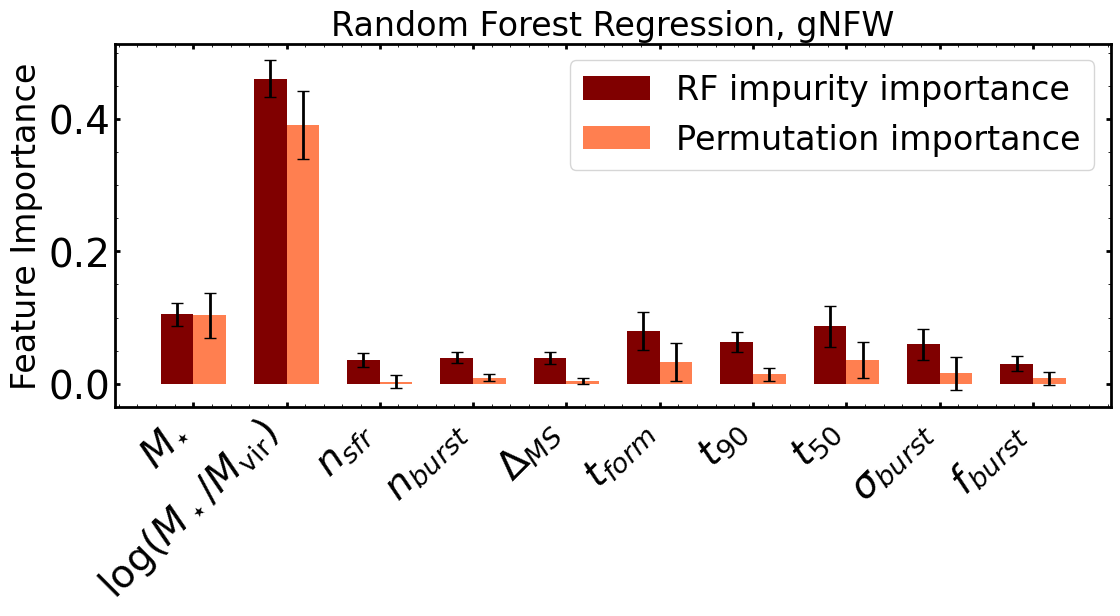}
      \centering
\caption{Bar chart showing the results from the RF regression predicting the inner slope, $\gamma$ (top left: DC14 fits; top right: DC14-scatter fits; bottom: gNFW) for the  sample using the following input features:  $M_\star$,  ( $M_\star/M_{\rm halo}$ for gNFW), $n_{\rm SFR}$, $n_{\rm burst}$, $\Delta_{\rm MS}$, $t_{\rm form}$, $t_{50}$, $t_{90}$, $\sigma_{\rm burst}$ and $f_{\rm burst}$. The dark-red bars show the mean impurity importance, while the orange bars show the permutation importance.  The error bars show the standard deviations across our MC realisations, used to account for measurement uncertainties.  }
\label{RFR}
\end{figure*}

We report our results of the RF regression for the DC14, DC14-scatter and gNFW fits in the top-left,  top-right and bottom sides of  Fig. \ref{RFR}, respectively.  
For the DC14 fits (Fig. \ref{RFR}-left),  $M_\star$ clearly emerges as the dominant predictive feature for the inner slope $\gamma$, with consistently high importance across both metrics. The mean decrease in impurity (MDI) assigns $M_\star$ a relative importance of $\sim 0.5$, while the permutation importance evaluated on the test set yields a lower but still significant contribution of $\sim 0.4$.  In contrast, all SFH-related parameters exhibit systematically lower importance.  Although some SFH features receive non-negligible MDI values (typically $\sim0.05-0.1$), this does not translate into additional predictive power when assessed through permutation importance, likely reflecting covariance between these input features.

As expected, the DC14-scatter results are fully consistent with those obtained for the baseline DC14 model, with stellar mass, $M_\star$, emerging as the dominant predictive feature for the inner slope (Fig.~\ref{RFR} top-right).  These results are fully consistent with the results obtained with the PCC analysis in Appendix \ref{ppcana}. 

For the gNFW fits  (Fig. \ref{RFR}-bottom), we obtain comparable results when using the same feature set, with no significant predictive contribution from the SFH parameters. However, when also adding  $M_\star/M_{\rm vir}$ to the feature set, we find that this quantity becomes the dominant predictor of $\gamma$ (MDI and permutation importance are $\sim$0.4-0.5), while both $M_\star$ and all SFH-related parameters contribute negligibly to the predictive performance.

The DC14 analysis is consistent with the model expectation that  $\gamma$ is primarily governed by the baryon-to-halo coupling. We note that the stellar-to-halo mass ratio is not included as an independent feature in the DC14-based RF analysis, since this dependence is explicitly encoded in the model's functional form; including it would, by construction, trivially dominate the predictive ranking. In contrast, for the gNFW parametrisation, no such dependence is imposed a priori. In this case, $M_\star/M_{\rm vir}$ (when included) emerges as the dominant predictor of $\gamma$, while stellar mass and all SFH-related parameters contribute negligibly. This highlights that the primary predictive signal for $\gamma$ is carried by the baryon-to-halo mass scaling rather than by the SFH diagnostics used in this study.

\section{Discussion}
\label{discussion}

\subsection{Stellar feedback-induced cores?}

While the PCC and RF analyses (Sect.~\ref{link}) indicate that stellar mass, or more fundamentally the stellar-to-halo mass ratio, is the primary driver of variations in the inner DM slope, with SFH-related quantities contributing little independent predictive power, we nevertheless examine how $\gamma$ and the SFH diagnostics vary jointly across the galaxy population. For the subsequent analysis, we show the inner slope values obtained from the DC14 fits, and note that our conclusions remain unchanged using the DC14-scatter or gNFW fits.

\begin{figure*}
   \centering
   \includegraphics[width=0.5\textwidth,angle=0,clip=true]{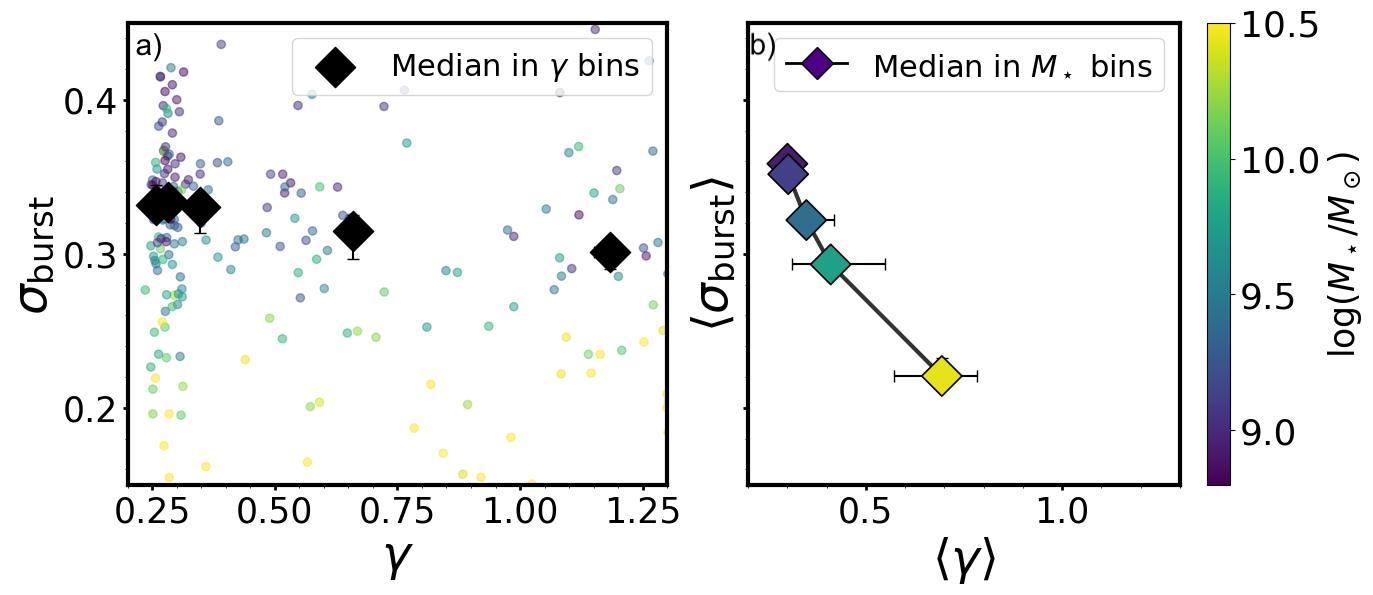}
      \centering
 \includegraphics[width=0.49\textwidth,angle=0,clip=true]{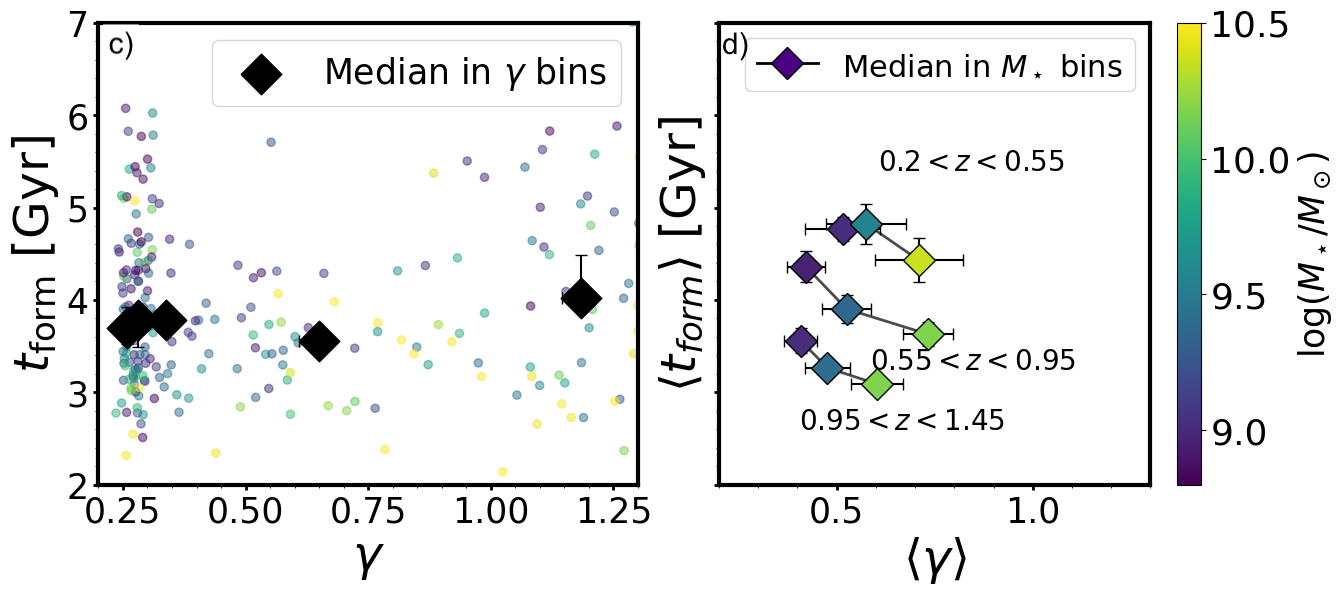}
   \centering      
   \includegraphics[width=0.49\textwidth,angle=0,clip=true]{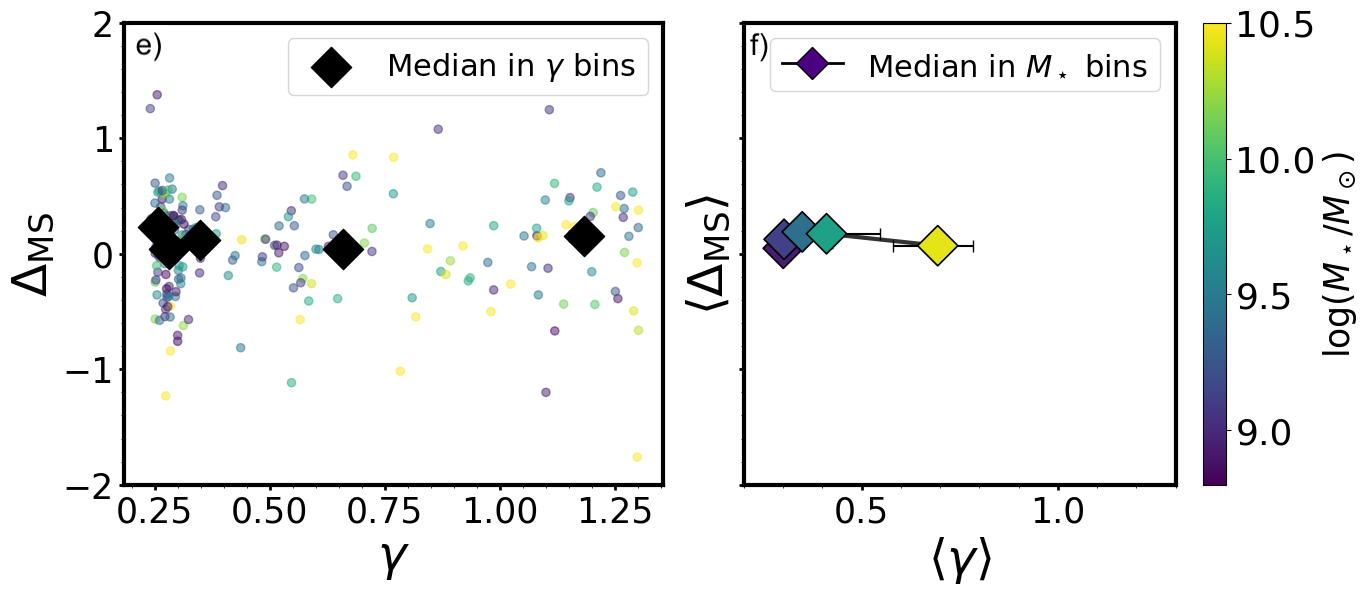}
 \centering
\caption{Relationship between the inner DM density slope, $\gamma$, and  representative SFH diagnostics for the mass-complete  sample. Panels (a), (c), and (e) show $\sigma_{\rm burst}$, formation time, $t_{\rm form}$ [Gyr] and the offset from the star-forming main sequence, $\Delta{\rm MS}$ (relative to the star-forming main sequence of \citealt{bog}), respectively, as a function of $\gamma$. Points are colour-coded by $M_\star$, with large black diamonds indicating medians in equal-occupancy $\gamma$ bins. Panels (b), (d), and (f) show instead the corresponding median $\gamma$ and SFH diagnostics in equal-occupancy $M_\star$ bins; the diamonds are colour-coded by the median $M_\star$ of each bin. For $t_{\rm form}$ (panel d), the measurements are shown separately for three redshift intervals ($0.2<z<0.55$, $0.55<z<0.95$, and $0.95<z<1.45$). In all panels, error bars denote the 16th--84th percentile intervals of the bootstrap distributions of the median, obtained by resampling galaxies within each bin. In some cases, the error bars are smaller than the data points.}

\label{gammatrends}
\end{figure*}

Figure~\ref{gammatrends}(a) shows the stochasticity parameter, $\sigma_{\rm burst}$, as a function of the inner density slope for the mass-complete sample, with individual points coloured by stellar mass. The black diamonds indicate the median  $\gamma$ and $\sigma_{\rm burst}$ values computed in five equal-occupancy $\gamma$ bins, obtained by dividing the sample into quantile intervals containing approximately the same number of galaxies. No clear trend is apparent between $\gamma$ and $\sigma_{\rm burst}$, consistent with the lack of a direct correlation between these two quantities, as demonstrated in Sect.~\ref{link}.

 Figure~\ref{gammatrends}(b) instead presents the relation after binning the sample by stellar mass. In this case, the diamonds show the median $\gamma$ and  $\sigma_{\rm burst}$ values in equal-occupancy $M_\star$ bins, colour-coded according to the median $M_\star$ in each bin.  In contrast to panel (a), the mass-binned relation reveals a clear sequence: lower-$M_\star$ bins have smaller median $\gamma$ values, corresponding to more cored inner profiles, and larger $\sigma_{\rm burst}$ values, corresponding to more stochastic or bursty SFHs. Higher-$M_\star$ bins, in contrast, show steeper inner profiles and lower $\sigma_{\rm burst}$ values, consistent with more continuous mass assembly along the star-forming main sequence.
 
A similar mass-dependent behaviour is observed for the burst fraction, $f_{\rm burst}$, which, like $\sigma_{\rm burst}$, directly probes the contribution of elevated SF episodes to stellar mass assembly. Lower-$M_\star$ bins are characterised by more cored inner slopes and higher $f_{\rm burst}$ values, whereas higher-$M_\star$ bins preferentially exhibit cuspier profiles and lower burst mass fractions. The mass-binned relation between $n_{\rm SFR}$ and $\gamma$ follows the same qualitative behaviour, with lower-$M_\star$ systems showing both shallower inner slopes and higher values of $n_{\rm SFR}$, consistent with either more bursty behaviour over the past 100 Myr compared to the average SFR over the history of the galaxy, or secular evolution along rising SFHs.

Figure~\ref{gammatrends}(c) presents the formation time in [Gyr] as a function of $\gamma$. As this panel illustrates, there is no clear correlation between $\gamma$ and $t_{\rm form}$, as already shown in Sect.~\ref{link}. 

Figure~\ref{gammatrends}(d) instead shows the median values of $\gamma$ and $t_{\rm form}$ computed in five stellar mass bins, separately for three redshift intervals. Across all three $z$ intervals, higher-$M_\star$ bins are consistently associated with both steeper inner density profiles and earlier mass assembly times, whereas lower-$M_\star$  bins preferentially host galaxies with shallower inner density profiles and later formation epochs. The same qualitative behaviour is observed when using alternative formation-time indicators, such as $t_{50}$ and $t_{90}$.  

We note that the trends in panels (b) and (d) should not be interpreted as evidence for a direct causal connection between $\sigma_{\rm burst}$ or $t_{\rm form}$ and the inner DM slope. Instead, these trends reflect the shared dependence of both the SFH diagnostics and $\gamma$ on $M_\star$. 

In contrast, Figure~\ref{gammatrends}(e) and (f) show the corresponding analysis for the offset from the star-forming main sequence, $\Delta_{\rm MS}$. As shown in panel (e), no clear correlation is found between $\gamma$ and $\Delta_{\rm MS}$, consistent with the results presented in Sect.~\ref{link} and in \cite{ciocan}. Similarly, the $M_\star$-binned analysis in panel (f) reveals no coherent sequence between the two quantities. The mass-binned analysis also reveals no clear trend between $\gamma$ and $n_{\rm burst}$. Unlike $\sigma_{\rm burst}$, $f_{\rm burst}$, and the formation-time diagnostics, both $\Delta_{\rm MS}$ and $n_{\rm burst}$ primarily probe the recent star-forming activity and may therefore be less sensitive to the cumulative SFH variability that is predicted to influence the long-term evolution of the central DM distribution (e.g. \citealt{gov}).  

Taken together, our results show that the inner DM slope and the SFH diagnostics probing the lifetime of the galaxy vary systematically along the stellar-mass sequence, while the PCC and RF analyses find little evidence that these SFH indicators provide independent predictive power on $\gamma$ beyond the dependence on stellar mass. Lower-$M_\star$ galaxies tend to have more extended and burstier SFHs and shallower, more cored inner profiles, whereas higher-$M_\star$ galaxies tend to have more continuous SFHs and steeper, cuspier profiles. These mass-binned population-level trends do not establish a direct causal connection between the SFH and the inner DM slope. They do, however, provide a useful context for comparing our results with the predictions of feedback-driven core-formation models. In these models, baryonic feedback can modify the central DM distribution through repeated gas outflows due to feedback processes and associated fluctuations in the gravitational potential (e.g. \citealt{nav}; \citealt{rg}; \citealt{gov}; \citealt{pon}; \citealt{tey}; \citealt{free}; \citealt{lifre}; \citealt{Jackson}).  Within this framework, feedback is expected to be most effective in lower-mass galaxies, where the energy injected by stellar processes is sufficient to perturb the central potential. Such galaxies typically experience more extended and bursty SFHs, which can drive repeated potential fluctuations and transform initially cuspy DM profiles into shallower inner distributions. More massive galaxies, in contrast, assemble their stellar content more rapidly and reside in deeper potential wells, making their central DM distributions less susceptible to this form of feedback-driven transformation and allowing steeper profiles to persist.

The NIHAO and FIRE-2 hydrodynamical simulations further supported this scenario: \cite{Sarrato} showed that variations in the burstiness and duration of SFHs contribute to the scatter in the relation between inner DM slopes and stellar-to-halo mass ratio. In particular, cored galaxies in the bright dwarf regime were found to have more bursty and more extended SFHs than their cuspier counterparts. Similarly, \cite{nu} analysed the Mochima cosmological zoom-in simulations and found that galaxies assembling most of their stellar mass at early times tend to preserve cuspy DM profiles, whereas galaxies with more extended SFHs are more likely to develop cores, in agreement with the trends observed in  Figure~\ref{gammatrends}, panels (b) and (d).  

Observational studies provide a complementary link: for example, \cite{ml} found that lower-$M_\star$ galaxies experience more numerous star-formation episodes (SFEs) and build their $M_\star$ over longer timescales, corresponding to later $t_{\rm form}$ (or $t_{90}$ in their analysis), than their more massive counterparts. The association between extended SFHs and a larger number of SFEs provides a possible link to the feedback-driven core-formation scenario, in which repeated SFEs generate successive fluctuations in the gravitational potential and cumulatively modify the central DM distribution. The tendency for galaxies with later $t_{\rm form}$ to exhibit shallower, more cored profiles is therefore qualitatively consistent with this picture. Although our sample does not reveal a direct one-to-one relation between $\gamma$ and the individual SFH diagnostics, the mass-binned association between $\gamma$ and formation time, together with the SFE--$t_{\rm form}$ relation reported by \cite{ml}, supports a scenario in which more extended and stochastic SFHs may favour the development of shallower DM profiles.  We stress, however, that this consistency is at the level of the $M_\star$-binned population trend, not a direct, independent effect:  while the mass sets the overall trend in $\gamma$, the SFH provides a plausible physical pathway through which this trend is realised.
 
Alternatively, the lack of a direct one-to-one correlation between the inner DM density slope and SFH indicators (beyond that set by the mass) could point to DM physics beyond collisionless $\Lambda$CDM. For example, self-interacting DM can produce central cores through dark matter self-scattering \citep{sidm}, while ultralight axion-like (fuzzy) DM can generate shallow central density profiles through quantum effects \citep{hu}. In such scenarios, the inner DM structure need not be directly coupled to the galaxy's SFH, providing alternative explanations for the absence of a detectable direct SFH--$\gamma$ correlation.



\subsection{Model limitations and caveats}
The observational results presented in this work indicate that the stellar-to-halo mass ratio is the primary parameter governing the inferred inner DM density slope, while reconstructed SFH properties provide little additional predictive power (Fig.~\ref{RFR}). The analysis recovers the same dependence in both the DC14 \citep{dc14} framework, where the inner slope is explicitly parameterised as a function of the stellar-to-halo mass ratio, and the independent gNFW framework, where $\gamma$ is treated as a completely free parameter (Fig.~\ref{gammaf}-left). The agreement between these two approaches indicates that the observed correlation is not imposed by the adopted halo parameterisation, but is instead supported by the data, providing observational support for the physical picture in which the stellar-to-halo mass ratio regulates the efficiency of core formation, as predicted by various hydrodynamical simulations \citep{dc14,toll,fre,read,Lazar,Jackson,aza}. The Bayesian model comparison further supports this interpretation, showing that the DC14 model provides statistically superior fits for a substantial fraction of the sample compared to gNFW (Fig.~\ref{bayes}).

The absence of a direct correlation between SFH indicators and the inner DM slope should nevertheless be interpreted with caution, because the timescales probed by our SFH diagnostics may not coincide with those governing feedback-driven core formation. In these scenarios, the outflows and potential fluctuations that transfer energy to the DM can occur on timescales much shorter than the dynamical time, whereas the subsequent redistribution of the DM and establishment of a core can proceed over longer relaxation timescales, typically of order a Gyr or longer \citep[e.g.][]{2016MNRAS.461.1745E,fre,Hashim23}. Thus, even if bursty SF is the physical driver of core formation, our SFH-indicators may not retain a direct signature of the burst episodes responsible for the process.  The lack of a residual correlation between our SFH diagnostics and $\gamma$ may therefore reflect, at least in part, a mismatch between the timescales probed by the SFH diagnostics and those over which the DM responds to feedback.

Several other caveats should also be considered when interpreting these results. First, the SFH diagnostics were inferred from Bayesian SED fitting and are therefore model-dependent. Although the non-parametric continuity model \citep{Leja19,Leja19b} used in this work is substantially more flexible than traditional parametric SFHs, and has been shown to accurately recover the SFHs of mock galaxies (see also \citealt{Wan} for a discussion), the inferred SFHs remain conditioned by the adopted stellar population synthesis models, dust attenuation prescription, and priors. Moreover, uncertainties in the underlying stellar evolution models, stellar spectral libraries, dust prescriptions, and the assumed IMF introduce systematic uncertainties in the derived stellar masses and SFRs, which can reach $\sim0.2$--$0.3$ dex \citep[e.g.][]{Mobasher}. These uncertainties naturally propagate into the SFH diagnostics employed throughout this work.

Second, the continuity prior regularises rapid variations in the SFR, implying that quantities such as $n_{\rm burst}$ should be regarded as conservative estimates of recent SF variability. Likewise, the temporal resolution achievable from broadband photometry limits the recovery of stochastic SF episodes occurring on timescales shorter than a few tens of Myr. Consequently, connections between very short-timescale burstiness and the inner DM density profile may remain inaccessible with the present methodology.

Finally, the robustness of the 3D disk--halo decomposition itself, including the assumed stellar and gas distributions, and adopted halo parametrisations, was extensively validated through mock observations and systematic tests presented in \cite{ciocan}. Nonetheless, a caveat of our approach is the assumption of purely circular motions in axisymmetric disks. Non-circular motion, such as inflows or outflows, can bias dynamical measurements and mimic a more cored distribution \citep[e.g.][]{Oman2019,Marasco}.  However, while non-circular motions are not explicitly modelled, the size of the sample, our selection criteria,  the generally low residuals of the best-fitting models, and the recovery of halo scaling relations (Fig. \ref{halos})  suggest that they are unlikely to systematically dominate the inferred halo properties for most galaxies. \footnote{As an additional test, we repeated our analysis using a stricter $v/\sigma>3$ selection instead of the fiducial $v/\sigma>1$ cut. We found that the main results remain unchanged, suggesting that they are robust to this more stringent kinematic selection.} 

Taken together, these considerations suggest that while systematic uncertainties may contribute to the scatter in the inferred halo and SFH properties, they are unlikely to alter the principal conclusion of this work: that stellar mass, or more fundamentally the stellar-to-halo mass ratio, encapsulates the dominant dependence of the inner DM density profile, with no detectable residual dependence on reconstructed SFH properties within the precision of the current observations and modelling framework.  


\section{Summary and conclusion}
\label{concl}
In this study, we investigate the potential connection between DM density profiles and SFHs for a sample of 375 intermediate-redshift SFGs ($0.1 < z < 1.43$), spanning a broad stellar mass range ($7< \log(M_\star/M_\odot) < 11$). We combined deep MUSE integral field spectroscopy from three surveys (MHUDF - \citealt{udf1,udf2}; MUSCATEL - Urrutia et al., in prep.; and lensing cluster fields- \citealt{Richard, Claeyssens}) with ancillary multi-wavelength photometry covering the UV to mid-infrared (\citealt{Johnson, ship}).
We performed a 3D disk–halo decomposition following \cite{nicolas, ciocan} that simultaneously accounts for stellar, gaseous, and DM components, including corrections for pressure support. The DM density profiles were parameterised using three models: DC14 \citep{dc14}, a DC14 variant incorporating scatter in the $\gamma$–$\log(X)$ relation (DC14-scatter), and a generalised NFW  profile \citep{genz, genz2, shach}. Stellar population parameters and SFH-diagnostics were derived using \texttt{BAGPIPES} \citep{Carnall18}, adopting a non-parametric continuity SFH model following \cite{Leja19, Leja19b}.  Our main findings are summarised below:

\begin{itemize}
    \item We performed a Bayesian comparison of the three DM halo parametrisations (DC14, DC14-scatter, and gNFW), finding that overall, the DC14-based models are more likely given the data compared to gNFW, while DC14 and DC14-scatter yield statistically indistinguishable fits for most galaxies (Fig. \ref{bayes}).

\item We recover the main halo scaling relations expected in $\Lambda$CDM-based models using the disk--halo decomposition (Fig.~\ref{halos}). The sample shows a strong anti-correlation between the DM fraction within $R_{\rm e}$ and stellar mass surface density, follows the expected halo structural relation between $M_{\rm vir}$ and $r_{\rm s,DMO}$, and broadly reproduces the stellar-to-halo mass relation, albeit with a larger scatter.
    
    \item  The inferred inner DM slopes show a coherent dependence on the stellar-to-halo mass ratio. Although the DC14-scatter model allows deviations from the fiducial DC14 $\gamma$--$M_\star/M_{\rm halo}$ relation, the inferred slopes exhibit only limited scatter around this relation. Moreover, the same trend is recovered with the gNFW model, which does not impose any dependency between $\gamma$ and $M_\star/M_{\rm halo}$ (Fig.~\ref{gammaf}), supporting a genuine connection between galaxy formation efficiency and inner DM structure.
    
    \item Across all models, the majority of galaxies exhibit shallow inner density profiles, with $\sim 60$--$70\%$ of systems classified as cored ($\gamma < 0.5$), and no strong systematic offsets between halo parametrisations  (Fig. \ref{gammaf}).
    
    \item A weak positive trend is observed between the inner DM slope and $M_\star$, with more massive galaxies exhibiting steeper inner density profiles (Fig.~\ref{gammaf}). However, this relation is weaker than the $\gamma$--$M_\star/M_{\rm halo}$ correlation, consistent with simulations indicating that the stellar-to-halo mass ratio is a more fundamental parameter governing the response of DM halos to galaxy formation processes.

   \item Random Forest analysis shows that the variation in the inner density slope is primarily driven by  $M_\star/M_{\rm halo}$ (or by stellar mass when using galaxy properties only) for all models,  accounting for the largest fraction of the predictive power. In contrast, SFH diagnostics such as $n_{\rm SFR}$, $f_{\rm burst}$, $\Delta_{\rm MS}$, $t_{\rm form}$, $t_{50}$, $t_{90}$, $\sigma_{\rm burst}$, and $n_{\rm burst}$ do not provide significant independent predictive power on $\gamma$ (Fig. \ref{RFR}). This conclusion is further supported by the partial correlation analysis (Appendix~\ref{ppcana}), which finds no residual correlation between SFH properties and $\gamma$ after controlling for stellar mass and redshift.

    \item Mass-binned population trends show that higher-$M_\star$ galaxies tend to be more cuspy,  less bursty and have earlier formation times. In contrast,  lower-mass systems are more often cored, have more bursty SFHs, and later mass assembly times (Fig. \ref{gammatrends}).  Nonetheless, these apparent correlations between DM slopes and SFH-diagnostics are  driven by their mutual dependence on $M_\star$ rather than indicating direct causal links.

\end{itemize}
Overall, our results indicate that the diversity of inner DM density profiles in intermediate-redshift galaxies is primarily linked to the efficiency of galaxy formation, as traced by the stellar-to-halo mass ratio. While population trends suggest that galaxies with more extended and bursty SFHs preferentially host shallower inner density profiles, these correlations are largely driven by their covariance with stellar mass rather than by an independent dependence of $\gamma$ on SFH properties. Future observations combining larger samples of spatially resolved galaxies with improved constraints on stellar populations and gas kinematics will be essential to further disentangle the interplay between galaxy formation histories and DM halo evolution.

\begin{acknowledgements}
 B.C.,  N.B.  and J.F. acknowledge support from the  ANR DARK grant (ANR-22-CE31-0006).
This research made use of the following open source software:  {\textsc{ GalPaK$^{\rm 3D}$}} \citep{galpak}, \texttt{BAGPIPES} \citep{Carnall18},  \texttt{Astropy} \citep{astropy}, \texttt{numpy} \citep{numpy}, \texttt{matplotlib} \citep{plot}, \texttt{MPDAF} \citep{mpdaf},  \texttt{CAMEL} \citep{camel},   \texttt{scikit-learn} \citep{Pedregosa} and \texttt{Roxy} \citep{roxy}.
\end{acknowledgements}


\bibliographystyle{aa}
\bibliography{references_fin}

@ARTICLE{alle,
       author = {{Allaert}, F. and {Gentile}, G. and {Baes}, M.},
        title = "{Testing baryon-induced core formation in {\ensuremath{\Lambda}}CDM: A comparison of the DC14 and coreNFW dark matter halo models on galaxy rotation curves}",
      journal = {\aap},
         year = 2017,
        month = sep,
       volume = {605},
          eid = {A55},
        pages = {A55},
          doi = {10.1051/0004-6361/201730402},
archivePrefix = {arXiv},
       eprint = {1706.08296},
 primaryClass = {astro-ph.GA},
       adsurl = {https://ui.adsabs.harvard.edu/abs/2017A&A...605A..55A}
}

@ARTICLE{2022MNRAS.511.4464A,
       author = {{Atek}, Hakim and {Furtak}, Lukas J. and {Oesch}, Pascal and {van Dokkum}, Pieter and {Reddy}, Naveen and {Contini}, Thierry and {Illingworth}, Garth and {Wilkins}, Stephen},
        title = "{The star formation burstiness and ionizing efficiency of low-mass galaxies}",
      journal = {\mnras},
         year = 2022,
        month = apr,
       volume = {511},
       number = {3},
        pages = {4464-4479},
          doi = {10.1093/mnras/stac360},
archivePrefix = {arXiv},
       eprint = {2202.04081},
 primaryClass = {astro-ph.GA},
       adsurl = {https://ui.adsabs.harvard.edu/abs/2022MNRAS.511.4464A}
}

@article{aza,
    author = {Azartash-Namin, B. and Engelhardt, A. and Munshi, F. and Keller, B. W. and Brooks, A. M. and others},
    title = {{Testing the Cuspy-core Problem in the FIRE-2 Simulations Using Rotation Curves of Low-mass Galaxies}},
    year = {2024},
    journal = {ApJ},
    volume = {970},
    articleno = {40A},
    doi = {10.3847/1538-4357/ad4b88},
}

@INPROCEEDINGS{bacon10,
       author = {{Bacon}, R. and {Accardo}, M. and {Adjali}, L. and {Anwand}, H. and {Bauer}, S. and {Biswas}, I. and {Blaizot}, J. and {Boudon}, D. and {Brau-Nogue}, S. and {Brinchmann}, J. and {Caillier}, P. and {Capoani}, L. and {Carollo}, C.~M. and {Contini}, T. and {Couderc}, P. and {Daguis{\'e}}, E. and {Deiries}, S. and {Delabre}, B. and {Dreizler}, S. and {Dubois}, J. and {Dupieux}, M. and {Dupuy}, C. and {Emsellem}, E. and {Fechner}, T. and {Fleischmann}, A. and {Fran{\c{c}}ois}, M. and {Gallou}, G. and {Gharsa}, T. and {Glindemann}, A. and {Gojak}, D. and {Guiderdoni}, B. and {Hansali}, G. and {Hahn}, T. and {Jarno}, A. and {Kelz}, A. and {Koehler}, C. and {Kosmalski}, J. and {Laurent}, F. and {Le Floch}, M. and {Lilly}, S.~J. and {Lizon}, J.-L. and {Loupias}, M. and {Manescau}, A. and {Monstein}, C. and {Nicklas}, H. and {Olaya}, J.-C. and {Pares}, L. and {Pasquini}, L. and {P{\'e}contal-Rousset}, A. and {Pell{\'o}}, R. and {Petit}, C. and {Popow}, E. and {Reiss}, R. and {Remillieux}, A. and {Renault}, E. and {Roth}, M. and {Rupprecht}, G. and {Serre}, D. and {Schaye}, J. and {Soucail}, G. and {Steinmetz}, M. and {Streicher}, O. and {Stuik}, R. and {Valentin}, H. and {Vernet}, J. and {Weilbacher}, P. and {Wisotzki}, L. and {Yerle}, N.},
        title = "{The MUSE second-generation VLT instrument}",
    booktitle = {Ground-based and Airborne Instrumentation for Astronomy III},
         year = 2010,
       editor = {{McLean}, Ian S. and {Ramsay}, Suzanne K. and {Takami}, Hideki},
       series = {Society of Photo-Optical Instrumentation Engineers (SPIE) Conference Series},
       volume = {7735},
        month = jul,
          eid = {773508},
        pages = {773508},
          doi = {10.1117/12.856027},
archivePrefix = {arXiv},
       eprint = {2211.16795},
 primaryClass = {astro-ph.IM},
       adsurl = {https://ui.adsabs.harvard.edu/abs/2010SPIE.7735E..08B}
}

@ARTICLE{mpdaf,
       author = {{Piqueras}, Laure and {Conseil}, Simon and {Shepherd}, Martin and {Bacon}, Roland and {Leclercq}, Floriane and {Richard}, Johan},
        title = "{MPDAF - A Python package for the analysis of VLT/MUSE data}",
      journal = {arXiv e-prints},
         year = 2017,
        month = oct,
          eid = {arXiv:1710.03554},
        pages = {arXiv:1710.03554},
          doi = {10.48550/arXiv.1710.03554},
archivePrefix = {arXiv},
       eprint = {1710.03554},
 primaryClass = {astro-ph.IM},
       adsurl = {https://ui.adsabs.harvard.edu/abs/2017arXiv171003554P}
}

@ARTICLE{udf1,
       author = {{Bacon}, Roland and {Conseil}, Simon and {Mary}, David and {Brinchmann}, Jarle and {Shepherd}, Martin and {Akhlaghi}, Mohammad and {Weilbacher}, Peter M. and {Piqueras}, Laure and {Wisotzki}, Lutz and {Lagattuta}, David and {Epinat}, Benoit and {Guerou}, Adrien and {Inami}, Hanae and {Cantalupo}, Sebastiano and {Courbot}, Jean Baptiste and {Contini}, Thierry and {Richard}, Johan and {Maseda}, Michael and {Bouwens}, Rychard and {Bouch{\'e}}, Nicolas and {Kollatschny}, Wolfram and {Schaye}, Joop and {Marino}, Raffaella Anna and {Pello}, Roser and {Herenz}, Christian and {Guiderdoni}, Bruno and {Carollo}, Marcella},
        title = "{The MUSE Hubble Ultra Deep Field Survey. I. Survey description, data reduction, and source detection}",
      journal = {\aap},
         year = 2017,
        month = dec,
       volume = {608},
          eid = {A1},
        pages = {A1},
          doi = {10.1051/0004-6361/201730833},
archivePrefix = {arXiv},
       eprint = {1710.03002},
 primaryClass = {astro-ph.GA},
       adsurl = {https://ui.adsabs.harvard.edu/abs/2017A&A...608A...1B}
}

@ARTICLE{udf2,
       author = {{Bacon}, Roland and {Brinchmann}, Jarle and {Conseil}, Simon and {Maseda}, Michael and {Nanayakkara}, Themiya and {Wendt}, Martin and {Bacher}, Raphael and {Mary}, David and {Weilbacher}, Peter M. and {Krajnovi{\'c}}, Davor and {Boogaard}, Leindert and {Bouch{\'e}}, Nicolas and {Contini}, Thierry and {Epinat}, Beno{\^\i}t and {Feltre}, Anna and {Guo}, Yucheng and {Herenz}, Christian and {Kollatschny}, Wolfram and {Kusakabe}, Haruka and {Leclercq}, Floriane and {Michel-Dansac}, L{\'e}o and {Pello}, Roser and {Richard}, Johan and {Roth}, Martin and {Salvignol}, Gregory and {Schaye}, Joop and {Steinmetz}, Matthias and {Tresse}, Laurence and {Urrutia}, Tanya and {Verhamme}, Anne and {Vitte}, Eloise and {Wisotzki}, Lutz and {Zoutendijk}, Sebastiaan L.},
        title = "{The MUSE Hubble Ultra Deep Field surveys: Data release II}",
      journal = {\aap},
         year = 2023,
        month = feb,
       volume = {670},
          eid = {A4},
        pages = {A4},
          doi = {10.1051/0004-6361/202244187},
archivePrefix = {arXiv},
       eprint = {2211.08493},
 primaryClass = {astro-ph.GA},
       adsurl = {https://ui.adsabs.harvard.edu/abs/2023A&A...670A...4B}
}

@ARTICLE{Bait,
       author = {{Bait}, Omkar and {Barway}, Sudhanshu and {Wadadekar}, Yogesh},
        title = "{On the interdependence of galaxy morphology, star formation and environment in massive galaxies in the nearby Universe}",
      journal = {\mnras},
         year = 2017,
        month = nov,
       volume = {471},
       number = {3},
        pages = {2687-2702},
          doi = {10.1093/mnras/stx1688},
archivePrefix = {arXiv},
       eprint = {1707.00568},
 primaryClass = {astro-ph.GA},
       adsurl = {https://ui.adsabs.harvard.edu/abs/2017MNRAS.471.2687B}
}

@ARTICLE{Baker,
       author = {{Baker}, William M. and {Maiolino}, Roberto and {Belfiore}, Francesco and {Curti}, Mirko and {Bluck}, Asa F.~L. and {Lin}, Lihwai and {Ellison}, Sara L. and {Thorp}, Mallory and {Pan}, Hsi-An},
        title = "{The metallicity's fundamental dependence on both local and global galactic quantities}",
      journal = {\mnras},
         year = 2023,
        month = feb,
       volume = {519},
       number = {1},
        pages = {1149-1170},
          doi = {10.1093/mnras/stac3594},
archivePrefix = {arXiv},
       eprint = {2210.03755},
 primaryClass = {astro-ph.GA},
       adsurl = {https://ui.adsabs.harvard.edu/abs/2023MNRAS.519.1149B}
}

@ARTICLE{roxy,
       author = {{Bartlett}, Deaglan J. and {Desmond}, Harry},
        title = "{Marginalised Normal Regression: Unbiased curve fitting in the presence of x-errors}",
      journal = {The Open Journal of Astrophysics},
         year = 2023,
        month = nov,
       volume = {6},
          eid = {42},
        pages = {42},
          doi = {10.21105/astro.2309.00948},
archivePrefix = {arXiv},
       eprint = {2309.00948},
 primaryClass = {stat.ME},
       adsurl = {https://ui.adsabs.harvard.edu/abs/2023OJAp....6E..42B}
}

@ARTICLE{Behroozi,
       author = {{Behroozi}, Peter and {Wechsler}, Risa H. and {Hearin}, Andrew P. and {Conroy}, Charlie},
        title = "{UNIVERSEMACHINE: The correlation between galaxy growth and dark matter halo assembly from z = 0-10}",
      journal = {\mnras},
         year = 2019,
        month = sep,
       volume = {488},
       number = {3},
        pages = {3143-3194},
          doi = {10.1093/mnras/stz1182},
archivePrefix = {arXiv},
       eprint = {1806.07893},
 primaryClass = {astro-ph.GA},
       adsurl = {https://ui.adsabs.harvard.edu/abs/2019MNRAS.488.3143B}
}

@ARTICLE{Bluck,
       author = {{Bluck}, Asa F.~L. and {Maiolino}, Roberto and {Brownson}, Simcha and {Conselice}, Christopher J. and {Ellison}, Sara L. and {Piotrowska}, Joanna M. and {Thorp}, Mallory D.},
        title = "{The quenching of galaxies, bulges, and disks since cosmic noon. A machine learning approach for identifying causality in astronomical data}",
      journal = {\aap},
         year = 2022,
        month = mar,
       volume = {659},
          eid = {A160},
        pages = {A160},
          doi = {10.1051/0004-6361/202142643},
archivePrefix = {arXiv},
       eprint = {2201.07814},
 primaryClass = {astro-ph.GA},
       adsurl = {https://ui.adsabs.harvard.edu/abs/2022A&A...659A.160B}
}

@ARTICLE{bog,
       author = {{Boogaard}, Leindert A. and {Brinchmann}, Jarle and {Bouch{\'e}}, Nicolas and {Paalvast}, Mieke and {Bacon}, Roland and {Bouwens}, Rychard J. and {Contini}, Thierry and {Gunawardhana}, Madusha L.~P. and {Inami}, Hanae and {Marino}, Raffaella A. and {Maseda}, Michael V. and {Mitchell}, Peter and {Nanayakkara}, Themiya and {Richard}, Johan and {Schaye}, Joop and {Schreiber}, Corentin and {Tacchella}, Sandro and {Wisotzki}, Lutz and {Zabl}, Johannes},
        title = "{The MUSE Hubble Ultra Deep Field Survey. XI. Constraining the low-mass end of the stellar mass - star formation rate relation at z < 1}",
      journal = {\aap},
         year = 2018,
        month = nov,
       volume = {619},
          eid = {A27},
        pages = {A27},
          doi = {10.1051/0004-6361/201833136},
archivePrefix = {arXiv},
       eprint = {1808.04900},
 primaryClass = {astro-ph.GA},
       adsurl = {https://ui.adsabs.harvard.edu/abs/2018A&A...619A..27B}
}

@ARTICLE{Bose,
       author = {{Bose}, Sownak and {Frenk}, Carlos S. and {Jenkins}, Adrian and {Fattahi}, Azadeh and {G{\'o}mez}, Facundo A. and {Grand}, Robert J.~J. and {Marinacci}, Federico and {Navarro}, Julio F. and {Oman}, Kyle A. and {Pakmor}, R{\"u}diger and {Schaye}, Joop and {Simpson}, Christine M. and {Springel}, Volker},
        title = "{No cores in dark matter-dominated dwarf galaxies with bursty star formation histories}",
      journal = {\mnras},
         year = 2019,
        month = jul,
       volume = {486},
       number = {4},
        pages = {4790-4804},
          doi = {10.1093/mnras/stz1168},
archivePrefix = {arXiv},
       eprint = {1810.03635},
 primaryClass = {astro-ph.GA},
       adsurl = {https://ui.adsabs.harvard.edu/abs/2019MNRAS.486.4790B}
}

@ARTICLE{galpak,
       author = {{Bouch{\'e}}, N. and {Carfantan}, H. and {Schroetter}, I. and {Michel-Dansac}, L. and {Contini}, T.},
        title = "{GalPak$^{3D}$: A Bayesian Parametric Tool for Extracting Morphokinematics of Galaxies from 3D Data}",
      journal = {\aj},
         year = 2015,
        month = sep,
       volume = {150},
       number = {3},
          eid = {92},
        pages = {92},
          doi = {10.1088/0004-6256/150/3/92},
archivePrefix = {arXiv},
       eprint = {1501.06586},
 primaryClass = {astro-ph.IM},
       adsurl = {https://ui.adsabs.harvard.edu/abs/2015AJ....150...92B}
}

@ARTICLE{nicolas1,
       author = {{Bouch{\'e}}, Nicolas F. and {Genel}, Shy and {Pellissier}, Alisson and {Dubois}, C{\'e}dric and {Contini}, Thierry and {Epinat}, Beno{\^\i}t and {Pillepich}, Annalisa and {Krajnovi{\'c}}, Davor and {Nelson}, Dylan and {Abril-Melgarejo}, Valentina and {Richard}, Johan and {Boogaard}, Leindert and {Maseda}, Michael and {Mercier}, Wilfried and {Bacon}, Roland and {Steinmetz}, Matthias and {Vogelsberger}, Mark},
        title = "{The MUSE Hubble Ultra Deep Field Survey. XVI. The angular momentum of low-mass star-forming galaxies: A cautionary tale and insights from TNG50}",
      journal = {\aap},
         year = 2021,
        month = oct,
       volume = {654},
          eid = {A49},
        pages = {A49},
          doi = {10.1051/0004-6361/202040225},
archivePrefix = {arXiv},
       eprint = {2101.12250},
 primaryClass = {astro-ph.GA},
       adsurl = {https://ui.adsabs.harvard.edu/abs/2021A&A...654A..49B}
}

@ARTICLE{nicolas,
       author = {{Bouch{\'e}}, Nicolas F. and {Bera}, Samuel and {Krajnovi{\'c}}, Davor and {Emsellem}, Eric and {Mercier}, Wilfried and {Schaye}, Joop and {Epinat}, Beno{\^\i}t and {Richard}, Johan and {Zoutendijk}, Sebastiaan L. and {Abril-Melgarejo}, Valentina and {Brinchmann}, Jarle and {Bacon}, Roland and {Contini}, Thierry and {Boogaard}, Leindert and {Wisotzki}, Lutz and {Maseda}, Michael and {Steinmetz}, Matthias},
        title = "{The MUSE Extremely Deep Field: Evidence for SFR-induced cores in dark-matter dominated galaxies at z ≃ 1}",
      journal = {\aap},
         year = 2022,
        month = feb,
       volume = {658},
          eid = {A76},
        pages = {A76},
          doi = {10.1051/0004-6361/202141762},
archivePrefix = {arXiv},
       eprint = {2109.07545},
 primaryClass = {astro-ph.GA},
       adsurl = {https://ui.adsabs.harvard.edu/abs/2022A&A...658A..76B}
}

@PHDTHESIS{1978PhDT.......195B,
       author = {{Bosma}, A.},
        title = "{The distribution and kinematics of neutral hydrogen in spiral galaxies of various morphological types}",
       school = {University of Groningen, Netherlands},
         year = 1978,
        month = mar,
       adsurl = {https://ui.adsabs.harvard.edu/abs/1978PhDT.......195B}
}

@ARTICLE{br,
       author = {{Brooks}, Alyson M. and {Zolotov}, Adi},
        title = "{Why Baryons Matter: The Kinematics of Dwarf Spheroidal Satellites}",
      journal = {\apj},
         year = 2014,
        month = may,
       volume = {786},
       number = {2},
          eid = {87},
        pages = {87},
          doi = {10.1088/0004-637X/786/2/87},
archivePrefix = {arXiv},
       eprint = {1207.2468},
 primaryClass = {astro-ph.CO},
       adsurl = {https://ui.adsabs.harvard.edu/abs/2014ApJ...786...87B}
}

@ARTICLE{Brammer,
       author = {{Brammer}, G.~B. and {Marchesini}, D. and {Labb{\'e}}, I. and {Spitler}, L. and {Lange-Vagle}, D. and {Barker}, E.~A. and {Tanaka}, M. and {Fontana}, A. and {Galametz}, A. and {Ferr{\'e}-Mateu}, A. and {Kodama}, T. and {Lundgren}, B. and {Martis}, N. and {Muzzin}, A. and {Stefanon}, M. and {Toft}, S. and {van der Wel}, A. and {Vulcani}, B. and {Whitaker}, K.~E.},
        title = "{Ultra-deep K-band Imaging of the Hubble Frontier Fields}",
      journal = {The Messenger},
         year = 2016,
        month = sep,
       volume = {165},
        pages = {34-37},
       adsurl = {https://ui.adsabs.harvard.edu/abs/2016Msngr.165...34B}
}

@ARTICLE{Breiman,
       author = {{Breiman}, Leo},
        title = "{Random Forests.}",
      journal = {Machine Learning},
         year = 2001,
        month = jan,
       volume = {45},
        pages = {5-32},
          doi = {10.1023/A:1010933404324},
       adsurl = {https://ui.adsabs.harvard.edu/abs/2001MachL..45....5B}
}

@ARTICLE{Brocklehurst,
       author = {{Brocklehurst}, M.},
        title = "{Calculations of level populations for the low levels of hydrogenic ions in gaseous nebulae.}",
      journal = {\mnras},
         year = 1971,
        month = jan,
       volume = {153},
        pages = {471},
          doi = {10.1093/mnras/153.4.471},
       adsurl = {https://ui.adsabs.harvard.edu/abs/1971MNRAS.153..471B}
}

@ARTICLE{BC03,
       author = {{Bruzual}, G. and {Charlot}, S.},
        title = "{Stellar population synthesis at the resolution of 2003}",
      journal = {\mnras},
         year = 2003,
        month = oct,
       volume = {344},
       number = {4},
        pages = {1000-1028},
          doi = {10.1046/j.1365-8711.2003.06897.x},
archivePrefix = {arXiv},
       eprint = {astro-ph/0309134},
 primaryClass = {astro-ph},
       adsurl = {https://ui.adsabs.harvard.edu/abs/2003MNRAS.344.1000B}
}

@ARTICLE{Buchner,
       author = {{Buchner}, J. and {Georgakakis}, A. and {Nandra}, K. and {Hsu}, L. and {Rangel}, C. and {Brightman}, M. and {Merloni}, A. and {Salvato}, M. and {Donley}, J. and {Kocevski}, D.},
        title = "{X-ray spectral modelling of the AGN obscuring region in the CDFS: Bayesian model selection and catalogue}",
      journal = {\aap},
         year = 2014,
        month = apr,
       volume = {564},
          eid = {A125},
        pages = {A125},
          doi = {10.1051/0004-6361/201322971},
archivePrefix = {arXiv},
       eprint = {1402.0004},
 primaryClass = {astro-ph.HE},
       adsurl = {https://ui.adsabs.harvard.edu/abs/2014A&A...564A.125B}
}

@ARTICLE{bull,
       author = {{Bullock}, J.~S. and {Kolatt}, T.~S. and {Sigad}, Y. and {Somerville}, R.~S. and {Kravtsov}, A.~V. and {Klypin}, A.~A. and {Primack}, J.~R. and {Dekel}, A.},
        title = "{Profiles of dark haloes: evolution, scatter and environment}",
      journal = {\mnras},
         year = 2001,
        month = mar,
       volume = {321},
       number = {3},
        pages = {559-575},
          doi = {10.1046/j.1365-8711.2001.04068.x},
archivePrefix = {arXiv},
       eprint = {astro-ph/9908159},
 primaryClass = {astro-ph},
       adsurl = {https://ui.adsabs.harvard.edu/abs/2001MNRAS.321..559B}
}

@ARTICLE{bull17,
       author = {{Bullock}, James S. and {Boylan-Kolchin}, Michael},
        title = "{Small-Scale Challenges to the {\ensuremath{\Lambda}}CDM Paradigm}",
      journal = {\araa},
         year = 2017,
        month = aug,
       volume = {55},
       number = {1},
        pages = {343-387},
          doi = {10.1146/annurev-astro-091916-055313},
archivePrefix = {arXiv},
       eprint = {1707.04256},
 primaryClass = {astro-ph.CO},
       adsurl = {https://ui.adsabs.harvard.edu/abs/2017ARA&A..55..343B}
}

@ARTICLE{Carnall19,
       author = {{Carnall}, A.~C. and {McLure}, R.~J. and {Dunlop}, J.~S. and {Cullen}, F. and {McLeod}, D.~J. and {Wild}, V. and {Johnson}, B.~D. and {Appleby}, S. and {Dav{\'e}}, R. and {Amorin}, R. and {Bolzonella}, M. and {Castellano}, M. and {Cimatti}, A. and {Cucciati}, O. and {Gargiulo}, A. and {Garilli}, B. and {Marchi}, F. and {Pentericci}, L. and {Pozzetti}, L. and {Schreiber}, C. and {Talia}, M. and {Zamorani}, G.},
        title = "{The VANDELS survey: the star-formation histories of massive quiescent galaxies at 1.0 < z < 1.3}",
      journal = {\mnras},
         year = 2019,
        month = nov,
       volume = {490},
       number = {1},
        pages = {417-439},
          doi = {10.1093/mnras/stz2544},
archivePrefix = {arXiv},
       eprint = {1903.11082},
 primaryClass = {astro-ph.GA},
       adsurl = {https://ui.adsabs.harvard.edu/abs/2019MNRAS.490..417C}
}

@ARTICLE{Carnall18,
       author = {{Carnall}, A.~C. and {McLure}, R.~J. and {Dunlop}, J.~S. and {Dav{\'e}}, R.},
        title = "{Inferring the star formation histories of massive quiescent galaxies with BAGPIPES: evidence for multiple quenching mechanisms}",
      journal = {\mnras},
         year = 2018,
        month = nov,
       volume = {480},
       number = {4},
        pages = {4379-4401},
          doi = {10.1093/mnras/sty2169},
archivePrefix = {arXiv},
       eprint = {1712.04452},
 primaryClass = {astro-ph.GA},
       adsurl = {https://ui.adsabs.harvard.edu/abs/2018MNRAS.480.4379C}
}

@ARTICLE{ciocan,
       author = {{Ciocan}, B.~I. and {Bouch{\'e}}, N.~F. and {Fensch}, J. and {Mercier}, W. and {Krajnovi{\'c}}, D. and {Richard}, J. and {Contini}, T. and {Jeanneau}, A.},
        title = "{MUSE-DARK: I. Dark matter halo properties of intermediate-z star-forming galaxies}",
      journal = {\aap},
         year = 2026,
        month = apr,
       volume = {708},
          eid = {A112},
        pages = {A112},
          doi = {10.1051/0004-6361/202557396},
archivePrefix = {arXiv},
       eprint = {2506.19721},
 primaryClass = {astro-ph.GA},
       adsurl = {https://ui.adsabs.harvard.edu/abs/2026A&A...708A.112C}
}

@ARTICLE{ciocan2,
       author = {{Ciocan}, B.~I. and {Bouch{\'e}}, N.~F. and {Fensch}, J. and {Krajnovi{\'c}}, D. and {Freundlich}, J. and {Desmond}, H. and {Famaey}, B. and {Techi}, R.},
        title = "{MUSE-DARK: III. The evolution of the radial acceleration relation at intermediate redshifts}",
      journal = {\aap},
         year = 2026,
        month = may,
       volume = {709},
          eid = {L16},
        pages = {L16},
          doi = {10.1051/0004-6361/202659230},
archivePrefix = {arXiv},
       eprint = {2604.22613},
 primaryClass = {astro-ph.GA},
       adsurl = {https://ui.adsabs.harvard.edu/abs/2026A&A...709L..16C}
}

@ARTICLE{chabrier03,
       author = {{Chabrier}, Gilles},
        title = "{Galactic Stellar and Substellar Initial Mass Function}",
      journal = {\pasp},
         year = 2003,
        month = jul,
       volume = {115},
       number = {809},
        pages = {763-795},
          doi = {10.1086/376392},
archivePrefix = {arXiv},
       eprint = {astro-ph/0304382},
 primaryClass = {astro-ph},
       adsurl = {https://ui.adsabs.harvard.edu/abs/2003PASP..115..763C}
}

@ARTICLE{CF00,
       author = {{Charlot}, St{\'e}phane and {Fall}, S. Michael},
        title = "{A Simple Model for the Absorption of Starlight by Dust in Galaxies}",
      journal = {\apj},
         year = 2000,
        month = aug,
       volume = {539},
       number = {2},
        pages = {718-731},
          doi = {10.1086/309250},
archivePrefix = {arXiv},
       eprint = {astro-ph/0003128},
 primaryClass = {astro-ph},
       adsurl = {https://ui.adsabs.harvard.edu/abs/2000ApJ...539..718C}
}

@ARTICLE{Ciesla,
       author = {{Ciesla}, L. and {Elbaz}, D. and {Fensch}, J.},
        title = "{The SFR-M$_{{\ensuremath{*}}}$ main sequence archetypal star-formation history and analytical models}",
      journal = {\aap},
         year = 2017,
        month = dec,
       volume = {608},
          eid = {A41},
        pages = {A41},
          doi = {10.1051/0004-6361/201731036},
archivePrefix = {arXiv},
       eprint = {1706.08531},
 primaryClass = {astro-ph.GA},
       adsurl = {https://ui.adsabs.harvard.edu/abs/2017A&A...608A..41C}
}

@ARTICLE{Claeyssens,
       author = {{Claeyssens}, A. and {Richard}, J. and {Blaizot}, J. and {Garel}, T. and {Kusakabe}, H. and {Bacon}, R. and {Bauer}, F.~E. and {Guaita}, L. and {Jeanneau}, A. and {Lagattuta}, D. and {Leclercq}, F. and {Maseda}, M. and {Matthee}, J. and {Nanayakkara}, T. and {Pello}, R. and {Thai}, T.~T. and {Tuan-Anh}, P. and {Verhamme}, A. and {Vitte}, E. and {Wisotzki}, L.},
        title = "{The Lensed Lyman-Alpha MUSE Arcs Sample (LLAMAS). I. Characterisation of extended Lyman-alpha halos and spatial offsets}",
      journal = {\aap},
         year = 2022,
        month = oct,
       volume = {666},
          eid = {A78},
        pages = {A78},
          doi = {10.1051/0004-6361/202142320},
archivePrefix = {arXiv},
       eprint = {2201.04674},
 primaryClass = {astro-ph.GA},
       adsurl = {https://ui.adsabs.harvard.edu/abs/2022A&A...666A..78C}
}

@ARTICLE{Cole,
       author = {{Cole}, Justin W. and {Papovich}, Casey and {Finkelstein}, Steven L. and {Bagley}, Micaela B. and {Dickinson}, Mark and {Iyer}, Kartheik G. and {Yung}, L.~Y. Aaron and {Ciesla}, Laure and {Amor{\'\i}n}, Ricardo O. and {Arrabal Haro}, Pablo and {Bhatawdekar}, Rachana and {Calabr{\`o}}, Antonello and {Cleri}, Nikko J. and {de la Vega}, Alexander and {Dekel}, Avishai and {Endsley}, Ryan and {Gawiser}, Eric and {Giavalisco}, Mauro and {Hathi}, Nimish P. and {Hirschmann}, Michaela and {Holwerda}, Benne W. and {Kartaltepe}, Jeyhan S. and {Koekemoer}, Anton M. and {Lucas}, Ray A. and {Mascia}, Sara and {Mobasher}, Bahram and {P{\'e}rez-Gonz{\'a}lez}, Pablo G. and {Rodighiero}, Giulia and {Ronayne}, Kaila and {Tacchella}, Sandro and {Weiner}, Benjamin J. and {Wilkins}, Stephen M.},
        title = "{CEERS: Increasing Scatter along the Star-forming Main Sequence Indicates Early Galaxies Form in Bursts}",
      journal = {\apj},
         year = 2025,
        month = feb,
       volume = {979},
       number = {2},
          eid = {193},
        pages = {193},
          doi = {10.3847/1538-4357/ad9a6a},
archivePrefix = {arXiv},
       eprint = {2312.10152},
 primaryClass = {astro-ph.GA},
       adsurl = {https://ui.adsabs.harvard.edu/abs/2025ApJ...979..193C}
}

@ARTICLE{cresci09,
       author = {{Cresci}, G. and {Hicks}, E.~K.~S. and {Genzel}, R. and {F{\"o}rster Schreiber}, N.~M. and {Davies}, R. and {Bouch{\'e}}, N. and {Buschkamp}, P. and {Genel}, S. and {Shapiro}, K. and {Tacconi}, L. and {Sommer-Larsen}, J. and {Burkert}, A. and {Eisenhauer}, F. and {Gerhard}, O. and {Lutz}, D. and {Naab}, T. and {Sternberg}, A. and {Cimatti}, A. and {Daddi}, E. and {Erb}, D.~K. and {Kurk}, J. and {Lilly}, S.~L. and {Renzini}, A. and {Shapley}, A. and {Steidel}, C.~C. and {Caputi}, K.},
        title = "{The SINS Survey: Modeling the Dynamics of z \raisebox{-0.5ex}\textasciitilde 2 Galaxies and the High-z Tully-Fisher Relation}",
      journal = {\apj},
         year = 2009,
        month = may,
       volume = {697},
       number = {1},
        pages = {115-132},
          doi = {10.1088/0004-637X/697/1/115},
archivePrefix = {arXiv},
       eprint = {0902.4701},
 primaryClass = {astro-ph.CO},
       adsurl = {https://ui.adsabs.harvard.edu/abs/2009ApJ...697..115C}
}

@ARTICLE{Dalcanton,
       author = {{Dalcanton}, Julianne J. and {Stilp}, Adrienne M.},
        title = "{Pressure Support in Galaxy Disks: Impact on Rotation Curves and Dark Matter Density Profiles}",
      journal = {\apj},
         year = 2010,
        month = sep,
       volume = {721},
       number = {1},
        pages = {547-561},
          doi = {10.1088/0004-637X/721/1/547},
archivePrefix = {arXiv},
       eprint = {1007.2535},
 primaryClass = {astro-ph.CO},
       adsurl = {https://ui.adsabs.harvard.edu/abs/2010ApJ...721..547D}
}

@ARTICLE{Danhaive,
       author = {{Danhaive}, A. Lola and {Tacchella}, Sandro and {Bunker}, Andrew J. and {Curtis-Lake}, Emma and {de Graaff}, Anna and {D'Eugenio}, Francesco and {Duan}, Qiao and {Egami}, Eiichi and {Eisenstein}, Daniel J. and {Johnson}, Benjamin D. and {Maiolino}, Roberto and {McClymont}, William and {Rieke}, Marcia and {Robertson}, Brant and {Sun}, Fengwu and {Willmer}, Christopher N.~A. and {Wu}, Zihao and {Zhu}, Yongda},
        title = "{The dark side of early galaxies: GEKO uncovers dark-matter fractions at z {\ensuremath{\sim}} 4 {\ensuremath{-}} 6}",
      journal = {\mnras},
         year = 2026,
        month = mar,
       volume = {546},
       number = {3},
          eid = {stag119},
        pages = {stag119},
          doi = {10.1093/mnras/stag119},
archivePrefix = {arXiv},
       eprint = {2510.14779},
 primaryClass = {astro-ph.GA},
       adsurl = {https://ui.adsabs.harvard.edu/abs/2026MNRAS.546ag119D}
}

@ARTICLE{debok10,
       author = {{de Blok}, W.~J.~G.},
        title = "{The Core-Cusp Problem}",
      journal = {Advances in Astronomy},
         year = 2010,
        month = jan,
       volume = {2010},
          eid = {789293},
        pages = {789293},
          doi = {10.1155/2010/789293},
archivePrefix = {arXiv},
       eprint = {0910.3538},
 primaryClass = {astro-ph.CO},
       adsurl = {https://ui.adsabs.harvard.edu/abs/2010AdAst2010E...5D}
}

@ARTICLE{DC14,
       author = {{Di Cintio}, Arianna and {Brook}, Chris B. and {Macci{\`o}}, Andrea V. and {Stinson}, Greg S. and {Knebe}, Alexander and {Dutton}, Aaron A. and {Wadsley}, James},
        title = "{The dependence of dark matter profiles on the stellar-to-halo mass ratio: a prediction for cusps versus cores}",
      journal = {\mnras},
         year = 2014,
        month = jan,
       volume = {437},
       number = {1},
        pages = {415-423},
          doi = {10.1093/mnras/stt1891},
archivePrefix = {arXiv},
       eprint = {1306.0898},
 primaryClass = {astro-ph.CO},
       adsurl = {https://ui.adsabs.harvard.edu/abs/2014MNRAS.437..415D}
}

@ARTICLE{dutton,
       author = {{Dutton}, Aaron A. and {Macci{\`o}}, Andrea V.},
        title = "{Cold dark matter haloes in the Planck era: evolution of structural parameters for Einasto and NFW profiles}",
      journal = {\mnras},
         year = 2014,
        month = jul,
       volume = {441},
       number = {4},
        pages = {3359-3374},
          doi = {10.1093/mnras/stu742},
archivePrefix = {arXiv},
       eprint = {1402.7073},
 primaryClass = {astro-ph.CO},
       adsurl = {https://ui.adsabs.harvard.edu/abs/2014MNRAS.441.3359D}
}

@ARTICLE{2016MNRAS.461.2658D,
       author = {{Dutton}, Aaron A. and {Macci{\`o}}, Andrea V. and {Dekel}, Avishai and {Wang}, Liang and {Stinson}, Gregory and {Obreja}, Aura and {Di Cintio}, Arianna and {Brook}, Chris and {Buck}, Tobias and {Kang}, Xi},
        title = "{NIHAO IX: the role of gas inflows and outflows in driving the contraction and expansion of cold dark matter haloes}",
      journal = {\mnras},
         year = 2016,
        month = sep,
       volume = {461},
       number = {3},
        pages = {2658-2675},
          doi = {10.1093/mnras/stw1537},
archivePrefix = {arXiv},
       eprint = {1605.05323},
 primaryClass = {astro-ph.GA},
       adsurl = {https://ui.adsabs.harvard.edu/abs/2016MNRAS.461.2658D}
}

@ARTICLE{Eisenstein,
       author = {{Eisenstein}, Daniel J. and {Willott}, Chris and {Alberts}, Stacey and {Arribas}, Santiago and {Bonaventura}, Nina and {Bunker}, Andrew J. and {Cameron}, Alex J. and {Carniani}, Stefano and {Charlot}, Stephane and {Curtis-Lake}, Emma and {D'Eugenio}, Francesco and {Ferruit}, Pierre and {Giardino}, Giovanna and {Hainline}, Kevin and {Hausen}, Ryan and {Jakobsen}, Peter and {Johnson}, Benjamin D. and {Maiolino}, Roberto and {Rauscher}, Bernard J. and {Rieke}, Marcia and {Rieke}, George and {Rix}, Hans-Walter and {Robertson}, Brant and {Stark}, Daniel P. and {Tacchella}, Sandro and {Williams}, Christina C. and {Willmer}, Christopher N.~A. and {Baker}, William M. and {Baum}, Stefi and {Bhatawdekar}, Rachana and {Boyett}, Kristan and {Chen}, Zuyi and {Chevallard}, Jacopo and {Circosta}, Chiara and {Curti}, Mirko and {Danhaive}, A. Lola and {DeCoursey}, Christa and {Endsley}, Ryan and {de Graaff}, Anna and {Dressler}, Alan and {Egami}, Eiichi and {Helton}, Jakob M. and {Hviding}, Raphael E. and {Ji}, Zhiyuan and {Jones}, Gareth C. and {Kumari}, Nimisha and {L{\"u}tzgendorf}, Nora and {Laseter}, Isaac and {Looser}, Tobias J. and {Lyu}, Jianwei and {Maseda}, Michael V. and {Nelson}, Erica and {Parlanti}, Eleonora and {Perna}, Michele and {Pusk{\'a}s}, D{\'a}vid and {Rawle}, Tim and {Rodr{\'\i}guez Del Pino}, Bruno and {Rujopakarn}, Wiphu and {Sandles}, Lester and {Saxena}, Aayush and {Scholtz}, Jan and {Sharpe}, Katherine and {Shivaei}, Irene and {Silcock}, Maddie S. and {Simmonds}, Charlotte and {Skarbinski}, Maya and {Smit}, Renske and {Stone}, Meredith and {Suess}, Katherine A. and {Sun}, Fengwu and {Tang}, Mengtao and {Topping}, Michael W. and {{\"U}bler}, Hannah and {Villanueva}, Natalia C. and {Wallace}, Imaan E.~B. and {Whitler}, Lily and {Witstok}, Joris and {Woodrum}, Charity},
        title = "{Overview of the JWST Advanced Deep Extragalactic Survey (JADES)}",
      journal = {\apjs},
         year = 2026,
        month = mar,
       volume = {283},
       number = {1},
          eid = {6},
        pages = {6},
          doi = {10.3847/1538-4365/ae3163},
archivePrefix = {arXiv},
       eprint = {2306.02465},
 primaryClass = {astro-ph.GA},
       adsurl = {https://ui.adsabs.harvard.edu/abs/2026ApJS..283....6E}
}

@ARTICLE{2016MNRAS.461.1745E,
       author = {{El-Zant}, Amr A. and {Freundlich}, Jonathan and {Combes}, Fran{\c{c}}oise},
        title = "{From cusps to cores: a stochastic model}",
      journal = {\mnras},
         year = 2016,
        month = sep,
       volume = {461},
       number = {2},
        pages = {1745-1759},
          doi = {10.1093/mnras/stw1398},
archivePrefix = {arXiv},
       eprint = {1603.00526},
 primaryClass = {astro-ph.GA},
       adsurl = {https://ui.adsabs.harvard.edu/abs/2016MNRAS.461.1745E}
}

@ARTICLE{2019ApJ...881...71E,
       author = {{Emami}, Najmeh and {Siana}, Brian and {Weisz}, Daniel R. and {Johnson}, Benjamin D. and {Ma}, Xiangcheng and {El-Badry}, Kareem},
        title = "{A Closer Look at Bursty Star Formation with L $_{H{\ensuremath{\alpha}} }$ and L $_{UV}$ Distributions}",
      journal = {\apj},
         year = 2019,
        month = aug,
       volume = {881},
       number = {1},
          eid = {71},
        pages = {71},
          doi = {10.3847/1538-4357/ab211a},
archivePrefix = {arXiv},
       eprint = {1809.06380},
 primaryClass = {astro-ph.GA},
       adsurl = {https://ui.adsabs.harvard.edu/abs/2019ApJ...881...71E}
}

@ARTICLE{camel,
       author = {{Epinat}, B. and {Tasca}, L. and {Amram}, P. and {Contini}, T. and {Le F{\`e}vre}, O. and {Queyrel}, J. and {Vergani}, D. and {Garilli}, B. and {Kissler-Patig}, M. and {Moultaka}, J. and {Paioro}, L. and {Tresse}, L. and {Bournaud}, F. and {L{\'o}pez-Sanjuan}, C. and {Perret}, V.},
        title = "{MASSIV: Mass Assembly Survey with SINFONI in VVDS. II. Kinematics and close environment classification}",
      journal = {\aap},
         year = 2012,
        month = mar,
       volume = {539},
          eid = {A92},
        pages = {A92},
          doi = {10.1051/0004-6361/201117711},
archivePrefix = {arXiv},
       eprint = {1201.3329},
 primaryClass = {astro-ph.CO},
       adsurl = {https://ui.adsabs.harvard.edu/abs/2012A&A...539A..92E}
}

@ARTICLE{Faucher,
       author = {{Faucher-Gigu{\`e}re}, Claude-Andr{\'e}},
        title = "{A model for the origin of bursty star formation in galaxies}",
      journal = {\mnras},
         year = 2018,
        month = jan,
       volume = {473},
       number = {3},
        pages = {3717-3731},
          doi = {10.1093/mnras/stx2595},
archivePrefix = {arXiv},
       eprint = {1701.04824},
 primaryClass = {astro-ph.GA},
       adsurl = {https://ui.adsabs.harvard.edu/abs/2018MNRAS.473.3717F}
}

@ARTICLE{feroz,
       author = {{Feroz}, F. and {Hobson}, M.~P. and {Bridges}, M.},
        title = "{MULTINEST: an efficient and robust Bayesian inference tool for cosmology and particle physics}",
      journal = {\mnras},
         year = 2009,
        month = oct,
       volume = {398},
       number = {4},
        pages = {1601-1614},
          doi = {10.1111/j.1365-2966.2009.14548.x},
archivePrefix = {arXiv},
       eprint = {0809.3437},
 primaryClass = {astro-ph},
       adsurl = {https://ui.adsabs.harvard.edu/abs/2009MNRAS.398.1601F}
}

@ARTICLE{fp,
       author = {{Flores}, Ricardo A. and {Primack}, Joel R.},
        title = "{Observational and Theoretical Constraints on Singular Dark Matter Halos}",
      journal = {\apjl},
         year = 1994,
        month = may,
       volume = {427},
        pages = {L1},
          doi = {10.1086/187350},
archivePrefix = {arXiv},
       eprint = {astro-ph/9402004},
 primaryClass = {astro-ph},
       adsurl = {https://ui.adsabs.harvard.edu/abs/1994ApJ...427L...1F}
}

@ARTICLE{free,
       author = {{Freundlich}, Jonathan and {Dekel}, Avishai and {Jiang}, Fangzhou and {Ishai}, Guy and {Cornuault}, Nicolas and {Lapiner}, Sharon and {Dutton}, Aaron A. and {Macci{\`o}}, Andrea V.},
        title = "{A model for core formation in dark matter haloes and ultra-diffuse galaxies by outflow episodes}",
      journal = {\mnras},
         year = 2020,
        month = jan,
       volume = {491},
       number = {3},
        pages = {4523-4542},
          doi = {10.1093/mnras/stz3306},
archivePrefix = {arXiv},
       eprint = {1907.11726},
 primaryClass = {astro-ph.GA},
       adsurl = {https://ui.adsabs.harvard.edu/abs/2020MNRAS.491.4523F}
}

@ARTICLE{fre,
       author = {{Freundlich}, Jonathan and {Jiang}, Fangzhou and {Dekel}, Avishai and {Cornuault}, Nicolas and {Ginzburg}, Omry and {Koskas}, R{\'e}my and {Lapiner}, Sharon and {Dutton}, Aaron and {Macci{\`o}}, Andrea V.},
        title = "{The Dekel-Zhao profile: a mass-dependent dark-matter density profile with flexible inner slope and analytic potential, velocity dispersion, and lensing properties}",
      journal = {\mnras},
         year = 2020,
        month = dec,
       volume = {499},
       number = {2},
        pages = {2912-2933},
          doi = {10.1093/mnras/staa2790},
archivePrefix = {arXiv},
       eprint = {2004.08395},
 primaryClass = {astro-ph.GA},
       adsurl = {https://ui.adsabs.harvard.edu/abs/2020MNRAS.499.2912F}
}

@ARTICLE{Genzel06,
       author = {{Genzel}, R. and {Tacconi}, L.~J. and {Eisenhauer}, F. and {F{\"o}rster Schreiber}, N.~M. and {Cimatti}, A. and {Daddi}, E. and {Bouch{\'e}}, N. and {Davies}, R. and {Lehnert}, M.~D. and {Lutz}, D. and {Nesvadba}, N. and {Verma}, A. and {Abuter}, R. and {Shapiro}, K. and {Sternberg}, A. and {Renzini}, A. and {Kong}, X. and {Arimoto}, N. and {Mignoli}, M.},
        title = "{The rapid formation of a large rotating disk galaxy three billion years after the Big Bang}",
      journal = {\nat},
         year = 2006,
        month = aug,
       volume = {442},
       number = {7104},
        pages = {786-789},
          doi = {10.1038/nature05052},
archivePrefix = {arXiv},
       eprint = {astro-ph/0608344},
 primaryClass = {astro-ph},
       adsurl = {https://ui.adsabs.harvard.edu/abs/2006Natur.442..786G}
}

@ARTICLE{genz,
       author = {{Genzel}, R. and {F{\"o}rster Schreiber}, N.~M. and {{\"U}bler}, H. and {Lang}, P. and {Naab}, T. and {Bender}, R. and {Tacconi}, L.~J. and {Wisnioski}, E. and {Wuyts}, S. and {Alexander}, T. and {Beifiori}, A. and {Belli}, S. and {Brammer}, G. and {Burkert}, A. and {Carollo}, C.~M. and {Chan}, J. and {Davies}, R. and {Fossati}, M. and {Galametz}, A. and {Genel}, S. and {Gerhard}, O. and {Lutz}, D. and {Mendel}, J.~T. and {Momcheva}, I. and {Nelson}, E.~J. and {Renzini}, A. and {Saglia}, R. and {Sternberg}, A. and {Tacchella}, S. and {Tadaki}, K. and {Wilman}, D.},
        title = "{Strongly baryon-dominated disk galaxies at the peak of galaxy formation ten billion years ago}",
      journal = {\nat},
         year = 2017,
        month = mar,
       volume = {543},
       number = {7645},
        pages = {397-401},
          doi = {10.1038/nature21685},
archivePrefix = {arXiv},
       eprint = {1703.04310},
 primaryClass = {astro-ph.GA},
       adsurl = {https://ui.adsabs.harvard.edu/abs/2017Natur.543..397G}
}

@ARTICLE{genz2,
       author = {{Genzel}, R. and {Price}, S.~H. and {{\"U}bler}, H. and {F{\"o}rster Schreiber}, N.~M. and {Shimizu}, T.~T. and {Tacconi}, L.~J. and {Bender}, R. and {Burkert}, A. and {Contursi}, A. and {Coogan}, R. and {Davies}, R.~L. and {Davies}, R.~I. and {Dekel}, A. and {Herrera-Camus}, R. and {Lee}, M.-J. and {Lutz}, D. and {Naab}, T. and {Neri}, R. and {Nestor}, A. and {Renzini}, A. and {Saglia}, R. and {Schuster}, K. and {Sternberg}, A. and {Wisnioski}, E. and {Wuyts}, S.},
        title = "{Rotation Curves in z {\ensuremath{\sim}} 1-2 Star-forming Disks: Evidence for Cored Dark Matter Distributions}",
      journal = {\apj},
         year = 2020,
        month = oct,
       volume = {902},
       number = {2},
          eid = {98},
        pages = {98},
          doi = {10.3847/1538-4357/abb0ea},
archivePrefix = {arXiv},
       eprint = {2006.03046},
 primaryClass = {astro-ph.GA},
       adsurl = {https://ui.adsabs.harvard.edu/abs/2020ApJ...902...98G}
}

@ARTICLE{Guo,
       author = {{Guo}, Yicheng and {Rafelski}, Marc and {Faber}, S.~M. and {Koo}, David C. and {Krumholz}, Mark R. and {Trump}, Jonathan R. and {Willner}, S.~P. and {Amor{\'\i}n}, Ricardo and {Barro}, Guillermo and {Bell}, Eric F. and {Gardner}, Jonathan P. and {Gawiser}, Eric and {Hathi}, Nimish P. and {Koekemoer}, Anton M. and {Pacifici}, Camilla and {P{\'e}rez-Gonz{\'a}lez}, Pablo G. and {Ravindranath}, Swara and {Reddy}, Naveen and {Teplitz}, Harry I. and {Yesuf}, Hassen},
        title = "{The Bursty Star Formation Histories of Low-mass Galaxies at 0.4 < z < 1 Revealed by Star Formation Rates Measured From H{\ensuremath{\beta}} and FUV}",
      journal = {\apj},
         year = 2016,
        month = dec,
       volume = {833},
       number = {1},
          eid = {37},
        pages = {37},
          doi = {10.3847/1538-4357/833/1/37},
archivePrefix = {arXiv},
       eprint = {1604.05314},
 primaryClass = {astro-ph.GA},
       adsurl = {https://ui.adsabs.harvard.edu/abs/2016ApJ...833...37G}
}

@ARTICLE{Gilbank,
       author = {{Gilbank}, David G. and {Baldry}, Ivan K. and {Balogh}, Michael L. and {Glazebrook}, Karl and {Bower}, Richard G.},
        title = "{The local star formation rate density: assessing calibrations using [OII], H and UV luminosities}",
      journal = {\mnras},
         year = 2010,
        month = jul,
       volume = {405},
       number = {4},
        pages = {2594-2614},
          doi = {10.1111/j.1365-2966.2010.16640.x},
archivePrefix = {arXiv},
       eprint = {1002.3172},
 primaryClass = {astro-ph.CO},
       adsurl = {https://ui.adsabs.harvard.edu/abs/2010MNRAS.405.2594G}
}

@ARTICLE{gov,
       author = {{Governato}, F. and {Brook}, C. and {Mayer}, L. and {Brooks}, A. and {Rhee}, G. and {Wadsley}, J. and {Jonsson}, P. and {Willman}, B. and {Stinson}, G. and {Quinn}, T. and {Madau}, P.},
        title = "{Bulgeless dwarf galaxies and dark matter cores from supernova-driven outflows}",
      journal = {\nat},
         year = 2010,
        month = jan,
       volume = {463},
       number = {7278},
        pages = {203-206},
          doi = {10.1038/nature08640},
archivePrefix = {arXiv},
       eprint = {0911.2237},
 primaryClass = {astro-ph.CO},
       adsurl = {https://ui.adsabs.harvard.edu/abs/2010Natur.463..203G}
}

@ARTICLE{Hashim23,
       author = {{Hashim}, Mahmoud and {El-Zant}, Amr A. and {Freundlich}, Jonathan and {Read}, Justin I. and {Combes}, Fran{\c{c}}oise},
        title = "{Halo heating from fluctuating gas in a model dwarf}",
      journal = {\mnras},
         year = 2023,
        month = may,
       volume = {521},
       number = {1},
        pages = {772-789},
          doi = {10.1093/mnras/stad581},
archivePrefix = {arXiv},
       eprint = {2209.08631},
 primaryClass = {astro-ph.GA},
       adsurl = {https://ui.adsabs.harvard.edu/abs/2023MNRAS.521..772H}
}

@ARTICLE{Hausler,
       author = {{H{\"a}u{\ss}ler}, Boris and {Bamford}, Steven P. and {Vika}, Marina and {Rojas}, Alex L. and {Barden}, Marco and {Kelvin}, Lee S. and {Alpaslan}, Mehmet and {Robotham}, Aaron S.~G. and {Driver}, Simon P. and {Baldry}, I.~K. and {Brough}, Sarah and {Hopkins}, Andrew M. and {Liske}, Jochen and {Nichol}, Robert C. and {Popescu}, Cristina C. and {Tuffs}, Richard J.},
        title = "{MegaMorph - multiwavelength measurement of galaxy structure: complete S{\'e}rsic profile information from modern surveys}",
      journal = {\mnras},
         year = 2013,
        month = mar,
       volume = {430},
       number = {1},
        pages = {330-369},
          doi = {10.1093/mnras/sts633},
archivePrefix = {arXiv},
       eprint = {1212.3332},
 primaryClass = {astro-ph.CO},
       adsurl = {https://ui.adsabs.harvard.edu/abs/2013MNRAS.430..330H}
}

@ARTICLE{Heavens,
       author = {{Heavens}, Alan and {Panter}, Benjamin and {Jimenez}, Raul and {Dunlop}, James},
        title = "{The star-formation history of the Universe from the stellar populations of nearby galaxies}",
      journal = {\nat},
         year = 2004,
        month = apr,
       volume = {428},
       number = {6983},
        pages = {625-627},
          doi = {10.1038/nature02474},
archivePrefix = {arXiv},
       eprint = {astro-ph/0403293},
 primaryClass = {astro-ph},
       adsurl = {https://ui.adsabs.harvard.edu/abs/2004Natur.428..625H}
}

@ARTICLE{Hernquist,
       author = {{Hernquist}, Lars},
        title = "{An Analytical Model for Spherical Galaxies and Bulges}",
      journal = {\apj},
         year = 1990,
        month = jun,
       volume = {356},
        pages = {359},
          doi = {10.1086/168845},
       adsurl = {https://ui.adsabs.harvard.edu/abs/1990ApJ...356..359H}
}

@ARTICLE{hu,
       author = {{Hu}, Wayne and {Barkana}, Rennan and {Gruzinov}, Andrei},
        title = "{Fuzzy Cold Dark Matter: The Wave Properties of Ultralight Particles}",
      journal = {\prl},
         year = 2000,
        month = aug,
       volume = {85},
       number = {6},
        pages = {1158-1161},
          doi = {10.1103/PhysRevLett.85.1158},
archivePrefix = {arXiv},
       eprint = {astro-ph/0003365},
 primaryClass = {astro-ph},
       adsurl = {https://ui.adsabs.harvard.edu/abs/2000PhRvL..85.1158H}
}

@INPROCEEDINGS{2026IAUS..392..119H,
       author = {{Haubner}, Konstantin and {Lelli}, Federico and {Di Teodoro}, Enrico and {Duey}, Francis and {McGaugh}, Stacy and {Schombert}, James and {Hess}, Kelley M. and {Apertif Team}},
        title = "{BIG-SPARC: The new SPARC database}",
    booktitle = {IAU Symposium},
         year = 2026,
       editor = {{Pisano}, D.~J. and {Mogotsi}, K. Moses and {Healy}, Julia and {Blyth}, Sarah},
       series = {IAU Symposium},
       volume = {20},
        month = jul,
        pages = {119-125},
          doi = {10.1017/S1743921324002886},
archivePrefix = {arXiv},
       eprint = {2411.13329},
 primaryClass = {astro-ph.GA},
       adsurl = {https://ui.adsabs.harvard.edu/abs/2026IAUS..392..119H}
}

@ARTICLE{plot,
       author = {{Hunter}, John D.},
        title = "{Matplotlib: A 2D Graphics Environment}",
      journal = {Computing in Science and Engineering},
         year = 2007,
        month = jan,
       volume = {9},
       number = {3},
        pages = {90-95},
          doi = {10.1109/MCSE.2007.55},
       adsurl = {https://ui.adsabs.harvard.edu/abs/2007CSE.....9...90H}
}

@ARTICLE{Iyer,
       author = {{Iyer}, Kartheik G. and {Tacchella}, Sandro and {Genel}, Shy and {Hayward}, Christopher C. and {Hernquist}, Lars and {Brooks}, Alyson M. and {Caplar}, Neven and {Dav{\'e}}, Romeel and {Diemer}, Benedikt and {Forbes}, John C. and {Gawiser}, Eric and {Somerville}, Rachel S. and {Starkenburg}, Tjitske K.},
        title = "{The diversity and variability of star formation histories in models of galaxy evolution}",
      journal = {\mnras},
         year = 2020,
        month = oct,
       volume = {498},
       number = {1},
        pages = {430-463},
          doi = {10.1093/mnras/staa2150},
archivePrefix = {arXiv},
       eprint = {2007.07916},
 primaryClass = {astro-ph.GA},
       adsurl = {https://ui.adsabs.harvard.edu/abs/2020MNRAS.498..430I}
}

@ARTICLE{Jackson,
       author = {{Jackson}, R.~A. and {Kaviraj}, S. and {Yi}, S.~K. and {Peirani}, S. and {Dubois}, Y. and {Martin}, G. and {Devriendt}, J.~E.~G. and {Slyz}, A. and {Pichon}, C. and {Volonteri}, M. and {Kimm}, T. and {Kraljic}, K.},
        title = "{The formation of cores in galaxies across cosmic time - the existence of cores is not in tension with the {\ensuremath{\Lambda}}CDM paradigm}",
      journal = {\mnras},
         year = 2024,
        month = feb,
       volume = {528},
       number = {2},
        pages = {1655-1667},
          doi = {10.1093/mnras/stae056},
}

@ARTICLE{Jeanneau,
       author = {{Jeanneau}, Alexandre and {Richard}, Johan and {Bouch{\'e}}, Nicolas F. and {Krajnovi{\'c}}, Davor and {Ciocan}, Bianca-Iulia and {Freundlich}, Jonathan and {Epinat}, Beno{\^\i}t and {Contini}, Thierry},
        title = "{MUSE-DARK: II. 3D morpho-kinematic modelling of lensed galaxies: Tully-Fisher relation of z {\ensuremath{\sim}} 1 star-forming galaxies}",
      journal = {\aap},
         year = 2026,
        month = may,
       volume = {709},
          eid = {A120},
        pages = {A120},
          doi = {10.1051/0004-6361/202659953},
archivePrefix = {arXiv},
       eprint = {2603.28856},
 primaryClass = {astro-ph.GA},
       adsurl = {https://ui.adsabs.harvard.edu/abs/2026A&A...709A.120J}
}

@ARTICLE{Johnson,
       author = {{Johnson}, Benjamin D. and {Robertson}, Brant E. and {Eisenstein}, Daniel J. and {Tacchella}, Sandro and {Pusk{\'a}s}, D{\'a}vid and {Duan}, Qiao and {Wu}, Zihao and {Hainline}, Kevin and {Rieke}, Marcia and {Willott}, Chris and {Willmer}, Christopher N.~A. and {Trussler}, James A.~A. and {Alberts}, Stacey and {Arribas}, Santiago and {Baker}, William M. and {Bunker}, Andrew J. and {Cameron}, Alex J. and {Carniani}, Stefano and {Carreira}, Courtney and {Cargile}, Phillip A. and {Curtis-Lake}, Emma and {Egami}, Eiichi and {Hausen}, Ryan and {Helton}, Jakob M. and {Ji}, Zhiyuan and {Maiolino}, Roberto and {P{\'e}rez-Gonz{\'a}lez}, Pablo G. and {Rinaldi}, Pierluigi and {Sun}, Fengwu and {Sun}, Yang and {Villanueva}, Natalia C. and {Williams}, Christina C. and {Zhu}, Yongda},
        title = "{JWST Advanced Deep Extragalactic Survey (JADES) Data Release 5: NIRCam Imaging in GOODS-S and GOODS-N}",
      journal = {arXiv e-prints},
         year = 2026,
        month = jan,
          eid = {arXiv:2601.15954},
        pages = {arXiv:2601.15954},
          doi = {10.48550/arXiv.2601.15954},
archivePrefix = {arXiv},
       eprint = {2601.15954},
 primaryClass = {astro-ph.GA},
       adsurl = {https://ui.adsabs.harvard.edu/abs/2026arXiv260115954J}
}

@article{kr,
    author = {Kass, R. E. and Raftery, A. E.},
    title = {{Bayes Factors}},
    year = {1995},
    journal = {J. Am. Stat. Assoc.},
    volume = {90},
    pages = {773-795},
    doi = {01621459.1995.10476572},
}

@ARTICLE{Kauffmann,
       author = {{Kauffmann}, Guinevere},
        title = "{Quantitative constraints on starburst cycles in galaxies with stellar masses in the range {}10$^{8}$-{}10$^{10}$ M$_{☉}$}",
      journal = {\mnras},
         year = 2014,
        month = jul,
       volume = {441},
       number = {3},
        pages = {2717-2724},
          doi = {10.1093/mnras/stu752},
       adsurl = {https://ui.adsabs.harvard.edu/abs/2014MNRAS.441.2717K}
}

@ARTICLE{katz,
       author = {{Katz}, Harley and {Lelli}, Federico and {McGaugh}, Stacy S. and {Di Cintio}, Arianna and {Brook}, Chris B. and {Schombert}, James M.},
        title = "{Testing feedback-modified dark matter haloes with galaxy rotation curves: estimation of halo parameters and consistency with {\ensuremath{\Lambda}}CDM scaling relations}",
      journal = {\mnras},
         year = 2017,
        month = apr,
       volume = {466},
       number = {2},
        pages = {1648-1668},
          doi = {10.1093/mnras/stw3101},
archivePrefix = {arXiv},
       eprint = {1605.05971},
 primaryClass = {astro-ph.GA},
       adsurl = {https://ui.adsabs.harvard.edu/abs/2017MNRAS.466.1648K}
}

@ARTICLE{Kaushal,
       author = {{Kaushal}, Yasha and {Nersesian}, Angelos and {Bezanson}, Rachel and {van der Wel}, Arjen and {Leja}, Joel and {Carnall}, Adam and {Gallazzi}, Anna and {Zibetti}, Stefano and {Khullar}, Gourav and {Franx}, Marijn and {Muzzin}, Adam and {de Graaff}, Anna and {Pacifici}, Camilla and {Whitaker}, Katherine E. and {Bell}, Eric F. and {Martorano}, Marco},
        title = "{A Census of Star Formation Histories of Massive Galaxies at 0.6 < z < 1 from Spectrophotometric Modeling Using Bagpipes and Prospector}",
      journal = {\apj},
         year = 2024,
        month = jan,
       volume = {961},
       number = {1},
          eid = {118},
        pages = {118},
          doi = {10.3847/1538-4357/ad0c4e},
archivePrefix = {arXiv},
       eprint = {2307.03725},
 primaryClass = {astro-ph.GA},
       adsurl = {https://ui.adsabs.harvard.edu/abs/2024ApJ...961..118K}
}

@ARTICLE{Kennicutt,
       author = {{Kennicutt}, Jr., Robert C.},
        title = "{Star Formation in Galaxies Along the Hubble Sequence}",
      journal = {\araa},
         year = 1998,
        month = jan,
       volume = {36},
        pages = {189-232},
          doi = {10.1146/annurev.astro.36.1.189},
archivePrefix = {arXiv},
       eprint = {astro-ph/9807187},
 primaryClass = {astro-ph},
       adsurl = {https://ui.adsabs.harvard.edu/abs/1998ARA&A..36..189K}
}

@ARTICLE{Klypin,
       author = {{Klypin}, Anatoly and {Kravtsov}, Andrey V. and {Bullock}, James S. and {Primack}, Joel R.},
        title = "{Resolving the Structure of Cold Dark Matter Halos}",
      journal = {\apj},
         year = 2001,
        month = jun,
       volume = {554},
       number = {2},
        pages = {903-915},
          doi = {10.1086/321400},
archivePrefix = {arXiv},
       eprint = {astro-ph/0006343},
 primaryClass = {astro-ph},
       adsurl = {https://ui.adsabs.harvard.edu/abs/2001ApJ...554..903K}
}

@ARTICLE{kroupa,
       author = {{Kroupa}, Pavel},
        title = "{On the variation of the initial mass function}",
      journal = {\mnras},
         year = 2001,
        month = apr,
       volume = {322},
       number = {2},
        pages = {231-246},
          doi = {10.1046/j.1365-8711.2001.04022.x},
archivePrefix = {arXiv},
       eprint = {astro-ph/0009005},
 primaryClass = {astro-ph},
       adsurl = {https://ui.adsabs.harvard.edu/abs/2001MNRAS.322..231K}
}

@ARTICLE{maria,
       author = {{Koller}, Maria and {Maiolino}, Roberto and {Baker}, William M.},
        title = "{Both stellar mass and gravitational potential shape the gas-phase metallicity}",
      journal = {\mnras},
         year = 2026,
        month = jan,
       volume = {545},
       number = {1},
          eid = {staf2011},
        pages = {staf2011},
          doi = {10.1093/mnras/staf2011},
archivePrefix = {arXiv},
       eprint = {2509.18961},
 primaryClass = {astro-ph.GA},
       adsurl = {https://ui.adsabs.harvard.edu/abs/2026MNRAS.545f2011K}
}

@ARTICLE{Lazar,
       author = {{Lazar}, Alexandres and {Bullock}, James S. and {Boylan-Kolchin}, Michael and {Chan}, T.~K. and {Hopkins}, Philip F. and {Graus}, Andrew S. and {Wetzel}, Andrew and {El-Badry}, Kareem and {Wheeler}, Coral and {Straight}, Maria C. and {Kere{\v{s}}}, Du{\v{s}}an and {Faucher-Gigu{\`e}re}, Claude-Andr{\'e} and {Fitts}, Alex and {Garrison-Kimmel}, Shea},
        title = "{A dark matter profile to model diverse feedback-induced core sizes of {\ensuremath{\Lambda}}CDM haloes}",
      journal = {\mnras},
         year = 2020,
        month = sep,
       volume = {497},
       number = {2},
        pages = {2393-2417},
          doi = {10.1093/mnras/staa2101},
archivePrefix = {arXiv},
       eprint = {2004.10817},
 primaryClass = {astro-ph.GA},
       adsurl = {https://ui.adsabs.harvard.edu/abs/2020MNRAS.497.2393L}
}

@ARTICLE{Leja19b,
       author = {{Leja}, Joel and {Johnson}, Benjamin D. and {Conroy}, Charlie and {van Dokkum}, Pieter and {Speagle}, Joshua S. and {Brammer}, Gabriel and {Momcheva}, Ivelina and {Skelton}, Rosalind and {Whitaker}, Katherine E. and {Franx}, Marijn and {Nelson}, Erica J.},
        title = "{An Older, More Quiescent Universe from Panchromatic SED Fitting of the 3D-HST Survey}",
      journal = {\apj},
         year = 2019,
        month = jun,
       volume = {877},
       number = {2},
          eid = {140},
        pages = {140},
          doi = {10.3847/1538-4357/ab1d5a},
archivePrefix = {arXiv},
       eprint = {1812.05608},
 primaryClass = {astro-ph.GA},
       adsurl = {https://ui.adsabs.harvard.edu/abs/2019ApJ...877..140L}
}

@ARTICLE{Leja19,
       author = {{Leja}, Joel and {Carnall}, Adam C. and {Johnson}, Benjamin D. and {Conroy}, Charlie and {Speagle}, Joshua S.},
        title = "{How to Measure Galaxy Star Formation Histories. II. Nonparametric Models}",
      journal = {\apj},
         year = 2019,
        month = may,
       volume = {876},
       number = {1},
          eid = {3},
        pages = {3},
          doi = {10.3847/1538-4357/ab133c},
archivePrefix = {arXiv},
       eprint = {1811.03637},
 primaryClass = {astro-ph.GA},
       adsurl = {https://ui.adsabs.harvard.edu/abs/2019ApJ...876....3L}
}

@ARTICLE{li20,
       author = {{Li}, Pengfei and {Lelli}, Federico and {McGaugh}, Stacy and {Schombert}, James},
        title = "{A Comprehensive Catalog of Dark Matter Halo Models for SPARC Galaxies}",
      journal = {\apjs},
         year = 2020,
        month = mar,
       volume = {247},
       number = {1},
          eid = {31},
        pages = {31},
          doi = {10.3847/1538-4365/ab700e},
archivePrefix = {arXiv},
       eprint = {2001.10538},
 primaryClass = {astro-ph.GA},
       adsurl = {https://ui.adsabs.harvard.edu/abs/2020ApJS..247...31L}
}

@ARTICLE{lifre,
       author = {{Li}, Zhaozhou and {Dekel}, Avishai and {Mandelker}, Nir and {Freundlich}, Jonathan and {Fran{\c{c}}ois}, Thibaut L.},
        title = "{The response of dark matter haloes to gas ejection: CuspCore II}",
      journal = {\mnras},
         year = 2023,
        month = feb,
       volume = {518},
       number = {4},
        pages = {5356-5375},
          doi = {10.1093/mnras/stac3233},
archivePrefix = {arXiv},
       eprint = {2206.07069},
 primaryClass = {astro-ph.GA},
       adsurl = {https://ui.adsabs.harvard.edu/abs/2023MNRAS.518.5356L}
}

@INPROCEEDINGS{Lotz,
       author = {{Lotz}, Jennifer and {Mountain}, M. and {Grogin}, N.~A. and {Koekemoer}, A.~M. and {Capak}, P.~L. and {Mack}, J. and {Coe}, D.~A. and {Barker}, E.~A. and {Adler}, D.~S. and {Avila}, R.~J. and {Anderson}, J. and {Casertano}, S. and {Christian}, C.~A. and {Gonzaga}, S. and {Ferguson}, H.~C. and {Fruchter}, A.~S. and {Jenkner}, H. and {Jordan}, I.~J. and {Hammer}, D. and {Hilbert}, B. and {Lawton}, B.~L. and {Lee}, J.~C. and {Lucas}, R.~A. and {MacKenty}, J.~W. and {Mutchler}, M.~J. and {Ogaz}, S. and {Reid}, I.~N. and {Royle}, P. and {Robberto}, M. and {Sembach}, K. and {Smith}, L.~J. and {Sokol}, J. and {Surace}, J.~A. and {Taylor}, D. and {Tumlinson}, J. and {Viana}, A. and {Williams}, R.~E. and {Workman}, W.},
        title = "{The HST Frontier Fields}",
    booktitle = {American Astronomical Society Meeting Abstracts \#223},
         year = 2014,
       series = {American Astronomical Society Meeting Abstracts},
       volume = {223},
        month = jan,
          eid = {254.01},
        pages = {254.01},
       adsurl = {https://ui.adsabs.harvard.edu/abs/2014AAS...22325401L}
}

@ARTICLE{2014ARA&A..52..415M,
       author = {{Madau}, Piero and {Dickinson}, Mark},
        title = "{Cosmic Star-Formation History}",
      journal = {\araa},
         year = 2014,
        month = aug,
       volume = {52},
        pages = {415-486},
          doi = {10.1146/annurev-astro-081811-125615},
archivePrefix = {arXiv},
       eprint = {1403.0007},
 primaryClass = {astro-ph.CO},
       adsurl = {https://ui.adsabs.harvard.edu/abs/2014ARA&A..52..415M}
}

@ARTICLE{Marasco,
       author = {{Marasco}, A. and {Oman}, K.~A. and {Navarro}, J.~F. and {Frenk}, C.~S. and {Oosterloo}, T.},
        title = "{Bars in dark-matter-dominated dwarf galaxy discs}",
      journal = {\mnras},
         year = 2018,
        month = may,
       volume = {476},
       number = {2},
        pages = {2168-2176},
          doi = {10.1093/mnras/sty354},
archivePrefix = {arXiv},
       eprint = {1711.09914},
 primaryClass = {astro-ph.GA},
       adsurl = {https://ui.adsabs.harvard.edu/abs/2018MNRAS.476.2168M}
}

@ARTICLE{Mehta,
       author = {{Mehta}, Vihang and {Teplitz}, Harry I. and {Scarlata}, Claudia and {Wang}, Xin and {Alavi}, Anahita and {Colbert}, James and {Rafelski}, Marc and {Grogin}, Norman and {Koekemoer}, Anton and {Prichard}, Laura and {Windhorst}, Rogier and {Barber}, Justin M. and {Conselice}, Christopher J. and {Dai}, Y. Sophia and {Gardner}, Jonathan P. and {Gawiser}, Eric and {Guo}, Yicheng and {Hathi}, Nimish and {Arrabal Haro}, Pablo and {Hayes}, Matthew and {Iyer}, Kartheik G. and {Jansen}, Rolf A. and {Ji}, Zhiyuan and {Kurczynski}, Peter and {Kuschel}, Maxwell and {Lucas}, Ray A. and {Mantha}, Kameswara and {O'Connell}, Robert W. and {Ravindranath}, Swara and {Robertson}, Brant E. and {Rutkowski}, Michael and {Siana}, Brian and {Yung}, L.~Y. Aaron},
        title = "{A Spatially Resolved Analysis of Star Formation Burstiness by Comparing UV and H{\ensuremath{\alpha}} in Galaxies at z {\ensuremath{\sim}} 1 with UVCANDELS}",
      journal = {\apj},
         year = 2023,
        month = aug,
       volume = {952},
       number = {2},
          eid = {133},
        pages = {133},
          doi = {10.3847/1538-4357/acd9cf},
archivePrefix = {arXiv},
       eprint = {2211.02056},
 primaryClass = {astro-ph.GA},
       adsurl = {https://ui.adsabs.harvard.edu/abs/2023ApJ...952..133M}
}

@ARTICLE{Mobasher,
       author = {{Mobasher}, Bahram and {Dahlen}, Tomas and {Ferguson}, Henry C. and {Acquaviva}, Viviana and {Barro}, Guillermo and {Finkelstein}, Steven L. and {Fontana}, Adriano and {Gruetzbauch}, Ruth and {Johnson}, Seth and {Lu}, Yu and {Papovich}, Casey J. and {Pforr}, Janine and {Salvato}, Mara and {Somerville}, Rachel S. and {Wiklind}, Tommy and {Wuyts}, Stijn and {Ashby}, Matthew L.~N. and {Bell}, Eric and {Conselice}, Christopher J. and {Dickinson}, Mark E. and {Faber}, Sandra M. and {Fazio}, Giovanni and {Finlator}, Kristian and {Galametz}, Audrey and {Gawiser}, Eric and {Giavalisco}, Mauro and {Grazian}, Andrea and {Grogin}, Norman A. and {Guo}, Yicheng and {Hathi}, Nimish and {Kocevski}, Dale and {Koekemoer}, Anton M. and {Koo}, David C. and {Newman}, Jeffrey A. and {Reddy}, Naveen and {Santini}, Paola and {Wechsler}, Risa H.},
        title = "{A Critical Assessment of Stellar Mass Measurement Methods}",
      journal = {\apj},
         year = 2015,
        month = jul,
       volume = {808},
       number = {1},
          eid = {101},
        pages = {101},
          doi = {10.1088/0004-637X/808/1/101},
archivePrefix = {arXiv},
       eprint = {1505.01501},
 primaryClass = {astro-ph.GA},
       adsurl = {https://ui.adsabs.harvard.edu/abs/2015ApJ...808..101M}
}

@ARTICLE{ml,
       author = {{Mu{\~n}oz L{\'o}pez}, C. and {Krajnovi{\'c}}, D. and {Epinat}, B. and {Urrutia}, T. and {Pessa}, I. and {Contini}, T. and {Nanayakkara}, T. and {Pharo}, J. and {Gon{\c{c}}alves}, T.~S. and {Thi Thai}, T. and {Bouch{\'e}}, N.~F.},
        title = "{Multiple star-forming episodes of intermediate-redshift galaxies}",
      journal = {\aap},
         year = 2025,
        month = oct,
       volume = {702},
          eid = {A146},
        pages = {A146},
          doi = {10.1051/0004-6361/202555197},
archivePrefix = {arXiv},
       eprint = {2509.01710},
 primaryClass = {astro-ph.GA},
       adsurl = {https://ui.adsabs.harvard.edu/abs/2025A&A...702A.146M}
}

@ARTICLE{nfw,
       author = {{Navarro}, Julio F. and {Frenk}, Carlos S. and {White}, Simon D.~M.},
        title = "{A Universal Density Profile from Hierarchical Clustering}",
      journal = {\apj},
         year = 1997,
        month = dec,
       volume = {490},
       number = {2},
        pages = {493-508},
          doi = {10.1086/304888},
archivePrefix = {arXiv},
       eprint = {astro-ph/9611107},
 primaryClass = {astro-ph},
       adsurl = {https://ui.adsabs.harvard.edu/abs/1997ApJ...490..493N}
}

@ARTICLE{nav,
       author = {{Navarro}, Julio F. and {Eke}, Vincent R. and {Frenk}, Carlos S.},
        title = "{The cores of dwarf galaxy haloes}",
      journal = {\mnras},
         year = 1996,
        month = dec,
       volume = {283},
       number = {3},
        pages = {L72-L78},
          doi = {10.1093/mnras/283.3.L72},
archivePrefix = {arXiv},
       eprint = {astro-ph/9610187},
 primaryClass = {astro-ph},
       adsurl = {https://ui.adsabs.harvard.edu/abs/1996MNRAS.283L..72N}
}

@ARTICLE{Nedkova,
       author = {{Nedkova}, Kalina V. and {H{\"a}u{\ss}ler}, Boris and {Marchesini}, Danilo and {Dimauro}, Paola and {Brammer}, Gabriel and {Eigenthaler}, Paul and {Feinstein}, Adina D. and {Ferguson}, Henry C. and {Huertas-Company}, Marc and {Johnston}, Evelyn J. and {Kado-Fong}, Erin and {Kartaltepe}, Jeyhan S. and {Labb{\'e}}, Ivo and {Lange-Vagle}, Daniel and {Martis}, Nicholas S. and {McGrath}, Elizabeth J. and {Muzzin}, Adam and {Oesch}, Pascal and {Ordenes-Brice{\~n}o}, Yasna and {Puzia}, Thomas and {Shipley}, Heath V. and {Simmons}, Brooke D. and {Skelton}, Rosalind E. and {Stefanon}, Mauro and {van der Wel}, Arjen and {Whitaker}, Katherine E.},
        title = "{Extending the evolution of the stellar mass-size relation at z {\ensuremath{\leq}} 2 to low stellar mass galaxies from HFF and CANDELS}",
      journal = {\mnras},
         year = 2021,
        month = sep,
       volume = {506},
       number = {1},
        pages = {928-956},
          doi = {10.1093/mnras/stab1744},
archivePrefix = {arXiv},
       eprint = {2106.07663},
 primaryClass = {astro-ph.GA},
       adsurl = {https://ui.adsabs.harvard.edu/abs/2021MNRAS.506..928N}
}

@ARTICLE{Neistein,
       author = {{Neistein}, Eyal and {van den Bosch}, Frank C. and {Dekel}, Avishai},
        title = "{Natural downsizing in hierarchical galaxy formation}",
      journal = {\mnras},
         year = 2006,
        month = oct,
       volume = {372},
       number = {2},
        pages = {933-948},
          doi = {10.1111/j.1365-2966.2006.10918.x},
archivePrefix = {arXiv},
       eprint = {astro-ph/0605045},
 primaryClass = {astro-ph},
       adsurl = {https://ui.adsabs.harvard.edu/abs/2006MNRAS.372..933N}
}

@ARTICLE{nu,
       author = {{Nu{\~n}ez-Casti{\~n}eyra}, A. and {Nezri}, E. and {Mollitor}, P. and {Michel-Dansac}, L. and {Devriendt}, J. and {Teyssier}, R.},
        title = "{Cosmological simulations of the same spiral galaxy: satellite properties, the role of baryonic physics and star formation history in shaping dark matter cores/cusps}",
      journal = {\jcap},
         year = 2026,
        month = may,
       volume = {2026},
       number = {5},
          eid = {025},
        pages = {025},
          doi = {10.1088/1475-7516/2026/05/025},
archivePrefix = {arXiv},
       eprint = {2509.07470},
 primaryClass = {astro-ph.GA},
       adsurl = {https://ui.adsabs.harvard.edu/abs/2026JCAP...05..025N}
}

@ARTICLE{oh,
       author = {{Oh}, Se-Heon and {Hunter}, Deidre A. and {Brinks}, Elias and {Elmegreen}, Bruce G. and {Schruba}, Andreas and {Walter}, Fabian and {Rupen}, Michael P. and {Young}, Lisa M. and {Simpson}, Caroline E. and {Johnson}, Megan C. and {Herrmann}, Kimberly A. and {Ficut-Vicas}, Dana and {Cigan}, Phil and {Heesen}, Volker and {Ashley}, Trisha and {Zhang}, Hong-Xin},
        title = "{High-resolution Mass Models of Dwarf Galaxies from LITTLE THINGS}",
      journal = {\aj},
         year = 2015,
        month = jun,
       volume = {149},
       number = {6},
          eid = {180},
        pages = {180},
          doi = {10.1088/0004-6256/149/6/180},
archivePrefix = {arXiv},
       eprint = {1502.01281},
 primaryClass = {astro-ph.GA},
       adsurl = {https://ui.adsabs.harvard.edu/abs/2015AJ....149..180O}
}

@ARTICLE{Oman2019,
       author = {{Oman}, Kyle A. and {Marasco}, Antonino and {Navarro}, Julio F. and {Frenk}, Carlos S. and {Schaye}, Joop and {Ben{\'\i}tez-Llambay}, Alejandro},
        title = "{Non-circular motions and the diversity of dwarf galaxy rotation curves}",
      journal = {\mnras},
         year = 2019,
        month = jan,
       volume = {482},
       number = {1},
        pages = {821-847},
          doi = {10.1093/mnras/sty2687},
archivePrefix = {arXiv},
       eprint = {1706.07478},
 primaryClass = {astro-ph.GA},
       adsurl = {https://ui.adsabs.harvard.edu/abs/2019MNRAS.482..821O}
}

@ARTICLE{Pacifici,
       author = {{Pacifici}, Camilla and {Oh}, Sree and {Oh}, Kyuseok and {Lee}, Jaehyun and {Yi}, Sukyoung K.},
        title = "{Timing the Evolution of Quiescent and Star-forming Local Galaxies}",
      journal = {\apj},
         year = 2016,
        month = jun,
       volume = {824},
       number = {1},
          eid = {45},
        pages = {45},
          doi = {10.3847/0004-637X/824/1/45},
archivePrefix = {arXiv},
       eprint = {1604.02460},
 primaryClass = {astro-ph.GA},
       adsurl = {https://ui.adsabs.harvard.edu/abs/2016ApJ...824...45P}
}

@ARTICLE{Pedregosa,
       author = {{Pedregosa}, Fabian and {Varoquaux}, Ga{\"e}l and {Gramfort}, Alexandre and {Michel}, Vincent and {Thirion}, Bertrand and {Grisel}, Olivier and {Blondel}, Mathieu and {M{\"u}ller}, Andreas and {Nothman}, Joel and {Louppe}, Gilles and {Prettenhofer}, Peter and {Weiss}, Ron and {Dubourg}, Vincent and {Vanderplas}, Jake and {Passos}, Alexandre and {Cournapeau}, David and {Brucher}, Matthieu and {Perrot}, Matthieu and {Duchesnay}, {\'E}douard},
        title = "{Scikit-learn: Machine Learning in Python}",
      journal = {Journal of Machine Learning Research},
         year = 2011,
        month = oct,
       volume = {12},
        pages = {2825-2830},
          doi = {10.48550/arXiv.1201.0490},
archivePrefix = {arXiv},
       eprint = {1201.0490},
 primaryClass = {cs.LG},
       adsurl = {https://ui.adsabs.harvard.edu/abs/2011JMLR...12.2825P}
}

@ARTICLE{galfit,
       author = {{Peng}, Chien Y. and {Ho}, Luis C. and {Impey}, Chris D. and {Rix}, Hans-Walter},
        title = "{Detailed Decomposition of Galaxy Images. II. Beyond Axisymmetric Models}",
      journal = {\aj},
         year = 2010,
        month = jun,
       volume = {139},
       number = {6},
        pages = {2097-2129},
          doi = {10.1088/0004-6256/139/6/2097},
archivePrefix = {arXiv},
       eprint = {0912.0731},
 primaryClass = {astro-ph.CO},
       adsurl = {https://ui.adsabs.harvard.edu/abs/2010AJ....139.2097P}
}

@ARTICLE{2008ApJ...675..234P,
       author = {{P{\'e}rez-Gonz{\'a}lez}, Pablo G. and {Rieke}, George H. and {Villar}, Victor and {Barro}, Guillermo and {Blaylock}, Myra and {Egami}, Eiichi and {Gallego}, Jes{\'u}s and {Gil de Paz}, Armando and {Pascual}, Sergio and {Zamorano}, Jaime and {Donley}, Jennifer L.},
        title = "{The Stellar Mass Assembly of Galaxies from z = 0 to z = 4: Analysis of a Sample Selected in the Rest-Frame Near-Infrared with Spitzer}",
      journal = {\apj},
         year = 2008,
        month = mar,
       volume = {675},
       number = {1},
        pages = {234-261},
          doi = {10.1086/523690},
archivePrefix = {arXiv},
       eprint = {0709.1354},
 primaryClass = {astro-ph},
       adsurl = {https://ui.adsabs.harvard.edu/abs/2008ApJ...675..234P}
}

@ARTICLE{persic,
       author = {{Persic}, Massimo and {Salucci}, Paolo and {Stel}, Fulvio},
        title = "{The universal rotation curve of spiral galaxies {\textemdash} I. The dark matter connection}",
      journal = {\mnras},
         year = 1996,
        month = jul,
       volume = {281},
       number = {1},
        pages = {27-47},
          doi = {10.1093/mnras/278.1.27},
archivePrefix = {arXiv},
       eprint = {astro-ph/9506004},
 primaryClass = {astro-ph},
       adsurl = {https://ui.adsabs.harvard.edu/abs/1996MNRAS.281...27P}
}

@ARTICLE{Pessa,
       author = {{Pessa}, Ismael and {Wisotzki}, Lutz and {Urrutia}, Tanya and {Bouch{\'e}}, Nicolas F. and {Leclercq}, Floriane and {Augustin}, Ramona and {Guo}, Yucheng and {Kozlova}, Daria and {Kusakabe}, Haruka and {Pharo}, John},
        title = "{First statistical constraints on galactic-scale outflow properties traced by their extended Mg II emission with MUSE}",
      journal = {\aap},
         year = 2026,
        month = apr,
       volume = {708},
          eid = {A214},
        pages = {A214},
          doi = {10.1051/0004-6361/202557875},
archivePrefix = {arXiv},
       eprint = {2602.11280},
 primaryClass = {astro-ph.GA},
       adsurl = {https://ui.adsabs.harvard.edu/abs/2026A&A...708A.214P}
}

@ARTICLE{Planck,
       author = {{Planck Collaboration} and {Ade}, P.~A.~R. and {Aghanim}, N. and {Arnaud}, M. and {Ashdown}, M. and {Aumont}, J. and {Baccigalupi}, C. and {Banday}, A.~J. and {Barreiro}, R.~B. and {Bartlett}, J.~G. and {Bartolo}, N. and {Battaner}, E. and {Battye}, R. and {Benabed}, K. and {Beno{\^\i}t}, A. and {Benoit-L{\'e}vy}, A. and {Bernard}, J.-P. and {Bersanelli}, M. and {Bielewicz}, P. and {Bock}, J.~J. and {Bonaldi}, A. and {Bonavera}, L. and {Bond}, J.~R. and {Borrill}, J. and {Bouchet}, F.~R. and {Boulanger}, F. and {Bucher}, M. and {Burigana}, C. and {Butler}, R.~C. and {Calabrese}, E. and {Cardoso}, J.-F. and {Catalano}, A. and {Challinor}, A. and {Chamballu}, A. and {Chary}, R.-R. and {Chiang}, H.~C. and {Chluba}, J. and {Christensen}, P.~R. and {Church}, S. and {Clements}, D.~L. and {Colombi}, S. and {Colombo}, L.~P.~L. and {Combet}, C. and {Coulais}, A. and {Crill}, B.~P. and {Curto}, A. and {Cuttaia}, F. and {Danese}, L. and {Davies}, R.~D. and {Davis}, R.~J. and {de Bernardis}, P. and {de Rosa}, A. and {de Zotti}, G. and {Delabrouille}, J. and {D{\'e}sert}, F.-X. and {Di Valentino}, E. and {Dickinson}, C. and {Diego}, J.~M. and {Dolag}, K. and {Dole}, H. and {Donzelli}, S. and {Dor{\'e}}, O. and {Douspis}, M. and {Ducout}, A. and {Dunkley}, J. and {Dupac}, X. and {Efstathiou}, G. and {Elsner}, F. and {En{\ss}lin}, T.~A. and {Eriksen}, H.~K. and {Farhang}, M. and {Fergusson}, J. and {Finelli}, F. and {Forni}, O. and {Frailis}, M. and {Fraisse}, A.~A. and {Franceschi}, E. and {Frejsel}, A. and {Galeotta}, S. and {Galli}, S. and {Ganga}, K. and {Gauthier}, C. and {Gerbino}, M. and {Ghosh}, T. and {Giard}, M. and {Giraud-H{\'e}raud}, Y. and {Giusarma}, E. and {Gjerl{\o}w}, E. and {Gonz{\'a}lez-Nuevo}, J. and {G{\'o}rski}, K.~M. and {Gratton}, S. and {Gregorio}, A. and {Gruppuso}, A. and {Gudmundsson}, J.~E. and {Hamann}, J. and {Hansen}, F.~K. and {Hanson}, D. and {Harrison}, D.~L. and {Helou}, G. and {Henrot-Versill{\'e}}, S. and {Hern{\'a}ndez-Monteagudo}, C. and {Herranz}, D. and {Hildebrandt}, S.~R. and {Hivon}, E. and {Hobson}, M. and {Holmes}, W.~A. and {Hornstrup}, A. and {Hovest}, W. and {Huang}, Z. and {Huffenberger}, K.~M. and {Hurier}, G. and {Jaffe}, A.~H. and {Jaffe}, T.~R. and {Jones}, W.~C. and {Juvela}, M. and {Keih{\"a}nen}, E. and {Keskitalo}, R. and {Kisner}, T.~S. and {Kneissl}, R. and {Knoche}, J. and {Knox}, L. and {Kunz}, M. and {Kurki-Suonio}, H. and {Lagache}, G. and {L{\"a}hteenm{\"a}ki}, A. and {Lamarre}, J.-M. and {Lasenby}, A. and {Lattanzi}, M. and {Lawrence}, C.~R. and {Leahy}, J.~P. and {Leonardi}, R. and {Lesgourgues}, J. and {Levrier}, F. and {Lewis}, A. and {Liguori}, M. and {Lilje}, P.~B. and {Linden-V{\o}rnle}, M. and {L{\'o}pez-Caniego}, M. and {Lubin}, P.~M. and {Mac{\'\i}as-P{\'e}rez}, J.~F. and {Maggio}, G. and {Maino}, D. and {Mandolesi}, N. and {Mangilli}, A. and {Marchini}, A. and {Maris}, M. and {Martin}, P.~G. and {Martinelli}, M. and {Mart{\'\i}nez-Gonz{\'a}lez}, E. and {Masi}, S. and {Matarrese}, S. and {McGehee}, P. and {Meinhold}, P.~R. and {Melchiorri}, A. and {Melin}, J.-B. and {Mendes}, L. and {Mennella}, A. and {Migliaccio}, M. and {Millea}, M. and {Mitra}, S. and {Miville-Desch{\^e}nes}, M.-A. and {Moneti}, A. and {Montier}, L. and {Morgante}, G. and {Mortlock}, D. and {Moss}, A. and {Munshi}, D. and {Murphy}, J.~A. and {Naselsky}, P. and {Nati}, F. and {Natoli}, P. and {Netterfield}, C.~B. and {N{\o}rgaard-Nielsen}, H.~U. and {Noviello}, F. and {Novikov}, D. and {Novikov}, I. and {Oxborrow}, C.~A. and {Paci}, F. and {Pagano}, L. and {Pajot}, F. and {Paladini}, R. and {Paoletti}, D. and {Partridge}, B. and {Pasian}, F. and {Patanchon}, G. and {Pearson}, T.~J. and {Perdereau}, O. and {Perotto}, L. and {Perrotta}, F. and {Pettorino}, V. and {Piacentini}, F. and {Piat}, M. and {Pierpaoli}, E. and {Pietrobon}, D. and {Plaszczynski}, S. and {Pointecouteau}, E. and {Polenta}, G. and {Popa}, L. and {Pratt}, G.~W. and {Pr{\'e}zeau}, G.},
        title = "{Planck 2015 results. XIII. Cosmological parameters}",
      journal = {\aap},
         year = 2016,
        month = sep,
       volume = {594},
          eid = {A13},
        pages = {A13},
          doi = {10.1051/0004-6361/201525830},
archivePrefix = {arXiv},
       eprint = {1502.01589},
 primaryClass = {astro-ph.CO},
       adsurl = {https://ui.adsabs.harvard.edu/abs/2016A&A...594A..13P}
}

@ARTICLE{pon,
       author = {{Pontzen}, Andrew and {Governato}, Fabio},
        title = "{How supernova feedback turns dark matter cusps into cores}",
      journal = {\mnras},
         year = 2012,
        month = apr,
       volume = {421},
       number = {4},
        pages = {3464-3471},
          doi = {10.1111/j.1365-2966.2012.20571.x},
archivePrefix = {arXiv},
       eprint = {1106.0499},
 primaryClass = {astro-ph.CO},
       adsurl = {https://ui.adsabs.harvard.edu/abs/2012MNRAS.421.3464P}
}

@ARTICLE{pug,
       author = {{Puglisi}, Annagrazia and {Dudzevi{\v{c}}i{\={u}}t{\.{e}}}, Ugn{\.{e}} and {Swinbank}, Mark and {Gillman}, Steven and {Tiley}, Alfred L. and {Bower}, Richard G. and {Cirasuolo}, Michele and {Cortese}, Luca and {Glazebrook}, Karl and {Harrison}, Chris and {Ibar}, Edo and {Molina}, Juan and {Obreschkow}, Danail and {Oman}, Kyle A. and {Schaller}, Matthieu and {Shankar}, Francesco and {Sharples}, Ray M.},
        title = "{KURVS: the outer rotation curve shapes and dark matter fractions of z   1.5 star-forming galaxies}",
      journal = {\mnras},
         year = 2023,
        month = sep,
       volume = {524},
       number = {2},
        pages = {2814-2835},
          doi = {10.1093/mnras/stad1966},
archivePrefix = {arXiv},
       eprint = {2305.04382},
 primaryClass = {astro-ph.GA},
       adsurl = {https://ui.adsabs.harvard.edu/abs/2023MNRAS.524.2814P}
}

@ARTICLE{rg,
       author = {{Read}, J.~I. and {Gilmore}, G.},
        title = "{Mass loss from dwarf spheroidal galaxies: the origins of shallow dark matter cores and exponential surface brightness profiles}",
      journal = {\mnras},
         year = 2005,
        month = jan,
       volume = {356},
       number = {1},
        pages = {107-124},
          doi = {10.1111/j.1365-2966.2004.08424.x},
archivePrefix = {arXiv},
       eprint = {astro-ph/0409565},
 primaryClass = {astro-ph},
       adsurl = {https://ui.adsabs.harvard.edu/abs/2005MNRAS.356..107R}
}

@ARTICLE{read,
       author = {{Read}, J.~I. and {Agertz}, O. and {Collins}, M.~L.~M.},
        title = "{Dark matter cores all the way down}",
      journal = {\mnras},
         year = 2016,
        month = jul,
       volume = {459},
       number = {3},
        pages = {2573-2590},
          doi = {10.1093/mnras/stw713},
archivePrefix = {arXiv},
       eprint = {1508.04143},
 primaryClass = {astro-ph.GA},
       adsurl = {https://ui.adsabs.harvard.edu/abs/2016MNRAS.459.2573R}
}

@ARTICLE{read3,
       author = {{Read}, J.~I. and {Walker}, M.~G. and {Steger}, P.},
        title = "{Dark matter heats up in dwarf galaxies}",
      journal = {\mnras},
         year = 2019,
        month = mar,
       volume = {484},
       number = {1},
        pages = {1401-1420},
          doi = {10.1093/mnras/sty3404},
archivePrefix = {arXiv},
       eprint = {1808.06634},
 primaryClass = {astro-ph.GA},
       adsurl = {https://ui.adsabs.harvard.edu/abs/2019MNRAS.484.1401R}
}

@ARTICLE{Rezaee,
       author = {{Rezaee}, Saeed and {Reddy}, Naveen A. and {Topping}, Michael W. and {Shivaei}, Irene and {Shapley}, Alice E. and {Fetherolf}, Tara and {Kriek}, Mariska and {Coil}, Alison and {Mobasher}, Bahram and {Siana}, Brian and {Du}, Xinnan and {Khostovan}, Ali Ahmad and {Weldon}, Andrew and {Emami}, Najmeh and {Chartab}, Nima},
        title = "{Exploring the correlation between H{\ensuremath{\alpha}}-to-UV ratio and burstiness for typical star-forming galaxies at z   2}",
      journal = {\mnras},
         year = 2023,
        month = nov,
       volume = {526},
       number = {1},
        pages = {1512-1527},
          doi = {10.1093/mnras/stad2842},
archivePrefix = {arXiv},
       eprint = {2208.12371},
 primaryClass = {astro-ph.GA},
       adsurl = {https://ui.adsabs.harvard.edu/abs/2023MNRAS.526.1512R}
}

@ARTICLE{Richard,
       author = {{Richard}, Johan and {Claeyssens}, Ad{\'e}la{\"\i}de and {Lagattuta}, David and {Guaita}, Lucia and {Bauer}, Franz Erik and {Pello}, Roser and {Carton}, David and {Bacon}, Roland and {Soucail}, Genevi{\`e}ve and {Lyon}, Gonzalo Prieto and {Kneib}, Jean-Paul and {Mahler}, Guillaume and {Cl{\'e}ment}, Benjamin and {Mercier}, Wilfried and {Variu}, Andrei and {Tamone}, Am{\'e}lie and {Ebeling}, Harald and {Schmidt}, Kasper B. and {Nanayakkara}, Themiya and {Maseda}, Michael and {Weilbacher}, Peter M. and {Bouch{\'e}}, Nicolas and {Bouwens}, Rychard J. and {Wisotzki}, Lutz and {de la Vieuville}, Geoffroy and {Martinez}, Johany and {Patr{\'\i}cio}, Vera},
        title = "{An atlas of MUSE observations towards twelve massive lensing clusters}",
      journal = {\aap},
         year = 2021,
        month = feb,
       volume = {646},
          eid = {A83},
        pages = {A83},
          doi = {10.1051/0004-6361/202039462},
archivePrefix = {arXiv},
       eprint = {2009.09784},
 primaryClass = {astro-ph.GA},
       adsurl = {https://ui.adsabs.harvard.edu/abs/2021A&A...646A..83R}
}

@ARTICLE{1975ApJ...201..327R,
       author = {{Roberts}, M.~S. and {Whitehurst}, R.~N.},
        title = "{The rotation curve and geometry of M31 at large galactocentric distances.}",
      journal = {\apj},
         year = 1975,
        month = oct,
       volume = {201},
        pages = {327-346},
          doi = {10.1086/153889},
       adsurl = {https://ui.adsabs.harvard.edu/abs/1975ApJ...201..327R}
}

@ARTICLE{Robertson,
       author = {{Robertson}, Brant E. and {Johnson}, Benjamin D. and {Tacchella}, Sandro and {Eisenstein}, Daniel J. and {Hainline}, Kevin and {Alberts}, Stacey and {Arribas}, Santiago and {Baker}, William M. and {Bunker}, Andrew J. and {Cameron}, Alex J. and {Carniani}, Stefano and {Carreira}, Courtney and {Chevallard}, Jacopo and {Circosta}, Chiara and {Curtis-Lake}, Emma and {Danhaive}, A. Lola and {Duan}, Qiao and {Egami}, Eiichi and {Hausen}, Ryan and {Helton}, Jakob M. and {Ji}, Zhiyuan and {Maiolino}, Roberto and {P{\'e}rez-Gonz{\'a}lez}, Pablo G. and {Pusk{\'a}s}, D{\'a}vid and {Rieke}, Marcia and {Rinaldi}, Pierluigi and {Sun}, Fengwu and {Sun}, Yang and {{\"U}bler}, Hannah and {Trussler}, James A.~A. and {Villanueva}, Natalia C. and {Whitler}, Lily and {Williams}, Christina C. and {Willmer}, Christopher N.~A. and {Willott}, Chris and {Wu}, Zihao and {Zhu}, Yongda},
        title = "{JWST Advanced Deep Extragalactic Survey (JADES) Data Release 5: Photometric Catalog}",
      journal = {arXiv e-prints},
         year = 2026,
        month = jan,
          eid = {arXiv:2601.15956},
        pages = {arXiv:2601.15956},
          doi = {10.48550/arXiv.2601.15956},
archivePrefix = {arXiv},
       eprint = {2601.15956},
 primaryClass = {astro-ph.GA},
       adsurl = {https://ui.adsabs.harvard.edu/abs/2026arXiv260115956R}
}

@ARTICLE{astropy,
       author = {{Astropy Collaboration} and {Robitaille}, Thomas P. and {Tollerud}, Erik J. and {Greenfield}, Perry and {Droettboom}, Michael and {Bray}, Erik and {Aldcroft}, Tom and {Davis}, Matt and {Ginsburg}, Adam and {Price-Whelan}, Adrian M. and {Kerzendorf}, Wolfgang E. and {Conley}, Alexander and {Crighton}, Neil and {Barbary}, Kyle and {Muna}, Demitri and {Ferguson}, Henry and {Grollier}, Fr{\'e}d{\'e}ric and {Parikh}, Madhura M. and {Nair}, Prasanth H. and {Unther}, Hans M. and {Deil}, Christoph and {Woillez}, Julien and {Conseil}, Simon and {Kramer}, Roban and {Turner}, James E.~H. and {Singer}, Leo and {Fox}, Ryan and {Weaver}, Benjamin A. and {Zabalza}, Victor and {Edwards}, Zachary I. and {Azalee Bostroem}, K. and {Burke}, D.~J. and {Casey}, Andrew R. and {Crawford}, Steven M. and {Dencheva}, Nadia and {Ely}, Justin and {Jenness}, Tim and {Labrie}, Kathleen and {Lim}, Pey Lian and {Pierfederici}, Francesco and {Pontzen}, Andrew and {Ptak}, Andy and {Refsdal}, Brian and {Servillat}, Mathieu and {Streicher}, Ole},
        title = "{Astropy: A community Python package for astronomy}",
      journal = {\aap},
         year = 2013,
        month = oct,
       volume = {558},
          eid = {A33},
        pages = {A33},
          doi = {10.1051/0004-6361/201322068},
archivePrefix = {arXiv},
       eprint = {1307.6212},
 primaryClass = {astro-ph.IM},
       adsurl = {https://ui.adsabs.harvard.edu/abs/2013A&A...558A..33A}
}

@ARTICLE{1974ApJ...193..309R,
       author = {{Rogstad}, D.~H. and {Lockhart}, I.~A. and {Wright}, M.~C.~H.},
        title = "{Aperture-synthesis observations of H I in the galaxy M83.}",
      journal = {\apj},
         year = 1974,
        month = oct,
       volume = {193},
        pages = {309-319},
          doi = {10.1086/153164},
       adsurl = {https://ui.adsabs.harvard.edu/abs/1974ApJ...193..309R}
}

@ARTICLE{Rubin,
       author = {{Rubin}, Vera C. and {Ford}, Jr., W. Kent},
        title = "{Rotation of the Andromeda Nebula from a Spectroscopic Survey of Emission Regions}",
      journal = {\apj},
         year = 1970,
        month = feb,
       volume = {159},
        pages = {379},
          doi = {10.1086/150317},
       adsurl = {https://ui.adsabs.harvard.edu/abs/1970ApJ...159..379R}
}

@ARTICLE{Rubin2,
       author = {{Rubin}, V.~C. and {Ford}, Jr., W.~K. and {Thonnard}, N.},
        title = "{Extended rotation curves of high-luminosity spiral galaxies. IV. Systematic dynamical properties, Sa -> Sc.}",
      journal = {\apjl},
         year = 1978,
        month = nov,
       volume = {225},
        pages = {L107-L111},
          doi = {10.1086/182804},
       adsurl = {https://ui.adsabs.harvard.edu/abs/1978ApJ...225L.107R}
}

@ARTICLE{2024A&A...681A.121S,
       author = {{S{\'a}nchez-Menguiano}, Laura and {S{\'a}nchez Almeida}, Jorge and {S{\'a}nchez}, Sebasti{\'a}n F. and {Mu{\~n}oz-Tu{\~n}{\'o}n}, Casiana},
        title = "{Stellar mass is not the best predictor of galaxy metallicity. The gravitational potential-metallicity relation {\ensuremath{\Phi}}ZR}",
      journal = {\aap},
         year = 2024,
        month = jan,
       volume = {681},
          eid = {A121},
        pages = {A121},
          doi = {10.1051/0004-6361/202346708},
archivePrefix = {arXiv},
       eprint = {2312.02046},
 primaryClass = {astro-ph.GA},
       adsurl = {https://ui.adsabs.harvard.edu/abs/2024A&A...681A.121S}
}

@ARTICLE{shach,
       author = {{Nestor Shachar}, A. and {Price}, S.~H. and {F{\"o}rster Schreiber}, N.~M. and {Genzel}, R. and {Shimizu}, T.~T. and {Tacconi}, L.~J. and {{\"U}bler}, H. and {Burkert}, A. and {Davies}, R.~I. and {Dekel}, A. and {Herrera-Camus}, R. and {Lee}, L.~L. and {Liu}, D. and {Lutz}, D. and {Naab}, T. and {Neri}, R. and {Renzini}, A. and {Saglia}, R. and {Schuster}, K.~F. and {Sternberg}, A. and {Wisnioski}, E. and {Wuyts}, S.},
        title = "{RC100: Rotation Curves of 100 Massive Star-forming Galaxies at z = 0.6-2.5 Reveal Little Dark Matter on Galactic Scales}",
      journal = {\apj},
         year = 2023,
        month = feb,
       volume = {944},
       number = {1},
          eid = {78},
        pages = {78},
          doi = {10.3847/1538-4357/aca9cf},
archivePrefix = {arXiv},
       eprint = {2209.12199},
 primaryClass = {astro-ph.GA},
       adsurl = {https://ui.adsabs.harvard.edu/abs/2023ApJ...944...78N}
}

@ARTICLE{cosmos2020,
       author = {{Shuntov}, M. and {McCracken}, H.~J. and {Gavazzi}, R. and {Laigle}, C. and {Weaver}, J.~R. and {Davidzon}, I. and {Ilbert}, O. and {Kauffmann}, O.~B. and {Faisst}, A. and {Dubois}, Y. and {Koekemoer}, A.~M. and {Moneti}, A. and {Milvang-Jensen}, B. and {Mobasher}, B. and {Sanders}, D.~B. and {Toft}, S.},
        title = "{COSMOS2020: Cosmic evolution of the stellar-to-halo mass relation for central and satellite galaxies up to z {\ensuremath{\sim}} 5}",
      journal = {\aap},
         year = 2022,
        month = aug,
       volume = {664},
          eid = {A61},
        pages = {A61},
          doi = {10.1051/0004-6361/202243136},
archivePrefix = {arXiv},
       eprint = {2203.10895},
 primaryClass = {astro-ph.GA},
       adsurl = {https://ui.adsabs.harvard.edu/abs/2022A&A...664A..61S}
}

@ARTICLE{sal,
       author = {{Salucci}, P. and {Burkert}, A.},
        title = "{Dark Matter Scaling Relations}",
      journal = {\apjl},
         year = 2000,
        month = jul,
       volume = {537},
       number = {1},
        pages = {L9-L12},
          doi = {10.1086/312747},
archivePrefix = {arXiv},
       eprint = {astro-ph/0004397},
 primaryClass = {astro-ph},
       adsurl = {https://ui.adsabs.harvard.edu/abs/2000ApJ...537L...9S}
}

@ARTICLE{sal2,
       author = {{Salucci}, P. and {Lapi}, A. and {Tonini}, C. and {Gentile}, G. and {Yegorova}, I. and {Klein}, U.},
        title = "{The universal rotation curve of spiral galaxies - II. The dark matter distribution out to the virial radius}",
      journal = {\mnras},
         year = 2007,
        month = jun,
       volume = {378},
       number = {1},
        pages = {41-47},
          doi = {10.1111/j.1365-2966.2007.11696.x},
archivePrefix = {arXiv},
       eprint = {astro-ph/0703115},
 primaryClass = {astro-ph},
       adsurl = {https://ui.adsabs.harvard.edu/abs/2007MNRAS.378...41S}
}

@ARTICLE{Sarrato,
       author = {{Sarrato-Al{\'o}s}, Jorge and {Bullock}, James and {Di Cintio}, Arianna and {Brook}, Christopher and {Valenciano}, Fernando and {Macci{\`o}}, Andrea V.},
        title = "{The Impact of Star Formation Histories on the Inner Dark Matter Density Slopes of Galaxies}",
      journal = {arXiv e-prints},
         year = 2026,
        month = jan,
          eid = {arXiv:2601.17113},
        pages = {arXiv:2601.17113},
          doi = {10.48550/arXiv.2601.17113},
archivePrefix = {arXiv},
       eprint = {2601.17113},
 primaryClass = {astro-ph.GA},
       adsurl = {https://ui.adsabs.harvard.edu/abs/2026arXiv260117113S}
}

@ARTICLE{Sharma2,
       author = {{Sharma}, Gauri and {Salucci}, Paolo and {van de Ven}, Glenn},
        title = "{Observational evidence of evolving dark matter profiles at z {\ensuremath{\leq}} 1}",
      journal = {\aap},
         year = 2022,
        month = mar,
       volume = {659},
          eid = {A40},
        pages = {A40},
          doi = {10.1051/0004-6361/202141822},
archivePrefix = {arXiv},
       eprint = {2109.14224},
 primaryClass = {astro-ph.GA},
       adsurl = {https://ui.adsabs.harvard.edu/abs/2022A&A...659A..40S}
}

@ARTICLE{2026arXiv260112315S,
       author = {{Sharma}, Gauri and {Battisti}, Andrew J. and {Wisnioski}, Emily and {Mendel}, J. Trevor and {Bellstedt}, Sabine and {Lagos}, Claudia Del P. and {Foster}, Caroline and {Poci}, Adriano and {Harborne}, Katherine E. and {Bagge}, Ryan and {Barsanti}, Stefania and {Bland-Hawthorn}, Joss and {Breda}, Iris and {Croom}, Scott M. and {Glazebrook}, Karl and {Mai}, Yifan and {Sweet}, Sarah M. and {Thater}, Sabine and {Valenzuela}, Lucas M. and {van de Ven}, Glenn and {Yi}, Sukyoung and {Zafar}, Tayyaba and {Ziegler}, Bodo},
        title = "{The MAGPI Survey: co-evolution of baryons and dark matter in star-forming disk-like galaxies at $0.1 \lesssim z \lesssim 0.85$}",
      journal = {arXiv e-prints},
         year = 2026,
        month = jan,
          eid = {arXiv:2601.12315},
        pages = {arXiv:2601.12315},
          doi = {10.48550/arXiv.2601.12315},
archivePrefix = {arXiv},
       eprint = {2601.12315},
 primaryClass = {astro-ph.GA},
       adsurl = {https://ui.adsabs.harvard.edu/abs/2026arXiv260112315S}
}

@ARTICLE{ship,
       author = {{Shipley}, Heath V. and {Lange-Vagle}, Daniel and {Marchesini}, Danilo and {Brammer}, Gabriel B. and {Ferrarese}, Laura and {Stefanon}, Mauro and {Kado-Fong}, Erin and {Whitaker}, Katherine E. and {Oesch}, Pascal A. and {Feinstein}, Adina D. and {Labb{\'e}}, Ivo and {Lundgren}, Britt and {Martis}, Nicholas and {Muzzin}, Adam and {Nedkova}, Kalina and {Skelton}, Rosalind and {van der Wel}, Arjen},
        title = "{HFF-DeepSpace Photometric Catalogs of the 12 Hubble Frontier Fields, Clusters, and Parallels: Photometry, Photometric Redshifts, and Stellar Masses}",
      journal = {\apjs},
         year = 2018,
        month = mar,
       volume = {235},
       number = {1},
          eid = {14},
        pages = {14},
          doi = {10.3847/1538-4365/aaacce},
archivePrefix = {arXiv},
       eprint = {1801.09734},
 primaryClass = {astro-ph.GA},
       adsurl = {https://ui.adsabs.harvard.edu/abs/2018ApJS..235...14S}
}

@ARTICLE{Sparre,
       author = {{Sparre}, Martin and {Hayward}, Christopher C. and {Feldmann}, Robert and {Faucher-Gigu{\`e}re}, Claude-Andr{\'e} and {Muratov}, Alexander L. and {Kere{\v{s}}}, Du{\v{s}}an and {Hopkins}, Philip F.},
        title = "{(Star)bursts of FIRE: observational signatures of bursty star formation in galaxies}",
      journal = {\mnras},
         year = 2017,
        month = apr,
       volume = {466},
       number = {1},
        pages = {88-104},
          doi = {10.1093/mnras/stw3011},
archivePrefix = {arXiv},
       eprint = {1510.03869},
 primaryClass = {astro-ph.GA},
       adsurl = {https://ui.adsabs.harvard.edu/abs/2017MNRAS.466...88S}
}

@ARTICLE{sidm,
       author = {{Spergel}, David N. and {Steinhardt}, Paul J.},
        title = "{Observational Evidence for Self-Interacting Cold Dark Matter}",
      journal = {\prl},
         year = 2000,
        month = apr,
       volume = {84},
       number = {17},
        pages = {3760-3763},
          doi = {10.1103/PhysRevLett.84.3760},
archivePrefix = {arXiv},
       eprint = {astro-ph/9909386},
 primaryClass = {astro-ph},
       adsurl = {https://ui.adsabs.harvard.edu/abs/2000PhRvL..84.3760S}
}

@INPROCEEDINGS{AO,
       author = {{Str{\"o}bele}, S. and {La Penna}, P. and {Arsenault}, R. and {Conzelmann}, R.~D. and {Delabre}, B. and {Duchateau}, M. and {Dorn}, R. and {Fedrigo}, E. and {Hubin}, N. and {Quentin}, J. and {Jolley}, P. and {Kiekebusch}, M. and {Kirchbauer}, J.~P. and {Klein}, B. and {Kolb}, J. and {Kuntschner}, H. and {Le Louarn}, M. and {Lizon}, J.~L. and {Madec}, P.-Y. and {Pettazzi}, L. and {Soenke}, C. and {Tordo}, S. and {Vernet}, J. and {Muradore}, R.},
        title = "{GALACSI system design and analysis}",
    booktitle = {Adaptive Optics Systems III},
         year = 2012,
       editor = {{Ellerbroek}, Brent L. and {Marchetti}, Enrico and {V{\'e}ran}, Jean-Pierre},
       series = {Society of Photo-Optical Instrumentation Engineers (SPIE) Conference Series},
       volume = {8447},
        month = jul,
          eid = {844737},
        pages = {844737},
          doi = {10.1117/12.926110},
       adsurl = {https://ui.adsabs.harvard.edu/abs/2012SPIE.8447E..37S}
}

@ARTICLE{2020MNRAS.497..698T,
       author = {{Tacchella}, Sandro and {Forbes}, John C. and {Caplar}, Neven},
        title = "{Stochastic modelling of star-formation histories II: star-formation variability from molecular clouds and gas inflow}",
      journal = {\mnras},
         year = 2020,
        month = sep,
       volume = {497},
       number = {1},
        pages = {698-725},
          doi = {10.1093/mnras/staa1838},
archivePrefix = {arXiv},
       eprint = {2006.09382},
 primaryClass = {astro-ph.GA},
       adsurl = {https://ui.adsabs.harvard.edu/abs/2020MNRAS.497..698T}
}

@ARTICLE{tey,
       author = {{Teyssier}, Romain and {Pontzen}, Andrew and {Dubois}, Yohan and {Read}, Justin I.},
        title = "{Cusp-core transformations in dwarf galaxies: observational predictions}",
      journal = {\mnras},
         year = 2013,
        month = mar,
       volume = {429},
       number = {4},
        pages = {3068-3078},
          doi = {10.1093/mnras/sts563},
archivePrefix = {arXiv},
       eprint = {1206.4895},
 primaryClass = {astro-ph.CO},
       adsurl = {https://ui.adsabs.harvard.edu/abs/2013MNRAS.429.3068T}
}

@ARTICLE{toll,
       author = {{Tollet}, Edouard and {Macci{\`o}}, Andrea V. and {Dutton}, Aaron A. and {Stinson}, Greg S. and {Wang}, Liang and {Penzo}, Camilla and {Gutcke}, Thales A. and {Buck}, Tobias and {Kang}, Xi and {Brook}, Chris and {Di Cintio}, Arianna and {Keller}, Ben W. and {Wadsley}, James},
        title = "{NIHAO - IV: core creation and destruction in dark matter density profiles across cosmic time}",
      journal = {\mnras},
         year = 2016,
        month = mar,
       volume = {456},
       number = {4},
        pages = {3542-3552},
          doi = {10.1093/mnras/stv2856},
archivePrefix = {arXiv},
       eprint = {1507.03590},
 primaryClass = {astro-ph.GA},
       adsurl = {https://ui.adsabs.harvard.edu/abs/2016MNRAS.456.3542T}
}

@ARTICLE{1977A&A....54..661T,
       author = {{Tully}, R.~B. and {Fisher}, J.~R.},
        title = "{A new method of determining distances to galaxies.}",
      journal = {\aap},
         year = 1977,
        month = feb,
       volume = {54},
        pages = {661-673},
       adsurl = {https://ui.adsabs.harvard.edu/abs/1977A&A....54..661T}
}

@ARTICLE{vdh,
       author = {{van de Hulst}, H.~C. and {Raimond}, E. and {van Woerden}, H.},
        title = "{Rotation and density distribution of the Andromeda nebula derived from observations of the 21-cm line}",
      journal = {\bain},
         year = 1957,
        month = nov,
       volume = {14},
        pages = {1},
       adsurl = {https://ui.adsabs.harvard.edu/abs/1957BAN....14....1V}
}

@ARTICLE{numpy,
       author = {{Van Der Walt}, St{\'e}fan and {Colbert}, S. Chris and {Varoquaux}, Ga{\"e}l},
        title = "{The NumPy Array: A Structure for Efficient Numerical Computation}",
      journal = {Computing in Science and Engineering},
         year = 2011,
        month = mar,
       volume = {13},
       number = {2},
        pages = {22-30},
          doi = {10.1109/MCSE.2011.37},
archivePrefix = {arXiv},
       eprint = {1102.1523},
 primaryClass = {cs.MS},
       adsurl = {https://ui.adsabs.harvard.edu/abs/2011CSE....13b..22V}
}

@ARTICLE{morph,
       author = {{van der Wel}, A. and {Bell}, E.~F. and {H{\"a}ussler}, B. and {McGrath}, E.~J. and {Chang}, Yu-Yen and {Guo}, Yicheng and {McIntosh}, D.~H. and {Rix}, H.-W. and {Barden}, M. and {Cheung}, E. and {Faber}, S.~M. and {Ferguson}, H.~C. and {Galametz}, A. and {Grogin}, N.~A. and {Hartley}, W. and {Kartaltepe}, J.~S. and {Kocevski}, D.~D. and {Koekemoer}, A.~M. and {Lotz}, J. and {Mozena}, M. and {Peth}, M.~A. and {Peng}, Chien Y.},
        title = "{Structural Parameters of Galaxies in CANDELS}",
      journal = {\apjs},
         year = 2012,
        month = dec,
       volume = {203},
       number = {2},
          eid = {24},
        pages = {24},
          doi = {10.1088/0067-0049/203/2/24},
archivePrefix = {arXiv},
       eprint = {1211.6954},
 primaryClass = {astro-ph.CO},
       adsurl = {https://ui.adsabs.harvard.edu/abs/2012ApJS..203...24V}
}

@ARTICLE{Wan,
       author = {{Wan}, Jenny T. and {Tacchella}, Sandro and {Johnson}, Benjamin D. and {Iyer}, Kartheik G. and {Speagle}, Joshua S. and {Maiolino}, Roberto},
        title = "{Stochastic prior for non-parametric star-formation histories}",
      journal = {\mnras},
         year = 2024,
        month = aug,
       volume = {532},
       number = {4},
        pages = {4002-4025},
          doi = {10.1093/mnras/stae1734},
archivePrefix = {arXiv},
       eprint = {2404.14494},
 primaryClass = {astro-ph.GA},
       adsurl = {https://ui.adsabs.harvard.edu/abs/2024MNRAS.532.4002W}
}

@ARTICLE{Weisz,
       author = {{Weisz}, Daniel R. and {Johnson}, Benjamin D. and {Johnson}, L. Clifton and {Skillman}, Evan D. and {Lee}, Janice C. and {Kennicutt}, Robert C. and {Calzetti}, Daniela and {van Zee}, Liese and {Bothwell}, Matthew S. and {Dalcanton}, Julianne J. and {Dale}, Daniel A. and {Williams}, Benjamin F.},
        title = "{Modeling the Effects of Star Formation Histories on H{\ensuremath{\alpha}} and Ultraviolet Fluxes in nearby Dwarf Galaxies}",
      journal = {\apj},
         year = 2012,
        month = jan,
       volume = {744},
       number = {1},
          eid = {44},
        pages = {44},
          doi = {10.1088/0004-637X/744/1/44},
archivePrefix = {arXiv},
       eprint = {1109.2905},
 primaryClass = {astro-ph.CO},
       adsurl = {https://ui.adsabs.harvard.edu/abs/2012ApJ...744...44W}
}

@article{Weldrake,
    author = {Weldrake, D. T. F. and de Blok, W. J. G. and Walter, F.},
    title = {{The dark matter halo of NGC 6822}},
    year = {2003},
    journal = {MNRAS},
    volume = {340},
    pages = {12-26},
    doi = {10.1046/j.1365-8711.2003.06246.x},
}

@ARTICLE{zhao,
       author = {{Zhao}, Hongsheng},
        title = "{Analytical models for galactic nuclei}",
      journal = {\mnras},
         year = 1996,
        month = jan,
       volume = {278},
       number = {2},
        pages = {488-496},
          doi = {10.1093/mnras/278.2.488},
archivePrefix = {arXiv},
       eprint = {astro-ph/9509122},
 primaryClass = {astro-ph},
       adsurl = {https://ui.adsabs.harvard.edu/abs/1996MNRAS.278..488Z}
}

\clearpage 

\begin{appendix}

\section{Sample of intermediate-z SFGs}
\label{app:snr}

The data selection followed \citet{ciocan}, to which we refer for full details. Briefly, we selected emission-line galaxies at $z<1.5$ from the main MUSE catalogues with sufficient quality for reliable kinematic modelling. We required a signal-to-noise in the brightest optical emission line ([O\,\textsc{ii}], H$\beta$, [O\,\textsc{iii}], or H$\alpha$) from the integrated spectrum of S/N$>15$. Furthermore, we imposed an effective signal-to-noise $\rm{S/N_{eff}} > 10$, where $\rm{S/N_{eff}}$ accounts for both line strength and spatial resolution as: $\rm{S/N_{eff}}= (R_{\rm{e}}/R_{\rm{PSF}})\cdot \rm{S/N_{max}}$. Here,  $R_{\rm{e}}$ is the stellar half-light radius, $R_{\rm{PSF}}=\rm{FWHM}/2$ is the radius of the MUSE PSF, and $\rm{S/N_{max}}$ is the S/N of the emission line in the brightest spaxel of the MUSE cube. We further imposed $R_{\rm e}/R_{\rm PSF} > 0.5$ to avoid unresolved systems. 

\begin{figure*}
\sidecaption
    \subfloat{{\includegraphics[width=12cm]{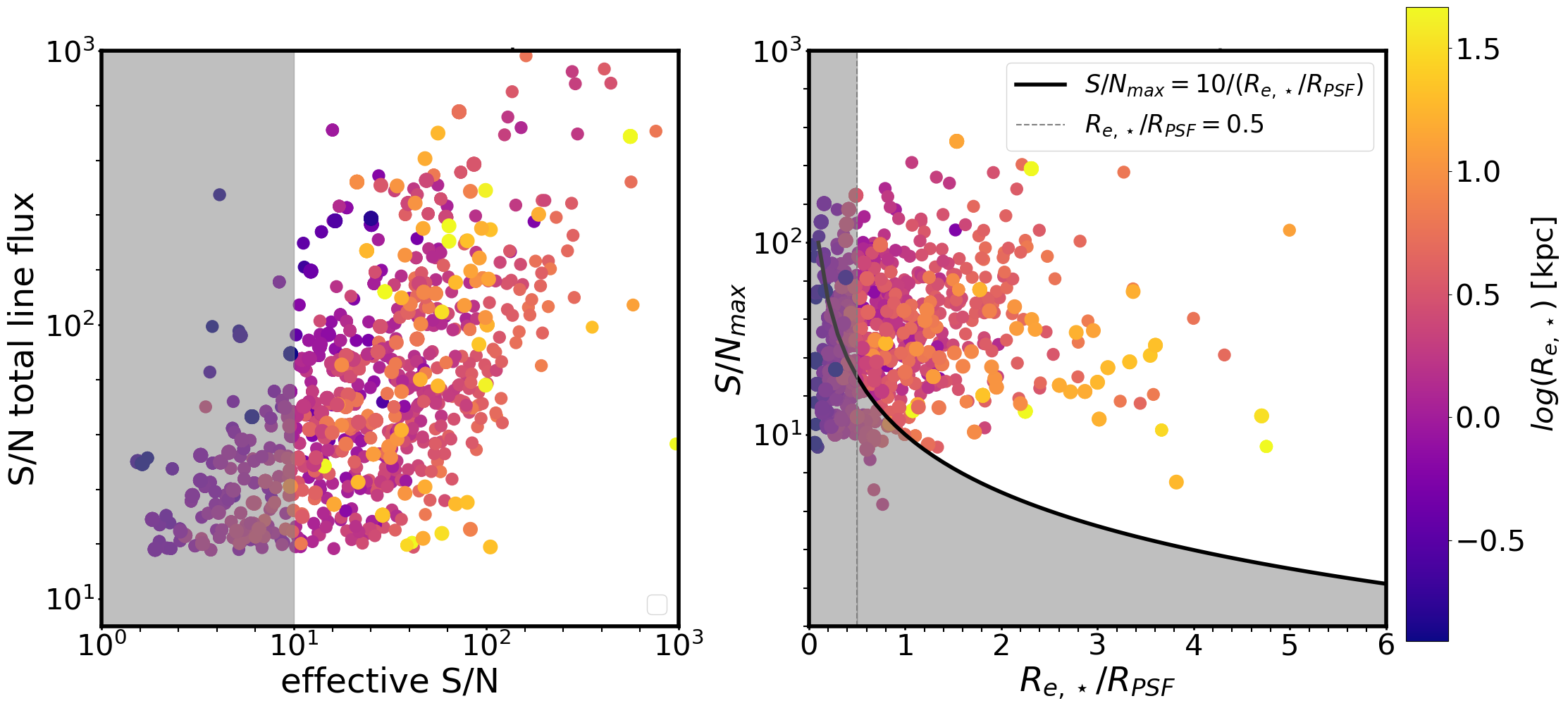} }}%
        \caption{\normalsize Left: Effective S/N as a function of the total line-flux S/N measured from the integrated spectrum. Right: Ratio between the stellar half-light radius and the PSF radius as a function of the S/N in the brightest spaxel of the MUSE cube. The grey shaded region indicates the selection cut, excluding galaxies with $\rm{S/N_{eff}} < 10$ and $R_{\rm e}/R_{\rm PSF} < 0.5$. Points are colour-coded by their half-light radius. }
\label{snr}
\end{figure*}

Figure \ref{snr} shows on the left the $\rm{S/N_{eff}}$ as a function of the $\rm{S/N}$ of the total line flux for the line emitters and on the right $R_{\rm{e}}/R_{\rm{PSF}}$ as a function of the S/N in the brightest spaxel. All galaxies falling in the grey-shaded region were excluded from the dynamical modelling.

Additionally, we restricted the sample to have inclinations $i > 30^\circ$, and excluded visually identified disturbed, interacting or merging systems (using HST or JWST imaging).  Finally, we retained only rotation-dominated systems with $v_{\rm max}/\sigma > 1$, where $\sigma$ is measured at $2R_{\rm e}$ (see Sect. ~\ref{gpk1}).

\begin{figure*}
   \centering
    \includegraphics[width=0.9\textwidth,angle=0,clip=true]{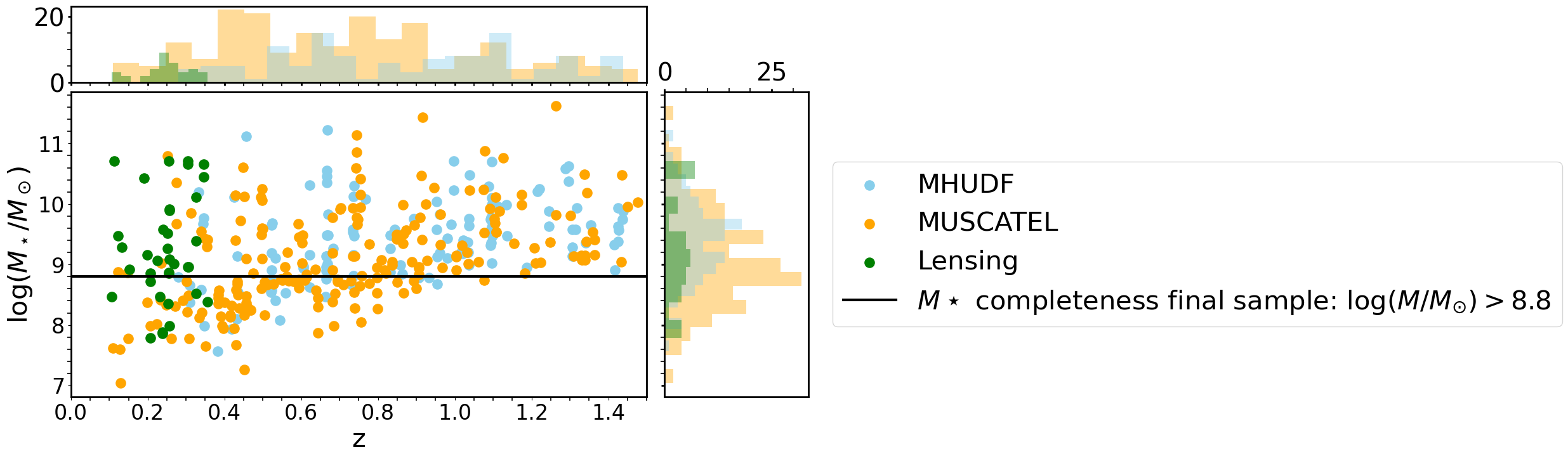}
   \centering   
\caption{Stellar mass as a function of redshift for the sample. Blue, orange, and green points and histograms correspond to galaxies from the MHUDF, MUSCATEL, and lensing cluster subsamples, respectively. The black line denotes completeness limit of
the sample, which is reasonably complete for $0.1 <z <1.43$ down to $\log(M_{\star}/M_{\odot})= 8.8$. }
    \label{mstar_z}%
\end{figure*}

The final sample analysed in this work comprises 375 SFGs meeting all the aforementioned selection criteria. Figure \ref{mstar_z} shows the stellar masses (from SED-fitting, see Sect.~\ref{SED}) as a function of the spectroscopic redshifts for our sample. The galaxies span $0.1 < z < 1.43$ and $7 < \log(M_{\star}/M_{\odot}) < 11.2$. The black line marks the adopted stellar-mass selection threshold used in this work. Within the redshift range probed, this corresponds to a constant lower limit of $\log(M_{\star}/M_{\odot}) = 8.8$, above which the sample is treated as complete.

\section{ \textsc{GalPaK$^{\rm 3D}$}: model parameters and priors}
\label{par}
While the URC model  described in Sect.~\ref{gpk1}  contains 12-13 free parameters, the disk–halo decomposition models, described in Sect.~ \ref{gpk2} comprise 14-15 parameters,
depending on the adopted DM halo prescription (DC14, DC14-scatter, gNFW). Table  \ref{gpkparams} lists the parameters used in our modelling together with the adopted priors.

\begin{table*}[ht]
 \centering
 \caption{Parameters used in the {\textsc{GalPaK$^{\rm 3D}$}} modelling, as described in Sects. \ref{gpk1} and \ref{gpk2}.}
    \begin{tabular}{|l|cccc|c|}
        \hline
        Parameters common to all models & URC & DC14 & DC14-scatter & gNFW & Priors \\
        \hline
        $x$, $y$, $z$ & \ding{51} & \ding{51} &  \ding{51} &\ding{51} & - \\
        Total Line Flux & \ding{51} & \ding{51}& \ding{51} & \ding{51} & -  \\
        Inclination ($i$) & \ding{51} & \ding{51} & \ding{51}& \ding{51} & $\mathcal{U}[i_{\rm HST}-5,i_{\rm HST}+5]$  \\
        Sérsic Index ($n_{\rm{gas}})$ & \ding{51} & \ding{51}& \ding{51}  & \ding{51} &$\mathcal{U}[0.5,5]$ \\
        Major-Axis PA & \ding{51} & \ding{51}& \ding{51} & \ding{51} &$\mathcal{U}[ -40,400]$  \\
        Effective Radius ($R_{\rm{e}}$) & \ding{51} & \ding{51} & \ding{51}& \ding{51} &  $\mathcal{U}[0.3,30]$\\
        Turnover radius ($r_{\rm{t}})$ & \ding{51} & \ding{55} & \ding{55}& \ding{55}   &$\mathcal{U}[0.01,3]$\\
        Maximum Velocity ($v_{\rm{max}}$) & \ding{51} & \ding{55} & \ding{55}& \ding{55}  &$\mathcal{U}[10,350$]   \\
        Virial Velocity ($v_{\rm{vir}}$) & \ding{55} & \ding{51} & \ding{51}& \ding{51}  &$\mathcal{U}[10,350] *$  \\
        Concentration ($c_{\rm{vir}}$) & \ding{55} & \ding{51}& \ding{51} & \ding{51} &  $\mathcal{U}[5,25]$\\
        Velocity Dispersion ($\sigma_0$) & \ding{51}& \ding{51} & \ding{51} & \ding{51} & $\mathcal{U}[0,200]$   \\
        HI density & \ding{55} & \ding{51} & \ding{51} & \ding{51}& $\mathcal{U}[0,15]$  \\
      
        Doublet Ratio ($r_{\rm{O2}})$ & \ding{51}/\ding{55} & \ding{51}/\ding{55}& \ding{51}/\ding{55} &\ding{51}/\ding{55}& $\mathcal{U}[0.35,1.5]$   \\
        $X$ & \ding{55} & \ding{51} & \ding{51}& \ding{55}& $\mathcal{U}[-3,-1.2]$   \\
        $\delta_{s}$ (DC14 scatter)  & \ding{55} & \ding{55} & \ding{51}& \ding{55} & $\mathcal{U}[-0.3,0.3]$  \\
        Stellar Mass ($M_{\star}$) & \ding{55}& \ding{55} & \ding{55} & \ding{51}& $\mathcal{U}[ \log(M_\star/M_\odot)_{\rm SED} -0.15,  \log(M_\star/M_\odot)_{\rm SED} + 0.15]$  \\
        $\gamma$ (gNFW) & \ding{55}& \ding{55} & \ding{55} & \ding{51}& $\mathcal{U}[0,1.5]$ \\
        \hline
    \end{tabular}
   \tablefoot{Checkmarks (\ding{51}) indicate parameters used for a given model;  crosses (\ding{55}) indicate parameters not used for a given model. Uniform priors are denoted as $\mathcal{U}[a,b]$.  *We note that for the gNFW we use $\Lambda$CDM priors, namely we impose the $M_{\star}-M_{\rm vir}$ relation from \cite{cosmos2020}.}

       \label{gpkparams}
\end{table*}

\section{\texttt{BAGPIPES}: model parameters and priors}
\label{appendix:sed}
Table  \ref{SEDpara} presents the parameters used in the SED modelling with \texttt{BAGPIPES} together with the adopted priors (Sect. \ref{SED}).

\begin{figure*}
   \centering
    \includegraphics[width=0.49\textwidth,angle=0,clip=true]{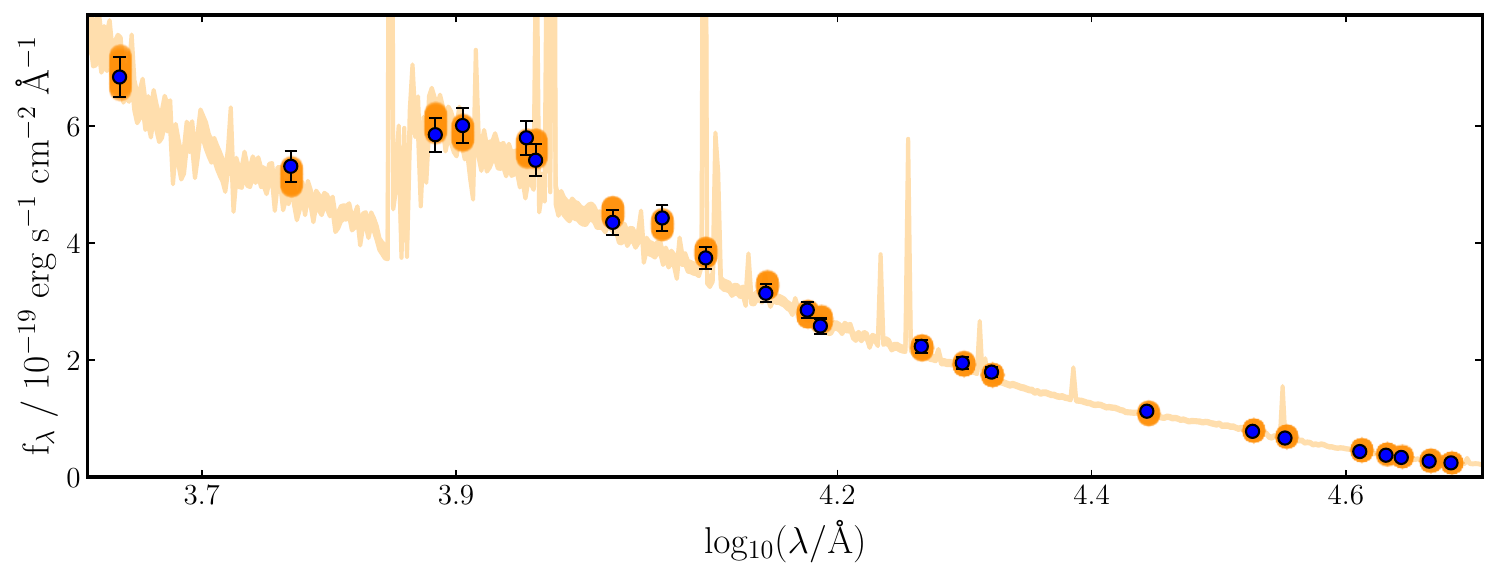}
        \includegraphics[width=0.49\textwidth,angle=0,clip=true]{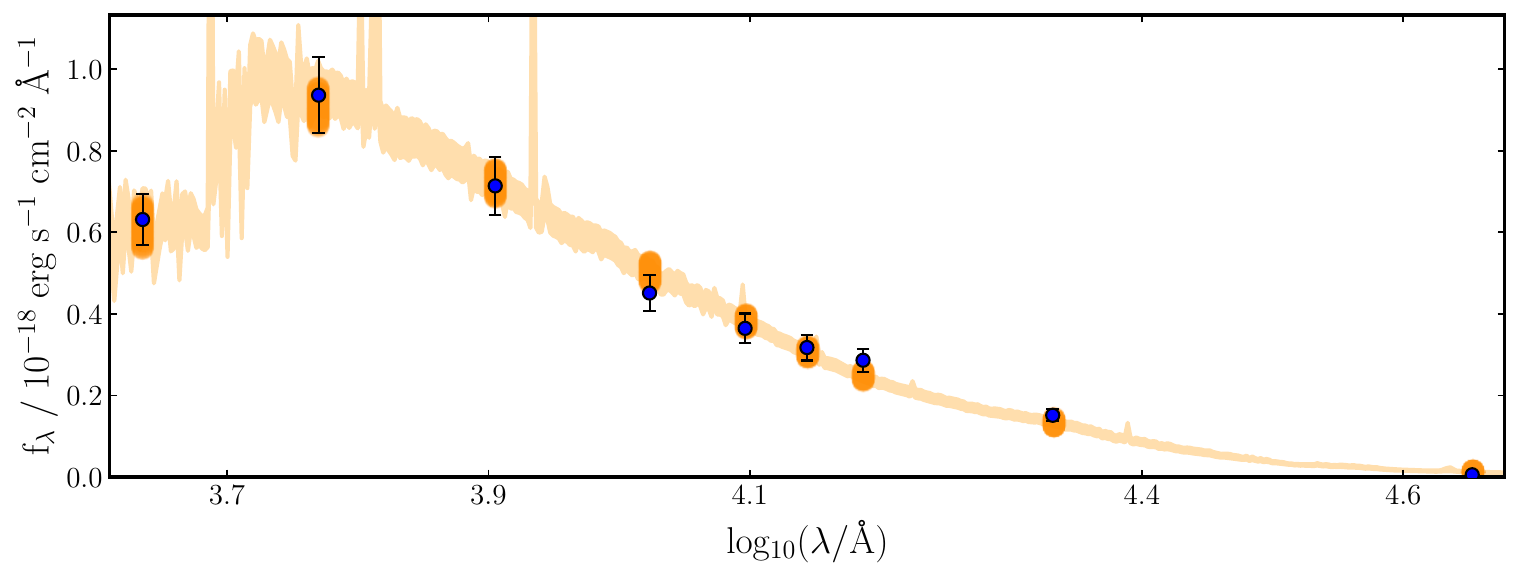}

    \centering
        \includegraphics[width=0.49\textwidth,angle=0,clip=true]{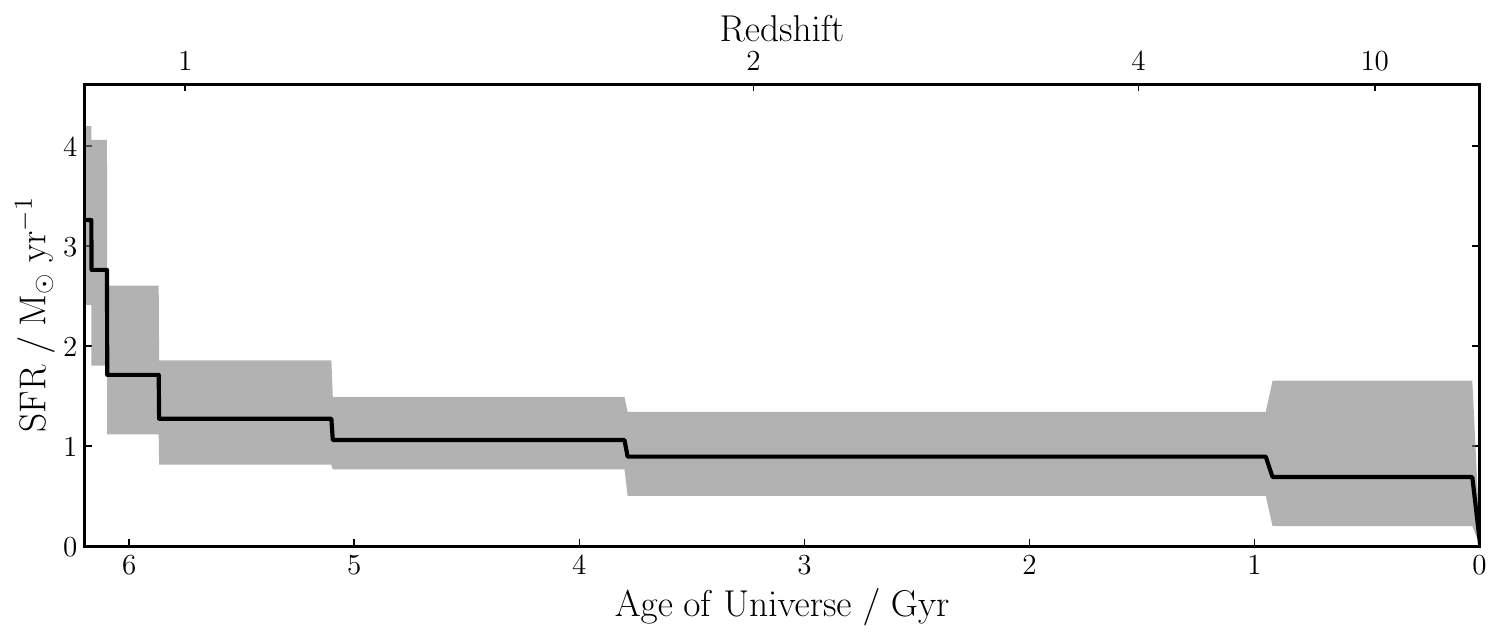}
            \includegraphics[width=0.49\textwidth,angle=0,clip=true]{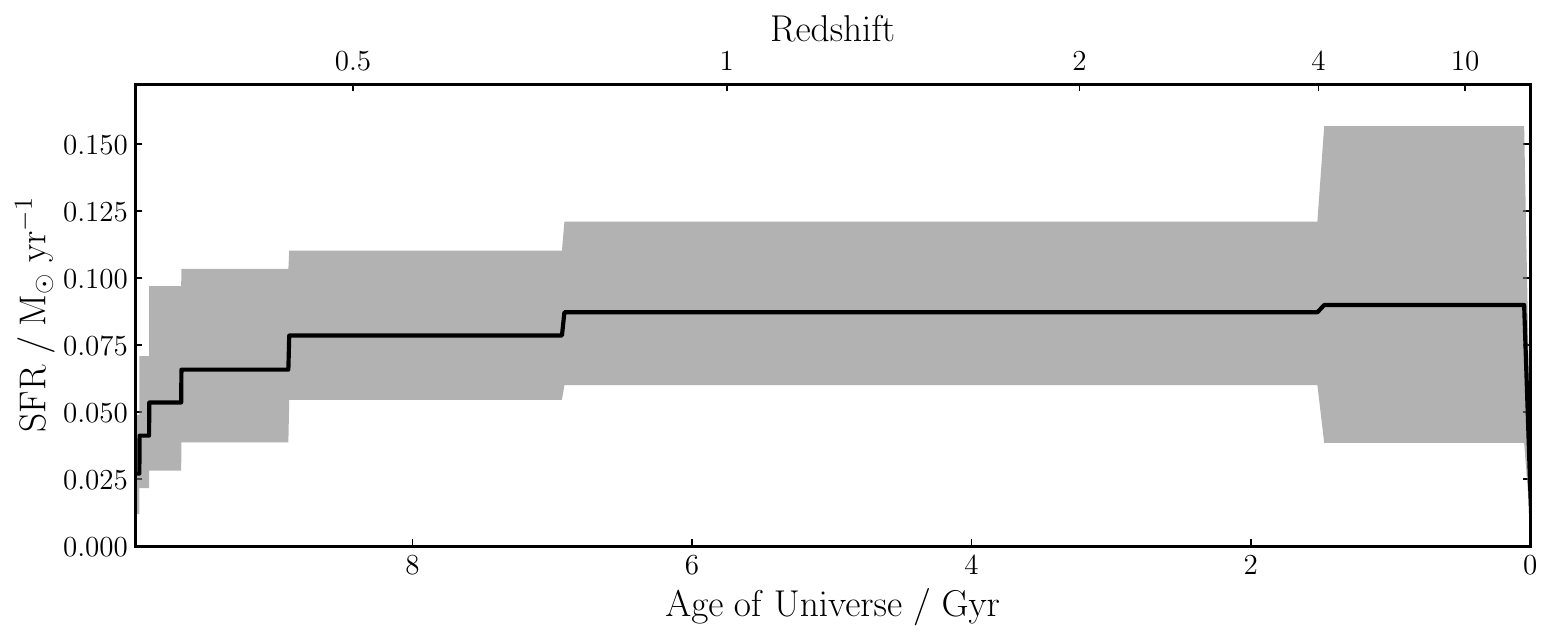}

   \centering   
\caption{Top: Example of \texttt{BAGPIPES} fits using the JADES photometric catalogue \citep{Eisenstein} for a $z\sim0.9$ galaxy from MHUDF (left), and the HFF-DeepSpace catalogue \citep{ship}  for a $z\sim0.3$ galaxy from MUSCATEL  (right) with the non-parametric continuity SFH model from \cite{Leja19}. The photometry is shown in blue, while the 16th to 84th percentile range for the posterior spectrum and photometry is shown in shaded orange. Bottom (left-right): Reconstructed SFH, i.e. SFR as a function of time for both galaxies.}
\label{SED_fit}
\end{figure*}

Figure \ref{SED_fit} shows examples of the SED fits for two galaxies from our sample  (top), together with the recovered SFH (bottom).

\begin{table*}
\centering
\caption{Parameters used in the \texttt{BAGPIPES} modelling, as described in Sect.~\ref{SED}.}
\label{SEDpara}
\begin{tabular}{|l|c|c|}
\hline
Parameter & Prior / Value \\
\hline
$\log_{10}(M_{\rm formed}/M_\odot)$ & $\mathcal{U}[0,13]$ \\
$Z/Z_\odot$ & log-uniform $\mathcal{U}[0.01,5]$ \\
$A_V$ & $\mathcal{U}[0,3]$ \\
Dust slope $n$ & $\mathcal{G}(0.7,0.3)$, trunc. to $[0.3,2.5]$ \\
$\eta$ (CF00 birth cloud factor) & 2 \\
$t_{\rm BC}$ [Myr] & 10 \\
$\log(\rm{SFR}_{n}/\rm{SFR}_{n+1})$ & $t_v(0.3,2)$\\
$\log U$ & $\mathcal{U}[-4,-2]$ \\
 $z$ & fixed ($z_{\rm spec}$) \\
\hline
\end{tabular}
\tablefoot{Uniform priors are denoted as $\mathcal{U}[a,b]$ and Gaussian priors as $\mathcal{G}(\mu,\sigma)$, while  $t_{v}(\sigma,\nu)$ represents a Student-t
distribution with scale parameter $\sigma$ and $\nu$ degrees of freedom.}
\label{SEDpara}
\end{table*}

\section{Disk-halo decomposition}
\label{appendix:decomp}
\begin{figure}
   \centering
    \includegraphics[width=0.5\textwidth,angle=0,clip=true]{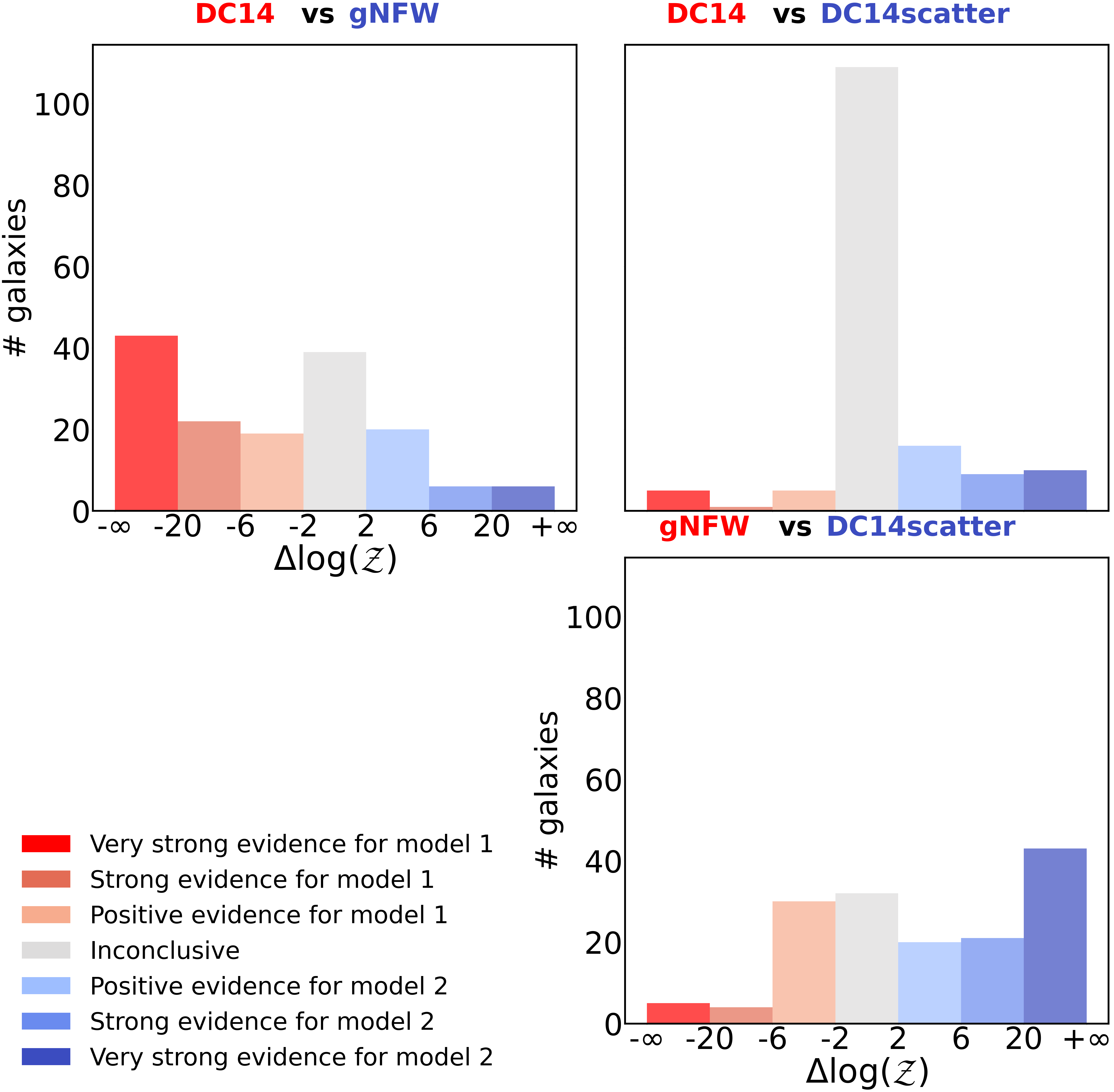}
   \centering   
\caption{Same as Fig. \ref{bayes} but for the highest-S/N  sample ($\rm{S/N_{eff}}>50$,167 SFGs). Red bins indicate galaxies with strong to very strong evidence in favour of the first model, while blue bins correspond to strong to very strong evidence in favour of the second model. Intermediate red and blue bins represent moderate evidence for either model. Grey bins show systems where both models perform comparably. In each panel, model 1 and model 2 refer to the first and second models listed in the title, respectively.}
\label{bayes2}
\end{figure}

Figure \ref{bayes2} shows the same as Fig.~ \ref{bayes}, but for the highest-S/N  sample ($\rm{S/N_{eff}}>50$; 167 SFGs).  This Fig. indicates that DC14 (53\% positive evidence) and DC14-scatter (51\% positive evidence) are both the most likely models given the data when compared to gNFW (15\% and 21\% positive evidence, respectively), with the remaining galaxies showing no statistically significant distinction between the models. We further note that the fraction of galaxies for which the models are indistinguishable decreases in the high-S/N subsample. Overall, the Bayes factor distributions indicate that DC14 and DC14-scatter yield broadly consistent levels of evidence and are generally favoured over gNFW in the highest-S/N regime. We note that we repeated the Bayesian analysis using the same set of priors for all halo models, namely priors on inclination only for the DC14(-scatter) and gNFW fits, with no priors on $v_{\rm vir}$ in the gNFW model. The results remain unchanged: the DC14-based models are still favoured over gNFW given the data. We find, however, that without the $M_\star$ and $v_{\rm vir}$ priors, the gNFW fits yield unphysical solutions for a substantial fraction of the sample. This motivates the inclusion of these priors in our fiducial gNFW fits shown in the main text.

\begin{figure*}
\centering
\includegraphics[width=9cm]{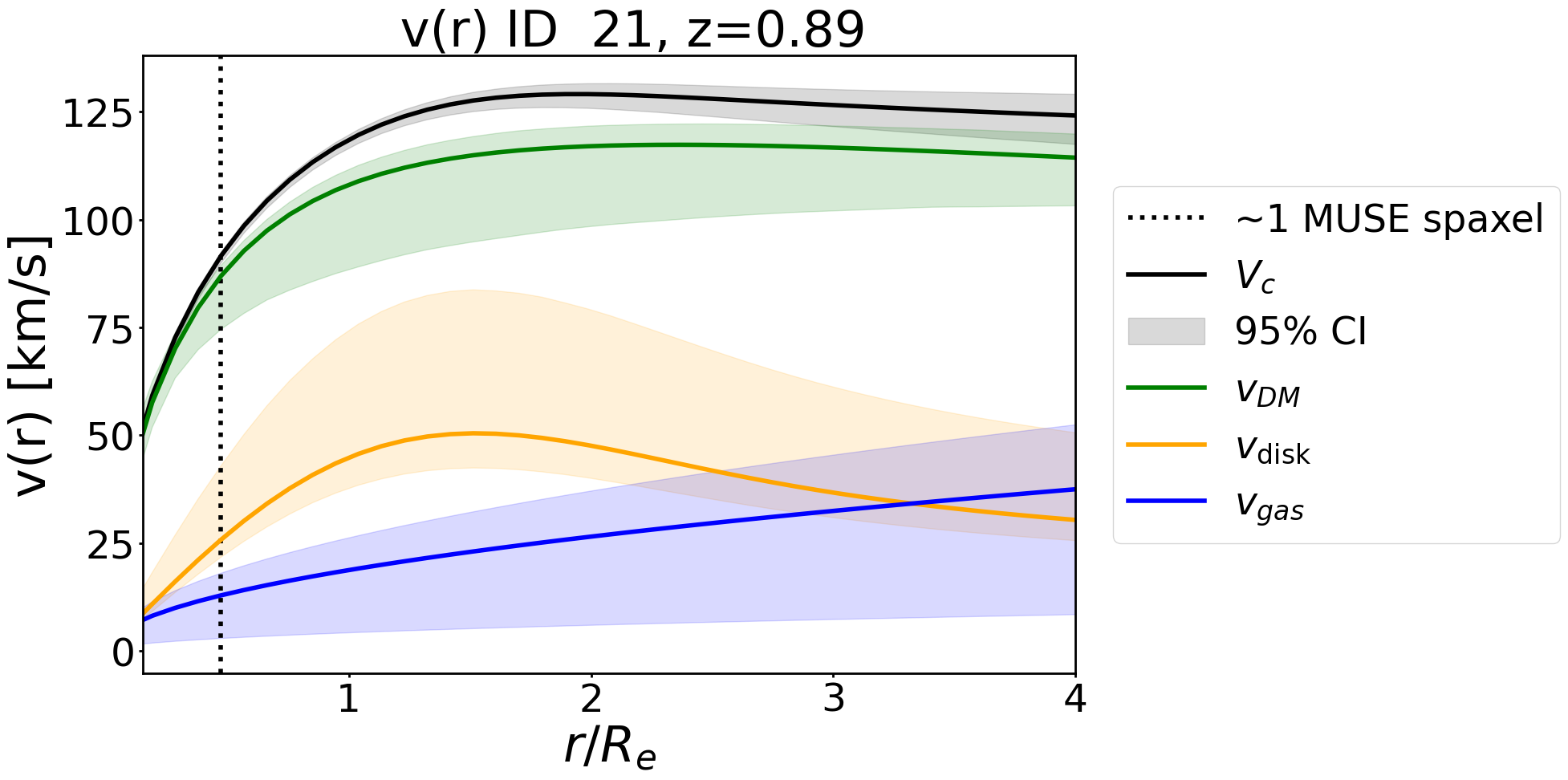} 
\centering%
\includegraphics[width=5cm]{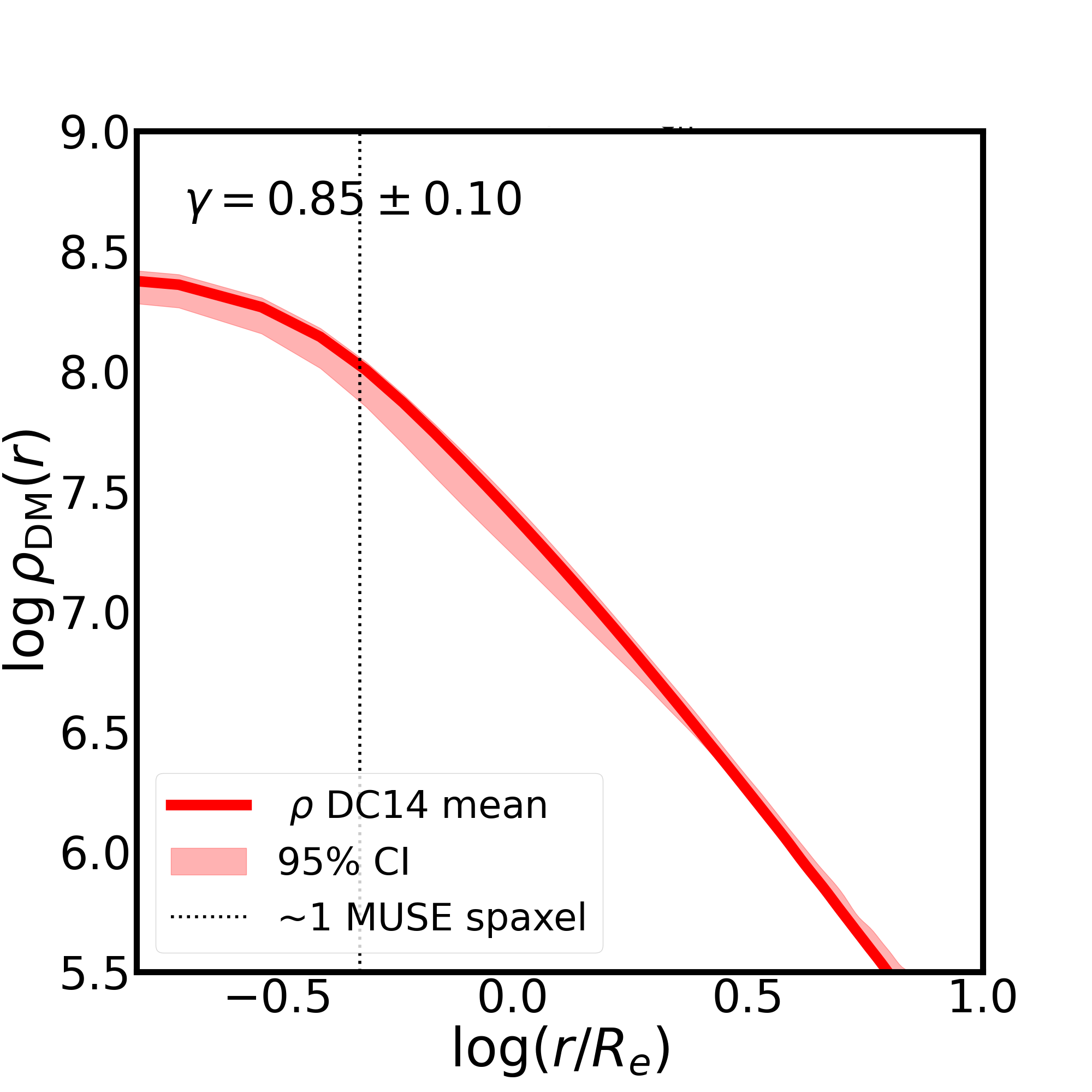} 
\centering
\includegraphics[width=9cm]{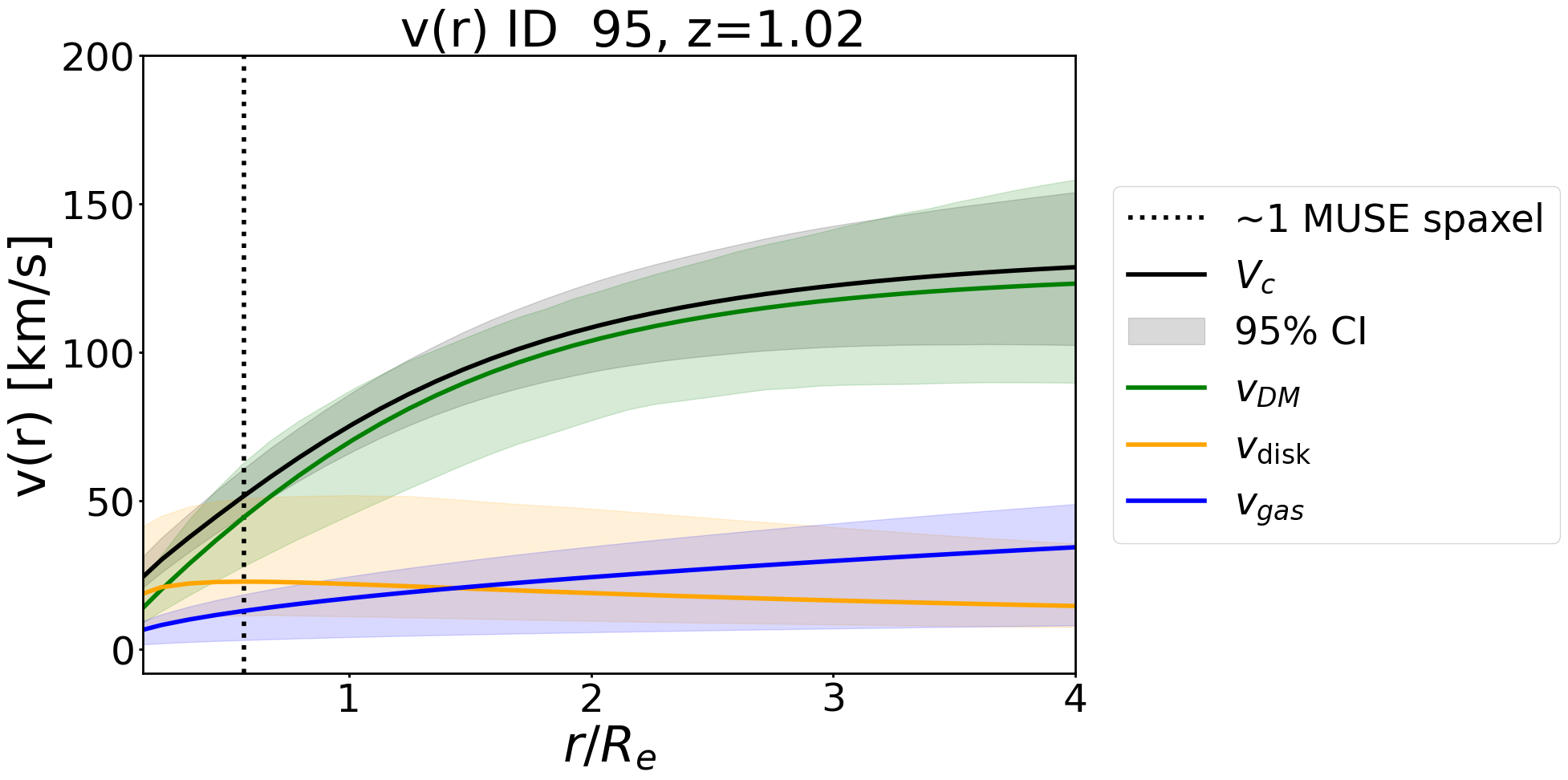} 
\centering%
\includegraphics[width=5cm]{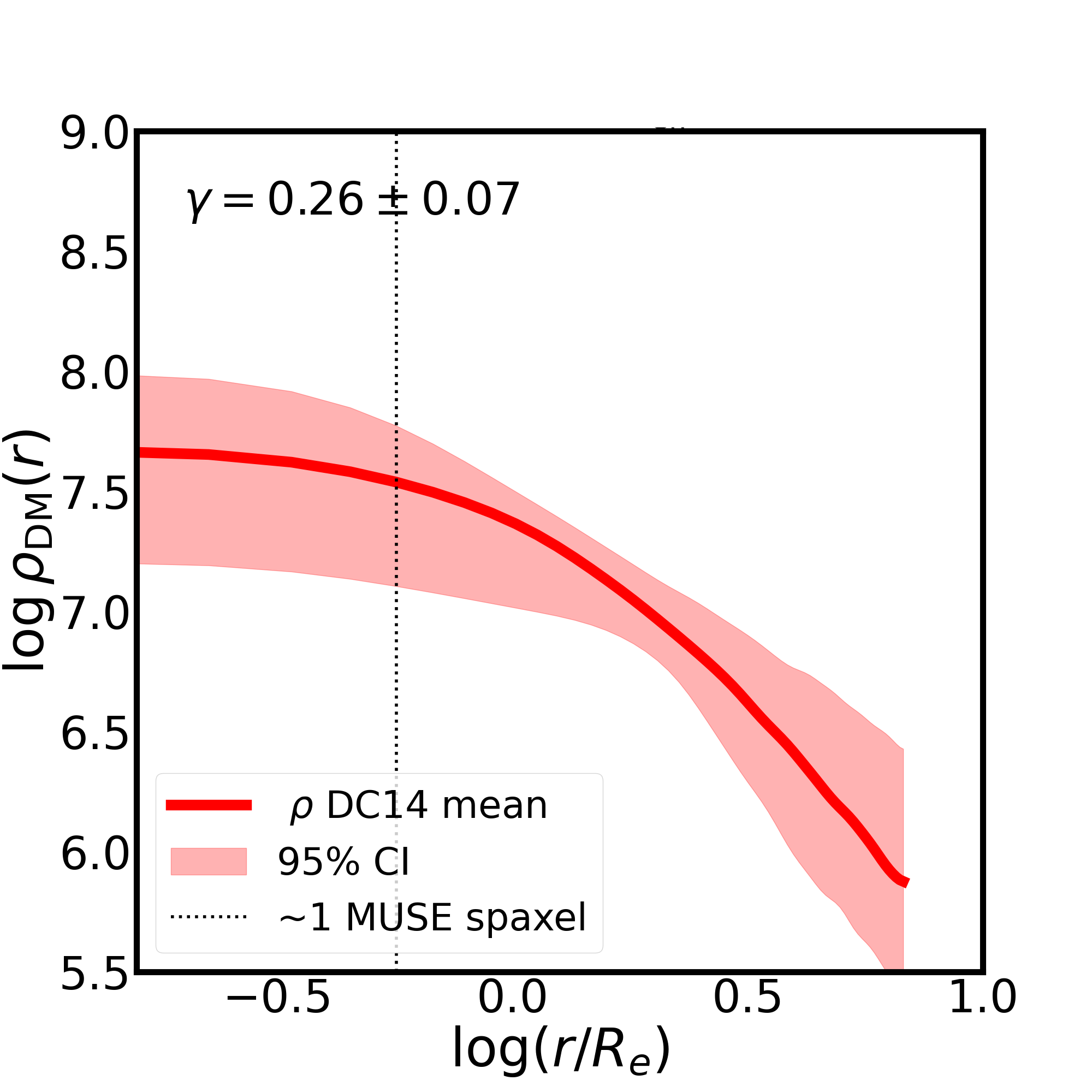} 
 \caption{\normalsize Left (top and bottom): Examples of the 3D disk--halo decomposition using the DC14 density profile. The solid black line shows the circular velocity $v_{\rm c}(r)$, i.e. $v_{\perp}(r)$ corrected for pressure support. The green curve corresponds to the DM component, while the orange and blue curves represent the disk and \HI\ components, respectively.  All velocities are corrected for inclination and instrumental effects, including seeing. 
        Right (top and bottom): Corresponding DM density profiles for the same galaxies. The y-axis shows $\rho_{\rm DM}(r)$ in units of $\rm M_{\odot}\,kpc^{-3}$, and the x-axis is $r/R_e$. The inferred inner slopes, $\gamma$, are reported in the upper-left corner of each panel.  In all panels, the shaded regions indicate the 95\% confidence intervals. Vertical dashed black lines indicate the 0.2'' MUSE-spaxel scale.}
\label{decomp}
\end{figure*}

Figure~\ref{decomp} shows the RC decomposition (left) and the corresponding density profiles (right) obtained using the DC14 model, for two galaxies from our sample. All velocities shown are intrinsic, having been corrected for inclination and other instrumental effects, including seeing. This Fig. demonstrates that the RCs are strongly dominated by the DM component. Consequently, the RC data retain sensitivity to the halo shape parameters, $\alpha(X)$, $\beta(X)$, and $\gamma(X)$, which, in the DC14 framework, depend on the ratio $M_{\rm \star}/M_{\rm vir}$. This dependence provides an independent constraint on $M_{\rm \star}/M_{\rm vir}$, thereby helping to break the traditional disk--halo degeneracy. 

 \section{Star Formation Histories }
 \label{sfh}
 
 In this Appendix, we investigate whether our SFH-diagnostics (defined in Sect. \ref{SFHdiagn} and \ref{SFHdiagn2}) correlate with $M_\star$ and $z$. 

We find that $n_{\rm SFR}$ increases systematically towards higher redshift and lower stellar mass, indicating that galaxies at earlier cosmic times and with lower masses form stars at rates increasingly elevated relative to their past-averaged activity. The redshift dependence is the stronger of the two, with a statistically significant Spearman rank correlation of $\rho = 0.51$ ($p = 1.1 \times 10^{-19}$), while the anticorrelation with $M_\star$ is weaker but still significant ($\rho = -0.24$, $p = 3.9 \times 10^{-5}$). This behaviour is consistent with galaxies being observed preferentially during the rising phase of their SFHs at higher redshift and lower masses \citep{Ciesla}, although it may also reflect the increased SF burstiness. 

We further examine the dependence of the burstiness indicator, $n_{\rm burst}$, on redshift and stellar mass. Similar to  $n_{\rm SFR}$, we find a statistically significant positive correlation between $n_{\rm burst}$ and $z$, with $\rho = 0.45$ ($p = 1.8 \times 10^{-14}$), indicating that galaxies at higher redshift exhibit more frequently rapidly rising or recently enhanced SF relative to their 200 Myr average. In contrast, we find no robust dependence of $n_{\rm burst}$ on stellar mass. Across the full sample, the correlation is consistent with zero, $\rho = 0.01$ ($p = 0.82$), and remains weak and statistically insignificant when restricting to the mass-complete subsample, $\rho = 0.10$ ($p = 0.10$). These results indicate that, within the statistical power of the present dataset, the relative enhancement or suppression of recent SF compared to the 200 Myr average does not systematically depend on $M_\star$. 

This lack of a strong mass dependence is noteworthy in the context of previous work. Simulations and observational studies such as \cite{Weisz, Guo, Sparre} generally report a moderate mass dependence, with lower-mass galaxies exhibiting lower $n_{\rm burst}$ values compared to higher-mass systems. In contrast, some observational analyses (e.g. \citealt{2022MNRAS.511.4464A,  Cole}) report the opposite trend, finding larger fractions of low-$M_\star$ galaxies with higher $n_{\rm burst}$ values. The absence of a significant correlation in our sample therefore suggests that differing methodologies and SFR tracers may play a dominant role in shaping reported trends.

\begin{figure*}
   \centering
    \includegraphics[width=0.5\textwidth,angle=0,clip=true]{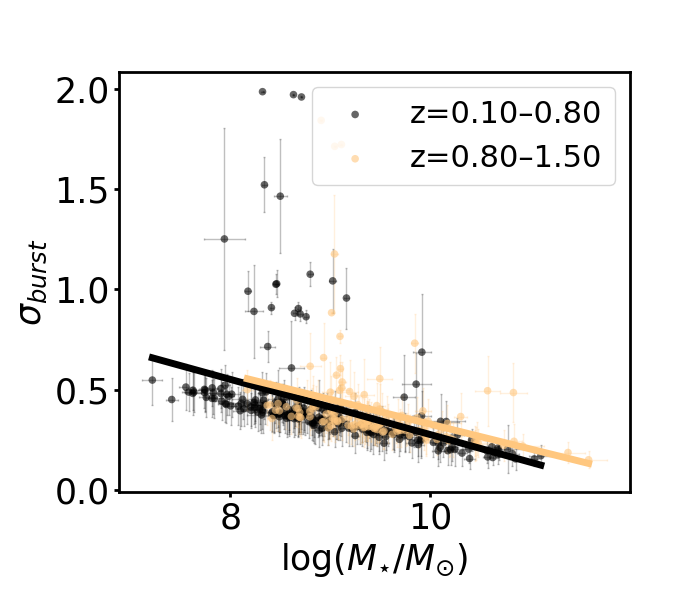}
   \centering   
\caption{$\sigma_{\rm burst}$, defined as the mass-weighted root-mean-square deviation of the SFH from the main sequence (i.e. SFH stochasticity parameter), as a function of stellar mass for the full  sample. The data points are colour-coded by redshift bin, corresponding to the intervals used in the Marginalised Normal Regression analysis (solid lines)  with \texttt{ROXY} \citep{roxy}, with a sequential colour map that traces increasing redshift (from black to orange). The error bars show the $1\sigma$ errors from \texttt{BAGPIPES}.}
\label{sigmafburst}
\end{figure*}

Figure~\ref{sigmafburst} shows the SFH stochasticity parameter, $\sigma_{\rm burst}$, as a function of stellar mass. We find a strong anti-correlation with $M_\star$ in both redshift bins, with Spearman rank coefficients ranging from $\rho=-0.65$ to $-0.85$ ($p=2.16\times10^{-18}$ to $10^{-56}$). In addition, $\sigma_{\rm burst}$ exhibits a weaker but statistically significant positive correlation with redshift ($\rho=0.22$, $p=5.3\times10^{-4}$), indicating that galaxies at earlier cosmic epochs tend to have more stochastic SFHs. Together, these results suggest that bursty SF is most prevalent in low-mass galaxies and becomes increasingly important towards higher redshift. 

Similarly, we find that $f_{\rm burst}$, defined as the fraction of stellar mass formed during starbursts (i.e. a measure of burst-dominated stellar mass assembly)  increases systematically towards higher redshift and lower stellar mass. The redshift dependence is the stronger of the two, with a statistically significant Spearman rank correlation of $\rho=0.24$ ($p=1.4\times10^{-4}$), indicating that galaxies at earlier cosmic epochs form a larger fraction of their stellar mass during elevated SF episodes. We also find a weaker but statistically significant anti-correlation with stellar mass ($\rho=-0.148$, $p=2.27\times10^{-2}$), suggesting that burst-dominated stellar mass assembly is more common in lower-mass galaxies. These trends are qualitatively consistent with previous theoretical and observational studies showing that SF becomes increasingly stochastic toward lower stellar masses and earlier cosmic epochs, as expected from the stronger impact of discrete gas accretion events and feedback cycles in low-mass galaxies (e.g. \citealt{Kauffmann, Sparre, 2020MNRAS.497..698T, Cole}).

\begin{figure*}
   \centering
    \includegraphics[width=0.5\textwidth,angle=0,clip=true]{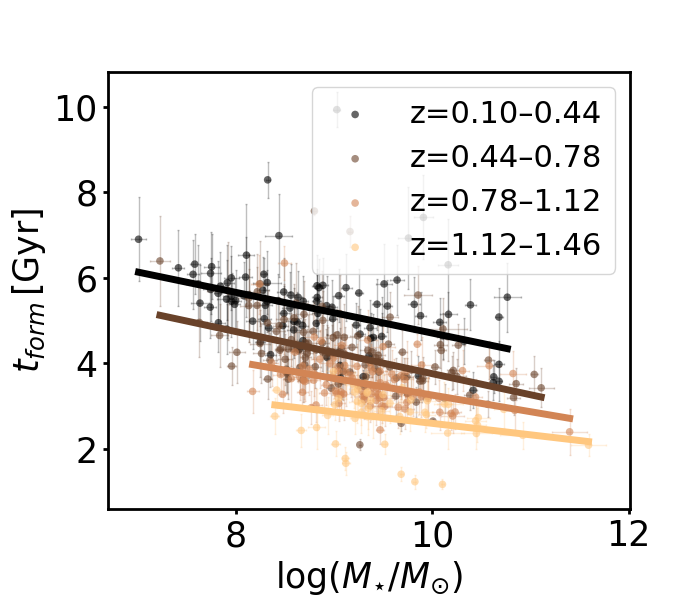}
   \centering   
\caption{$t_{\rm form}$, defined as the mass-weighted average formation epoch of the stellar component (measured from the Big Bang), as a function of stellar mass for the full  sample. The data points are colour-coded by redshift bin, corresponding to the intervals adopted in the Marginalised Normal Regression analysis (solid lines) using \texttt{ROXY} \citep{roxy}, with a sequential colour map tracing increasing redshift (from black to orange). The error bars show the $1\sigma$ errors from \texttt{BAGPIPES}.}
\label{tform}
\end{figure*}

Finally, we show in Figure~\ref{tform}  the mass-weighted average formation epoch of the stellar mass in each galaxy, $t_{\rm form}$, as a function of stellar mass for the entire  sample, with data points colour-coded by redshift bin. We find a clear anticorrelation between $t_{\rm form}$ and stellar mass across all redshift bins: more massive galaxies exhibit systematically earlier mass-assembly epochs than their lower-mass counterparts. This trend is consistently significant in all redshift intervals, with Spearman rank coefficients ranging from $\rho \simeq -0.27$ to $\rho \simeq -0.51$ ($p \lesssim 5 \times 10^{-2}$ to $p \ll 10^{-5}$), and persists over the full redshift range probed by the sample, although its normalisation evolves with redshift. We note that our results remain unchanged when using alternative formation-time indicators, such as $t_{50}$ or $t_{90}$. These observed trends are consistent with the so-called galaxy downsizing scenario, in which the most massive galaxies form their stellar populations earlier (and on shorter timescales), while lower-mass systems continue to assemble their stellar mass at later cosmic times. Our results are in agreement with previous observational studies \citep{Heavens,Pacifici, Kaushal} and with theoretical predictions \citep{Neistein,Behroozi}.

In summary, some of the  SFH diagnostics used throughout this work exhibit systematic dependencies on both stellar mass and redshift. While $n_{\rm SFR}$, $\sigma_{\rm burst}$, and $f_{\rm burst}$ increase towards higher redshift and lower stellar masses, $n_{\rm burst}$ primarily evolves with $z$ and shows no significant dependence on $M_\star$ in the  sample. Independently, more massive galaxies formed the bulk of their stellar populations at earlier cosmic times, consistent with the downsizing scenario. These results demonstrate that the SFH quantities employed in the PCC (Sect.~\ref{ppcana}) and RF (Sect.~\ref{RF}) analyses are not independent of the fundamental galaxy properties that shape galaxy evolution. Consequently, throughout the main analysis, we explicitly account for stellar-mass trends when assessing the relationship between SFHs and the inner  DM density profile.

\section{Partial correlation coefficient analysis}
\label{ppcana}

To investigate the direct relationships between physical variables while accounting for mutual dependencies, we performed a PCC analysis (following a methodology similar to that adopted in, e.g. \citealt{Bait, Baker}). This approach quantifies the strength of association between two variables after statistically removing the influence of additional correlated parameters, thereby isolating intrinsic relationships from spurious correlations induced by shared dependencies.

To assess the relative roles of different parameters in shaping the inner DM slope, $\gamma$, we applied the PCC framework to a set of SFH-related quantities, namely $n_{\rm SFR}$, $n_{\rm burst}$, $\Delta_{\rm MS}$, $t_{\rm form}$, $t_{50}$, $t_{90}$, $\sigma_{\rm burst}$ and $f_{\rm burst}$, together with $M_\star$. Stellar mass was included as a control variable because Fig.~\ref{gammaf} indicates a tentative systematic trend between $\gamma$ and $M_\star$, which is expected within the DC14 framework, where the inner slope depends explicitly on the stellar-to-halo mass ratio, and is also partly observed in the gNFW framework. Moreover, some SFH indicators considered here exhibit correlations with stellar mass (Appendix \ref{sfh}), making $M_\star$ an important confounding variable when evaluating whether SFH parameters carry independent information on $\gamma$. We additionally controlled for redshift, since several of the SFH diagnostics evolve significantly with redshift and  may otherwise induce residual redshift-driven correlations with $\gamma$. The PCC analysis thus accounts simultaneously for the effects of both $M_\star$ and $z$ when assessing the relation between $\gamma$ and the SFH diagnostics.

For each SFH parameter $x$, we computed two complementary partial correlations: (i) between $\gamma$ and $x$ while controlling for both $M_\star$ and redshift, and (ii) between $\gamma$ and $M_\star$ while controlling for both $x$ and redshift. These were obtained using a regression-based residual approach, in which the linear dependence of each variable on the conditioning parameters was first removed, and the Pearson correlation was then computed on the resulting residuals. For example, the partial correlation between $\gamma$ and $x$ at fixed $M_\star$ and redshift was obtained by correlating the residuals of both variables after subtracting their best-fitting linear dependence on $M_\star$ and $z$. Conversely, the partial correlation between $\gamma$ and $M_\star$ at fixed $x$ and $z$ was obtained by removing the linear dependence of both variables on $x$ and $z$ before computing their Pearson correlation.

To account for observational uncertainties, we propagated measurement errors using an MC approach. We generated 1000 realisations of the dataset by perturbing all variables within their measurement uncertainties under the assumption of Gaussian errors, and recomputed the PCCs for each realisation. The final values and uncertainties were taken as the median and standard deviation of the resulting distributions.

\begin{figure*}
   \centering
    \includegraphics[width=0.45\textwidth,angle=0,clip=true]{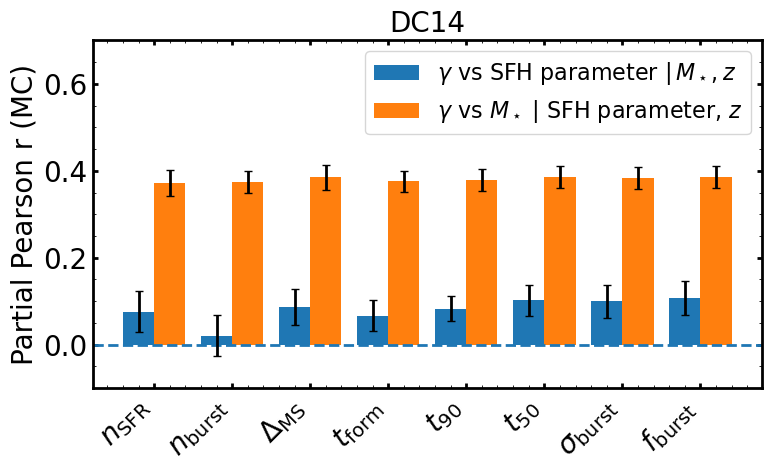}
   \centering  
    \includegraphics[width=0.45\textwidth,angle=0,clip=true]{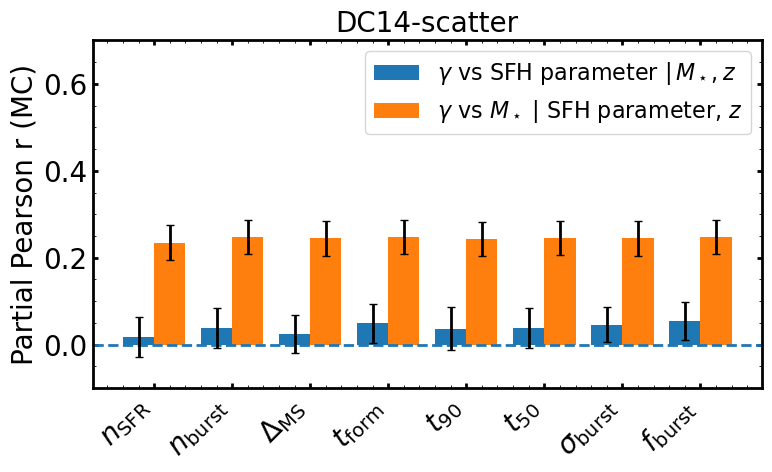}
   \centering   
    \includegraphics[width=0.45\textwidth,angle=0,clip=true]{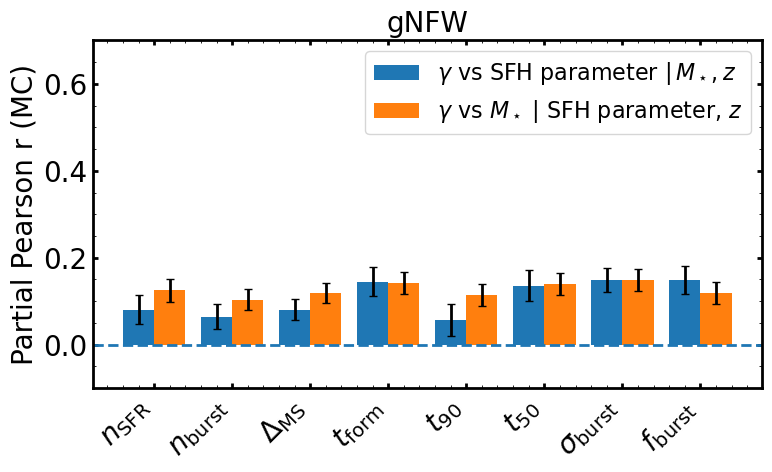}
	 \centering
\caption{Partial correlation analysis of the inner slope $\gamma$ (DC14 fits--top left, DC14-scatter fits--top right, gNFW fits--bottom) with SFH parameters ($n_{\rm SFR}$, $n_{\rm burst}$, $\Delta_{\rm MS}$, $t_{\rm form}$, $t_{50}$, $t_{90}$, $\sigma_{\rm burst}$ and $f_{\rm burst}$) and stellar mass $M_\star$ and $z$. Blue bars show the partial correlation coefficient $r$ between $\gamma$ and individual SFH parameters at fixed $M_\star$ and redshift, testing whether SFH retains a residual linear association with $\gamma$ after accounting for stellar mass and redshift. Orange bars show the partial correlation between $\gamma$ and $M_\star$ at fixed SFH parameter and redshift, testing whether stellar mass remains associated with $\gamma$ after accounting for the corresponding SFH parameter and cosmic evolution. The error bars show the standard deviations across our MC realisations, used to account for measurement uncertainties.}

\label{pcc}
\end{figure*}

The resulting partial correlation coefficients are shown in Fig.~\ref{pcc} for the DC14 fits (top-left), DC14-scatter (top-right) and gNFW fits (bottom). For the DC14 fits, we find a moderate partial correlation between $\gamma$ and stellar mass ($r \sim 0.4$, orange bars), whereas all SFH parameters exhibit negligible partial correlations ($r < 0.15$, blue bars). These results remain after controlling simultaneously for redshift, indicating that, at fixed $M_\star$ and redshift, there is no significant residual dependence of $\gamma$ on the considered SFH parameters. In contrast, a non-negligible correlation between $\gamma$ and $M_\star$ persists even after controlling for both the SFH diagnostic and redshift. This suggests that stellar mass most robustly traces the variance in $\gamma$.

As expected, the DC14-scatter results are fully consistent with the baseline DC14 case, showing a somewhat weaker partial correlation between $\gamma$ and $M_\star$ ($r \sim 0.25$), while remaining negligible for all SFH parameters after controlling for redshift.

For the gNFW fits, we find a weaker mass dependence, with only a low-to-moderate partial correlation ($r < 0.2$) between $\gamma$ and $M_\star$ and, similarly, a low-to-moderate partial correlation between $\gamma$ and any SFH parameter, with no clear dominance of a single driver. These results are obtained after simultaneously accounting for redshift, and therefore indicate that neither the mass nor the SFH correlations show a strong residual dependence on $\gamma$. This differs from the DC14 results, where the dependence on $M_\star$ is more clearly isolated. In the gNFW case, the comparable partial correlations likely reflect the greater freedom of the parametrisation. Overall, this suggests that while the DC14 framework naturally separates mass-driven effects from SFH-driven variations in $\gamma$, the gNFW parametrisation exhibits stronger degeneracies between these quantities, limiting its ability to disentangle their relative contributions.

We note that a caveat of this analysis is that PCC is sensitive only to monotonic relationships and, in the present implementation, removes linear dependencies on the conditioning variables. We nevertheless used it as an initial exploratory tool because the nature of the correlations between our parameters is not known a priori and may be either linear or more complex. To overcome this limitation and capture potential nonlinear dependencies, we also used RF in the main text.

\end{appendix}

\end{document}